\documentclass[fleqn,usenatbib]{mnras}

\usepackage[T1]{fontenc}

\DeclareRobustCommand{\VAN}[3]{#2}
\let\VANthebibliography\thebibliography
\def\thebibliography{\DeclareRobustCommand{\VAN}[3]{##3}\VANthebibliography}

\usepackage{graphicx}	
\usepackage{amsmath}	
\usepackage{amssymb}	
\usepackage{multirow}
\usepackage{subfig}
\usepackage{comment}

\usepackage{newtxtext,newtxmath}
\usepackage{bm}

\title[TNG in the HSC-SSP]{IllustrisTNG in the HSC-SSP: image data release and the major role of mini mergers as drivers of asymmetry and star formation}

\author[Bottrell et al.]{Connor Bottrell,$^{1,2,3}$\thanks{E-mail: connor.bottrell@icrar.org}
Hassen M. Yesuf,$^{1,4}$
Gerg\"o Popping,$^{5}$ \newauthor
Kiyoaki Christopher Omori,$^{6}$ 
Shenli Tang,$^{1,3,7,8}$ 
Xuheng Ding,$^{1,3}$ 
Annalisa Pillepich,$^{9}$\newauthor
Dylan Nelson,$^{10}$
Lukas Eisert,$^{9}$
Hua Gao,$^{11}$
Andy D. Goulding,$^{12}$
Boris S. Kalita,$^{1,3,4}$\newauthor
Wentao Luo,$^{13,14}$
Jenny E. Greene,$^{12}$
Jingjing Shi$^{1,3}$, \&
John D. Silverman$^{1,3,8}$\\
$^{1}$Kavli Institute for the Physics and Mathematics of the Universe (WPI), UTIAS, University of Tokyo, Kashiwa, Chiba 277-8583, Japan\\
$^{2}$International Centre for Radio Astronomy Research, University of Western Australia, 35 Stirling Hwy, Crawley, WA 6009, Australia\\
$^{3}$Center for Data-Driven Discovery, Kavli IPMU (WPI), UTIAS, The University of Tokyo, Kashiwa, Chiba 277-8583, Japan\\
$^{4}$Kavli Institute for Astronomy and Astrophysics, Peking University, Beijing 100871, People’s Republic of China\\
$^{5}$European Southern Observatory, Karl-Schwarzschild-Str. 2, D-85748, Garching, Germany\\
$^{6}$Division of Particle and Astrophysical Science, Nagoya University, Furo-cho, Chikusa-ku, Nagoya 464–8602, Japan\\
$^{7}$Institute for Cosmic Ray Research, The University of Tokyo, 5-1-5 Kashiwanoha, Kashiwa, Chiba 277-8582, Japan\\
$^{8}$Department of Astronomy, School of Science, The University of Tokyo, 7-3-1 Hongo, Bunkyo, Tokyo 113-0033, Japan\\
$^{9}$Max-Planck-Institut f{\"u}r Astronomie, K{\"o}nigstuhl 17, 69117 Heidelberg, Germany\\
$^{10}$Universit{\"a}t Heidelberg, Zentrum f{\"u}r Astronomie, Institut f{\"u}r Theoretische Astrophysik, Albert-Ueberle-Str. 2, 69120 Heidelberg, Germany\\
$^{11}$Institute for Astronomy, University of Hawaii, 2680 Woodlawn Drive, Honolulu HI 96822, USA\\
$^{12}$Department of Astrophysical Sciences, Princeton University, 4 Ivy Lane, Princeton, NJ 08544\\
$^{13}$Institute of Deep Space Sciences, Deep Space Exploration Laboratory, Hefei 230026, China\\
$^{14}$School of Physical Sciences, University of Science and Technology of China, Hefei 230026, China
}

\date{Accepted XXX. Received YYY; in original form ZZZ}

\pubyear{2015}

\begin{document}
\def \nuprocess{$\nu$-process}
\def \nodata{. . .}
\def \degree{$^{\circ}$}
\def \Msolar{M$_{\odot}$}
\def \alphafe{[$\alpha$/Fe]}
\def \na{New Astronomy}
\def \HI{H\ion{I}}
\def \sion{\ion{II}}
\def \vninety{v$_{90}$}
\def \Lbol{L$_{\rm bol}$}
\def \Mstar{M_{\star}}
\def \logMstar{\log(M_{\star}/\mathrm{M}_{\odot})}
\def \logRimp{log(R$_{\rm imp}$/kpc)}
\def \logLbol{log(L$_{\rm AGN}$/erg s$^{-1}$)}
\def \logsSFR{log(\mathrm{sSFR / yr}^{-1})}
\def \kms{km s$^{-1}$}
\def \zabs{z$_{\rm abs}$}
\def \zem{z$_{\rm em}$}
\def \Rimp{$\rho_{\rm imp}$}
\def \Rvir{$\rho_{\rm vir}$}
\def \deltaEW{${\rm \Delta log(EW/m\AA)}$}
\def \RadRat{$f_{\rm AGN}/f_{\rm HM01}$}
\def \mnfe{[Mn/Fe]$_{\rm DC}$}
\def \Mgeqw{W$_{0}^{2796}$}
\def \Feeqw{W$_{0}^{2600}$}
\def \fracMgFe{ ${\rm{W}_{0}^{2796}}$/${\rm{W}_{0}^{2600}}$}
\def \omegaDLA{$\Omega_{\rm H \textsc{i}}$}
\def \ndla{30}
\def \npdla{46}
\def \nlpdla{41}
\def \nxpdla{5}
\def \nmdla{27}
\def \nlmdla{21}
\def \nxmdla{6}
\def \CosmoZ{$\langle Z/Z_{\odot} \rangle$}
\def \fNX{$f(N,X)$}
\newcommand{\gimtwod}{\textsc{gim2d}}
\newcommand{\sersic}{s\'{e}rsic} 
\newcommand{\Sersic}{S\'{e}rsic}
\newcommand{\sextractor}{\textsc{SExtractor}}
\def \rifs{\texttt{RealSim-IFS}}
\def \realsim{\texttt{RealSim}}
\def \tpost{T_{\mathrm{PM}}}
\def \dtc{\Delta T_{\mathrm{Coalescence}}}
\def \dtcoal{\Delta T_{\mathrm{Coal}}}
\def \skirt{\texttt{SKIRT}}
\def \deltasfms{\Delta \langle \log(\mathrm{SFR})\rangle_{\mathrm{MS}}}
\def \dlogsfms{\Delta \langle \log(\mathrm{SSFR})\rangle_{\mathrm{MS}}}
\def \fgas{f_{\mathrm{gas}}}
\def \mstar{M_{\star}}
\def \logsfr{\log({\mathrm{SFR}})}
\def \dlogsfr{\Delta\log({\mathrm{SFR}})}
\def \logssfr{\log({\mathrm{SSFR}})}
\def \dra{\Delta R_A}
\def \dsfr{\Delta \mathrm{SFR}}
\def \dtcoffset{\dtc^\mathrm{Offset}}
\def \dtcoaloffset{\dtcoal^\mathrm{Offset}}
\def \dsfms{\Delta \mathrm{SFMS}}

\label{firstpage}
\pagerange{\pageref{firstpage}--\pageref{lastpage}}
\maketitle

\begin{abstract}
At fixed galaxy stellar mass, there is a clear observational connection between structural asymmetry and offset from the star forming main sequence, $\dsfms$. Herein, we use the TNG50 simulation to investigate the relative roles of major mergers (stellar mass ratios $\mu\geq0.25$), minor ($0.1 \leq \mu < 0.25$), and mini mergers ($0.01 \leq \mu < 0.1$) in driving this connection amongst star forming galaxies (SFGs). We use dust radiative transfer post-processing with SKIRT to make a large, public collection of synthetic Hyper Suprime-Cam Subaru Strategic Program (HSC-SSP) images of simulated TNG galaxies over $0.1\leq z \leq 0.7$ with $\logMstar\geq9$ ($\sim750$k images). Using their instantaneous SFRs, known merger histories/forecasts, and HSC-SSP asymmetries, we show (1) that TNG50 SFGs qualitatively reproduce the observed trend between $\dsfms$ and asymmetry and (2) a strikingly similar trend emerges between $\dsfms$ and the time-to-coalescence for mini mergers. Controlling for redshift, stellar mass, environment, and gas fraction, we show that individual mini merger events yield small enhancements in SFRs and asymmetries that are sustained on long timescales (at least $\sim3$ Gyr after coalescence, on average) -- in contrast to major/minor merger remnants which peak at much greater amplitudes but are consistent with controls only $\sim1$ Gyr after coalescence. Integrating the boosts in SFRs and asymmetries driven by $\mu\geq0.01$ mergers since $z=0.7$ in TNG50 SFGs, we show that mini mergers are responsible for (i) $55$ per cent of all merger-driven star formation and (ii) $70$ per cent of merger-driven asymmetric structure. Due to their relative frequency and prolonged boost timescales, mini mergers dominate over their minor and major counterparts in driving star formation and asymmetry in SFGs.
\end{abstract}

\begin{keywords}
galaxies: general --galaxies: photometry --galaxies: interactions --galaxies: structure --galaxies: star formation --methods: numerical 
\end{keywords}


\section{Introduction}

The defining and most ubiquitous form of data in extragalactic astronomy are images. Using information extracted from images to diagnose the viability of theoretical models is a cornerstone for improving our understanding of how observed galaxies assembled and evolved. However, galaxy formation models do not always lend themselves to even-handed diagnosis at the image-level. To conduct a faithful validation of these models against reality, it is necessary to forward-model theoretical predictions into images that are then directly compatible with real imaging data. 

The class of models that is (1) most appropriate for forward-modelling to image-space and (2) supports comparisons between theoretical predictions and galaxy \emph{populations} is the cosmological hydrodynamical simulation (e.g. \citealt{2014MNRAS.444.1518V,2014MNRAS.444.1453D,2015MNRAS.446..521S,2016MNRAS.455.2778F,2018MNRAS.473.4077P,2019MNRAS.486.2827D,2021A&A...651A.109D,2023MNRAS.522.3831F,2022arXiv221010059H}). These simulations numerically solve equations for hydrodynamics and gravity to explicitly track the co-evolution of dark and baryonic material across cosmic time. While the details of the numerical methods and galaxy physics models used in these simulations can differ greatly, each crucially predicts the evolution of the distribution, dynamics, and characteristics of gas and stars in and around galaxies. As a result, simulation predictions can be easily forward modelled, at zeroth order, by creating a 2D projection of gas/star surface density for a given galaxy. Alternatively, ray-tracing radiative transfer (RT) can be used to self-consistently generate a realization of what an observer would see if they had captured an image of the simulated galaxy with their instrument (e.g. \citealt{2015MNRAS.447.2753T,2017MNRAS.470..771T,2019MNRAS.483.4140R,2020MNRAS.494.5636W,2022MNRAS.510.3321P,2023ApJ...946...71C}). A number of RT tools exist for this purpose (e.g. \citealt{2006MNRAS.372....2J,2011A&A...536A..79R,2020A&C....3100381C,2021ApJS..252...12N}).


There are several key areas in which synthetic images (and other observables) of galaxies generated from cosmological simulations are useful, including: (1) direct comparison of galaxy formation theory with observations at the image level (e.g. \citealt{2015MNRAS.454.1886S,2016MNRAS.462.1470L,2017MNRAS.467.1033B,2017MNRAS.467.2879B,2019MNRAS.483.4140R,2019MNRAS.489.1859H,2021MNRAS.501.4359Z,2022MNRAS.511.2544D,2023arXiv230207277V}; Eisert et al. in prep); (2) interpretation of observed galaxy features, scaling relations, and sensitivity limits (e.g. \citealt{2021ApJ...919..139W,2021MNRAS.507..886T,2022MNRAS.516.4354W,2022MNRAS.509.2654V,2022MNRAS.515.3406M,2023ApJ...950...56H,2023ApJ...946...71C,2023A&A...672A..51R,2023arXiv230409202Z}); (3) learning mappings between galaxy images and the historical/meta information for galaxies that is only accessible from the simulation (e.g. \citealt{2019MNRAS.490.5390B,2020ApJ...895..115F,2021MNRAS.506..677C,2021MNRAS.504..372B,2023MNRAS.519.2199E,2021ApJ...912...45N,2022MNRAS.515.3938S,2023MNRAS.522....1N,2023MNRAS.519.4920G,2023MNRAS.523.5408A,2023arXiv230915539O}); and (4) as input/priors for training generative and score-based models (e.g. \citealt{2020MNRAS.496.2346M,2022mla..confE..48A,2022MNRAS.515..652H}; Adam et al. in prep). In each of these areas, the important advantage of RT-generated images is that they directly model emission, i.e., light. The process of generating light from simulation data can incorporate information about stellar populations, birth clouds and nebular emission, intervening gas/dust, etc. This higher level of fidelity increases the scope for contrasting real and synthetic observations. Meanwhile, it also enables more refined connections between observables (e.g. wavelength-dependent morphologies) and the \emph{information} about galaxy formation/assembly histories that those observables encode (e.g. \citealt{2020MNRAS.493.4551O,2023RASTI...2...78F}).

One such connection is the correlation between the star formation rates (SFRs) and total stellar masses, $\mstar$, of star forming galaxies (SFGs) -- known as the star forming main sequence (SFMS, e.g. \citealt{2004MNRAS.351.1151B,2007ApJS..173..267S,2007ApJ...660L..43N}). The existence of this relation over a large dynamic range in $\Mstar$ and the relatively tight vertical stratification ($\sim 0.3$ dex scatter) suggests that observed SFGs exist in a dynamic state of quasi-equilibrium between \emph{in-situ} stellar assembly, \emph{ex-situ} accretion of gas and stars, and negative, regulatory feedback (e.g. \citealt{2010ApJ...718.1001B,2011MNRAS.415...11D,2016MNRAS.455.2592R,2019MNRAS.487..456B}). However, at fixed stellar mass, a detailed account of which physical processes drive offsets from the main sequence, $\dsfms$, remain elusive. Numerical simulations and observational work agree that $\dsfms$ is sensitive to gas fraction, phase distribution, and depletion times (e.g. \citealt{2015ApJ...800...20G,2015ApJ...812L..23S,2018ApJ...867...92S,2016MNRAS.457.2790T,2016MNRAS.462.1749S,2018MNRAS.477L..16T,2020ApJ...892...87W,2020ApJ...895...25W,2020MNRAS.493L..39E,2023ApJ...947...61S}). Meanwhile, $\dsfms$ correlates with several structural characteristics -- revealing that, at fixed stellar mass, morphological information extracted from images is predictive of the vertical stratification of galaxy SFRs (e.g. \citealt{2001AJ....122.1861S,2003MNRAS.341...54K,2004MNRAS.351.1151B,2011ApJ...742...96W,2012ApJ...760..131C,2015MNRAS.448..237W,2020MNRAS.492...96B,2020ApJ...889...14Y,2021ApJ...923..205Y}).

In particular, \cite{2021ApJ...923..205Y} conducted a comprehensive analysis (and ranking) of the connections between various structural parameters and $\dsfms$ at fixed stellar mass for SFGs at $z\sim0.1$. These included residual asymmetry (defined later in Equations \ref{eq:asym} and \ref{eq:rasym}; \citealt{1994ApJ...432...75A,1995ApJ...451L...1S,2002ApJS..142....1S}); S\'ersic index \citep{1963BAAA....6...41S,2005PASA...22..118G}; Concentration \citep{1994ApJ...432...75A}; half-light radius $R_{50}$; effective mass surface density $\mu=0.5\Mstar/(\pi R^2_{50})$, and stellar velocity dispersion. Their analysis offered two key insights with respect to the utility of each structural parameter as a predictor for $\dsfms$ at fixed stellar mass: morphological asymmetry (1) exhibits the smallest degree of correlation with any other structural parameter and (2) is the best predictor of $\dsfms$ of the structural parameters considered. Crucially, asymmetry was the only structural parameter to which $\dsfms$ responded monotonically, on average, at any given $\Mstar$.

In a turbulent interstellar gas medium, the spatial locations of star formation sites are generally not symmetric under azimuthal rotation and are locally bright in optical bands (e.g. \citealt{1959ApJ...129..243S,1998ApJ...498..541K,2013ApJ...776....1K,2017MNRAS.465.1682H,2019ApJ...884L..33L,2023MNRAS.519.1452W}). Consequently, at least some measured asymmetry in SFGs is expected to be unrelated to processes that are driving star formation and, rather, symptomatic of it (e.g. trivially tracing galaxy colour). This \emph{non-structural} contribution to measured asymmetries may be mitigated in redder optical bands which more closely probe the stellar mass distribution (e.g. the Sloan Digital Sky Survey $i$-band; \citealt{1996AJ....111.1748F,1998AJ....116.3040G,2008ApJ...677..186R}). Contributors to \emph{structural} asymmetry include broader morphological characteristics such as stellar warps and lopsidedness -- which are observationally ubiquitous amongst local Universe SFGs and correlate with SFRs (e.g. \citealt{1980MNRAS.193..313B,1994A&A...288..365B,2013ApJ...772..135Z,2021ApJ...923..205Y,2023MNRAS.526..567D}). This link suggests that the phenomena that drive differences in asymmetries between galaxies are simultaneously responsible for the difference in their SFRs. 

Two plausible mechanisms for driving the observed connection between $\dsfms$ and asymmetry are (1) external gas inflows i.e. accretion and (2) galaxy-galaxy interactions and mergers (e.g. \citealt{2008A&ARv..15..189S,2009PhR...471...75J}). However, their contributions to asymmetry and star formation are not well constrained, observationally. Within sensitivity limits, the mass accretion rates of external atomic gas are a factor of $5-10$ times smaller than the current SFRs of local SFGs \citep{2014A&A...567A..68D,2021ApJ...923..220D}. And while similar-mass galaxy mergers are known triggers of enhanced asymmetry and SFRs \emph{in both merging galaxies and their remnants} (e.g. \citealt{1977egsp.conf..401T,1985AJ.....90..708K,1992ARA&A..30..705B,2007ApJ...666..212D,2013MNRAS.435.3627E,2016MNRAS.461.2589P,2020ApJ...895..100X,2022MNRAS.516.1462T,2022MNRAS.514.3294B,2023MNRAS.519.4920G}), their role in the vertical stratification of the SFMS and in the ubiquity of asymmetric structure is more contentious (e.g. \citealt{2008A&ARv..15..189S,2009PhR...471...75J,2013ApJ...772..135Z,2014MNRAS.443L..49P,2014MNRAS.437L..41K,2019MNRAS.485.5631C,2022ApJ...931...34F,2022A&A...661A..52P}). 

To have an important role in the relationship between asymmetry and $\dsfms$, a process must: (1) trigger asymmetry/SFR-enhancing events that occur frequently enough that a significant fraction of the SFG population are \emph{currently} affected and/or (2) yield changes that are sustained on timescales that are at least comparable to the time between events. In this context, the main problem invoked for \emph{major} mergers (mass ratios $\geq1:4$) is that they are rare in the local Universe (e.g. \citealt{2004ApJ...617L...9L,2007ApJ...666..212D,2008ApJ...685..235P,2009A&A...501..505L,2014MNRAS.444.3986R}). In contrast, sub-major mergers are relatively frequent. If sub-major mergers trigger asymmetry/SFR boosts that survive on timescales similar to the time between consecutive merger events, then they may have a more influential role in the relations between $\dsfms$ and asymmetry observed in SFGs. 

In this work, we seek to characterise the roles of mergers in various mass ratio regimes and their relative impacts on the relation between SFG asymmetries and SFRs. Specifically, we examine the relative roles of mergers in three mass ratio regimes: major $\mu\geq0.25$; minor $0.1\leq \mu<0.25$, and mini $0.01\leq \mu<0.1$; where $\mu$ is the stellar mass ratio between the less massive and more massive galaxy. To do this, we use dust radiative transfer to forward-model an exceptionally large number of galaxies from $0.1\leq z \leq0.7$ in the TNG cosmological magneto-hydrodynamical simulations \citep{2018MNRAS.473.4077P,2019ComAC...6....2N} into realistic synthetic images ($\sim750$k images in total). We make the imaging data publicly available so that they may be used in the full scope of applications laid out at the beginning of this section. The imaging data includes both idealized photometry (pristine, noiseless, and high-resolution) and survey-realistic Hyper Suprime-Cam Subaru Strategic Program images made using final-depth images from the third data release (HSC-SSP PDR3 \citealt{2022PASJ...74..247A}) -- which we choose for its premier combination of depth, resolution, and large survey footprint. The survey-realistic images are central to this investigation on two counts: (1) they allow us to investigate the sensitivity of measured asymmetries to \emph{known} major/minor/mini merger histories/forecasts under realistic observational conditions and (2) the connections we establish to observables are directly compatible with observations. 

In Section \ref{sec:data} we detail the forward-modelling of TNG galaxies and the synthetic images. Section \ref{sec:measurements} then describes our surface brightness fitting and asymmetry measurements, characterization of TNG galaxy merger histories and environments, SFMS fitting, and the control-matching procedures used to isolate the effects of mergers. Readers who are more interested in the scientific component of this work are directed to Section \ref{sec:results} -- which lays out our results. We first establish whether the simulations \emph{can} be used to interpret the observed relationship between $\dsfms$ and asymmetry. With the justification that the observed trend is reproduced, we then investigate the relative roles major/minor/mini mergers in driving this trend. We discuss our results in context of observations and other simulations in Section \ref{sec:discussion}. Our measurements and calculations adopt a Planck 2015 -based cosmology with $H_0 = 100 h$ km s$^{-1}$ Mpc$^{-1}$ with $h=0.6774$, $\Omega_m = 0.308$, and $\Omega_{\Lambda} = 0.692$ \citep{2016A&A...594A..13P}, where relevant.

\section{Simulation data and forward-modelling}\label{sec:data}

To interpret the observed relationship between asymmetry and star formation in galaxies, we forward-model galaxies from $0.1 \leq z \leq 0.7$ in the IllustrisTNG (TNG) cosmological magneto-hydrodynamical simulation suite \citep{2019ComAC...6....2N} into survey-realistic images from the HSC-SSP PDR3 Wide Layer \citep{2022PASJ...74..247A} using the Monte Carlo dusty radiative transfer software \texttt{SKIRT} \citep{2020A&C....3100381C}. With these in hand, we fit simple single-S\'ersic models to the 2D surface brightness distributions of each forward-modelled galaxy using the \texttt{GALIGHT} parametric decomposition tool \citep{2020ApJ...888...37D}. Non-parametric morphological statistics, including asymmetries \citep{1994ApJ...432...75A,1996ApJS..107....1A}, are computed for each galaxy as well as asymmetries derived from S\'ersic model-subtracted residuals \citep{1995ApJ...451L...1S}. We use these measured asymmetries alongside the instantaneous SFRs and merger histories/forecasts of TNG galaxies to investigate the degree to which mergers link asymmetry and SFRs.

\subsection{IllustrisTNG simulations}\label{sec:tng}

The TNG simulations\footnote{\url{https://www.tng-project.org}} are a suite of publicly available cosmological magneto-hydrodynamical simulations of galaxy formation \citep{2018MNRAS.475..648P,2018MNRAS.475..676S,2018MNRAS.475..624N,2018MNRAS.477.1206N,2018MNRAS.480.5113M}. The simulations track the co-evolution of gas, stars, dark matter, supermassive black holes, and magnetic fields from $z=127$ to $z=0$ using the AREPO moving-mesh hydrodynamic and gravity solver \citep{2010MNRAS.401..791S} and a galaxy formation model that is common to all simulations in the suite \citep{2017MNRAS.465.3291W,2018MNRAS.473.4077P} and builds upon the original Illustris model \citep{2013MNRAS.436.3031V,2014MNRAS.438.1985T}. In brief, the salient components of the TNG physical model are: (1) primordial/metal-line radiative gas cooling/heating in the presence of an evolving ultraviolet background; (2) stochastic star formation in a two-phase, effective interstellar medium (ISM) above a threshold gas density $n_{\mathrm{H}} \simeq 0.1$ cm$^{-3}$; (3) evolution of stellar populations and associated pressurization and chemical enrichment of their ambient ISM by supernovae (SNe) and asymptotic giant branch (AGB) stars; (4) galactic-scale outflows driven by SNe via a kinetic wind scheme; (5) seeding and growth of supermassive black holes (SMBH); (6) SMBH feedback in (i) a high-accretion rate ``thermal'' mode that continuously injects thermal energy into gas around black holes and (ii) a low-accretion rate ``kinetic'' mode that launches outflows by injecting kinetic energy in a pulsed, directed fashion; and (7) primordial seeding and evolution of magnetic fields assuming ideal magneto-hydrodynamics.

The TNG suite includes simulations run in three periodic, cubic volumes: $V_{\mathrm{box}} = (51.7^3,\;106.5^3,\;302.6^3)$ cMpc$^3$, respectively called TNG50, TNG100, and TNG300, using the same physical model. From small to large, these simulations have descending levels of resolution. In the context of galaxy structure, the resolution of the simulations can be broadly summarized by two quantities: (1) the Plummer-equivalent gravitational softening length for collisionless particles, $\epsilon_{\star,\mathrm{dm}} = (0.29,\;0.74,\;1.48)$, and (2) the mean size of star-forming gas cells $\bar{r}_{\mathrm{cell,sf}} = (0.138,\;0.355,\;0.715)$ ckpc. On scales smaller than $\epsilon_{\star,\mathrm{dm}}$, gravitational forces between stellar and dark matter particles are approximate. Meanwhile, $\bar{r}_{\mathrm{cell,sf}}$ approximates the minimum scale on which gas dynamics can be resolved by the simulations and below which galaxy astrophysics processes such as cooling/heating, turbulence, star formation, and feedback are sub-grid.

In this work, we focus on the TNG50 simulation \citep{2019MNRAS.490.3234N,2019MNRAS.490.3196P} due to its capacity to resolve structures on similar scales to HSC-SSP observations at $z\geq0.1$. The median seeing full-width at half-maximum (FWHM) in HSC-SSP $i$-band co-added images is 0.61 arcsec, i.e. $1.2$ kpc at $z=0.1$ \citep{2018PASJ...70S...1M}. TNG50 is the only simulation in the TNG suite that is capable of sampling the HSC-SSP seeing at the Nyquist rate down to $z=0.1$ for both $\epsilon_{\star,\mathrm{dm}}$ and $\bar{r}_{\mathrm{cell,sf}}$. In other words, the information that can be extracted on scales smaller than the HSC-SSP seeing are similar for observed HSC-SSP galaxies and seeing-convolved TNG50 galaxies for $z\geq0.1$. However, it is important to note that below approximately $\bar{r}_{\mathrm{cell,sf}}$ in the simulated data, there are no real physics or information to extract.

We select galaxies from TNG50 with total stellar masses $\logMstar\geq9$ from simulation snapshots in the redshift range $0.1\leq z \leq0.7$ for forward-modelling. This includes all galaxies from the $0.1\leq z \leq0.4$ snapshots and, additionally, the $z=(0.5,0.6,0.7)$ snapshots. The stellar mass cut only applies to the selection of our galaxy sample and does not apply to the progenitors or companions of these galaxies when characterizing their merger histories (Section \ref{sec:mergers}). The $\logMstar\geq9$ cut ensures that each galaxy in the TNG50 sample comprises at least $10^4$ stellar particles and, consequently, their stellar structures are (1) at least reasonably resolved and (2) not strongly affected by spurious dynamical heating by dark matter particles \citep{2021MNRAS.508.5114L,2023arXiv230605753L}. Beyond $z\sim0.7$, the completeness at $\logMstar\geq9$ drops precipitously in the HSC-SSP Wide Layer (see Figure 4 of \citealt{2020ApJ...904..128I}). We therefore take $z=0.7$ as the limiting redshift of our forward-modelling sample with $\logMstar\geq9$. Lastly, we use the \texttt{SubhaloFlag} field from the TNG group catalogues to omit ``galaxies'' whose origin is non-cosmological (see Section 5.2 of \citealt{2019ComAC...6....2N}).

Otherwise, our selection includes both centrals and satellites with no restrictions on galaxy properties. The resulting TNG50 sample comprises $58,737$ galaxies over $0.1 \leq z \leq 0.4$ and an additional $8,006$ from the $z=(0.5,0.6,0.7)$ snapshots. While the analysis component of this study focuses on the TNG50 simulations, we also generate images for $124,686$ TNG100 galaxies with $\logMstar\geq10$ and \texttt{SubhaloFlag=1} for $0.1\leq z \leq0.4$ as part of the image data release. 

\subsection{Radiative transfer with SKIRT}\label{skirt}

\begin{figure*}
\centering
	\includegraphics[width=\linewidth]{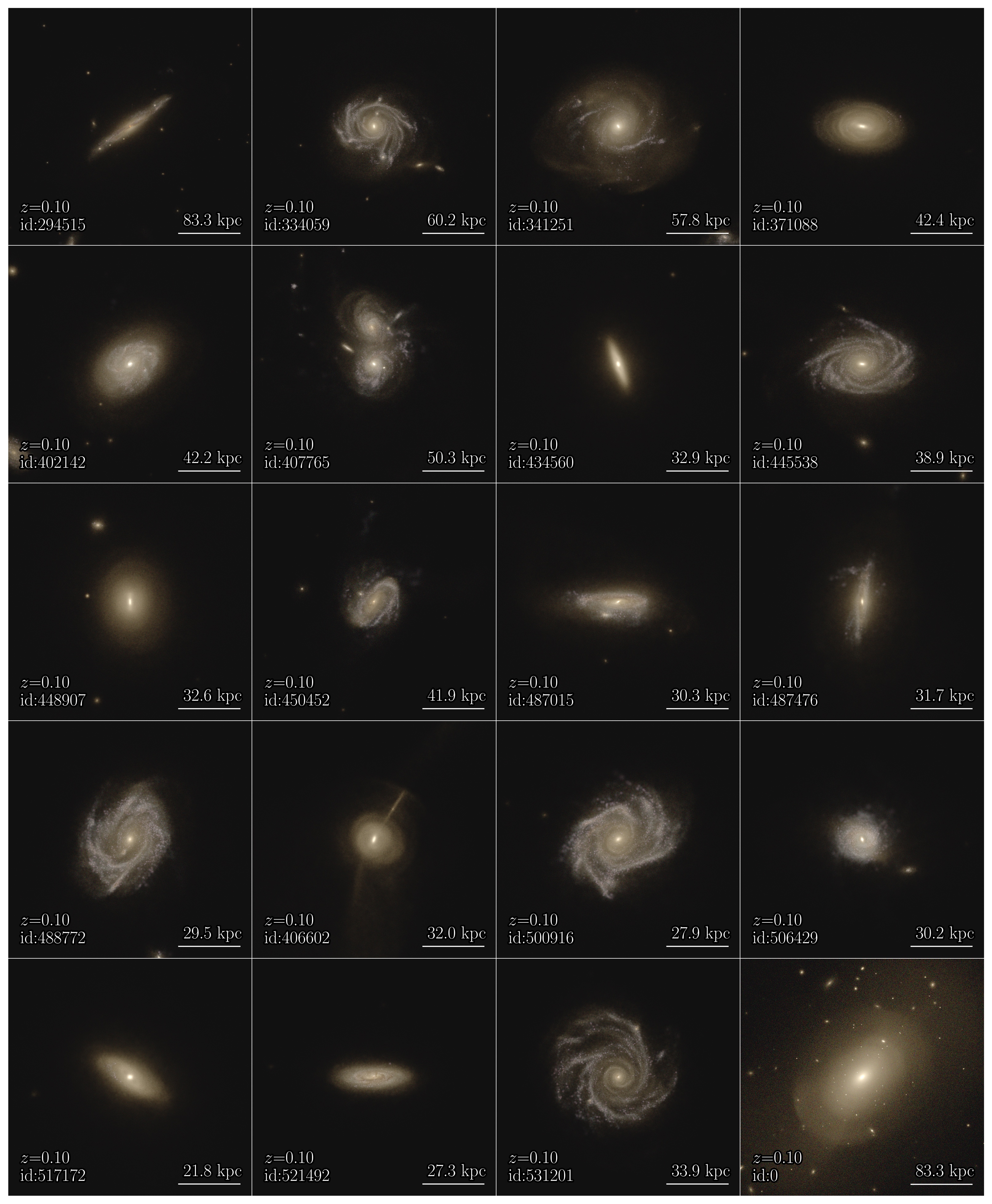}
   \caption[Idealized HSC $gri$ image mosaic $z=0.1$]{Mosaic of idealized HSC $gri$-colour images for 20 selected TNG50 galaxies with $\logMstar\geq10.5$ at $z=0.1$ (also viewable in the Infinite Gallery on the TNG project website: \url{https://www.tng-project.org/explore/gallery}). The idealized images are the direct output from \skirt{} and contain no seeing effects and no noise other than the Monte Carlo noise from the radiative transfer simulation. The spatial resolution of each image is 100 pc per pixel. \skirt{} was run using cutouts from the friends-of-friends group and therefore include light from each galaxy's group members, physical companions, and from intra-cluster light (ICL) in galaxy clusters. SubfindIDs are indicated at the lower left of each panel and the scale of each FoV is given on the lower right. The images are zoomed in by $150$ per cent from their original fields of view.}
    \label{fig:mosaic_idealized_91}
\end{figure*}

\begin{figure*}
\centering
	\includegraphics[width=\linewidth]{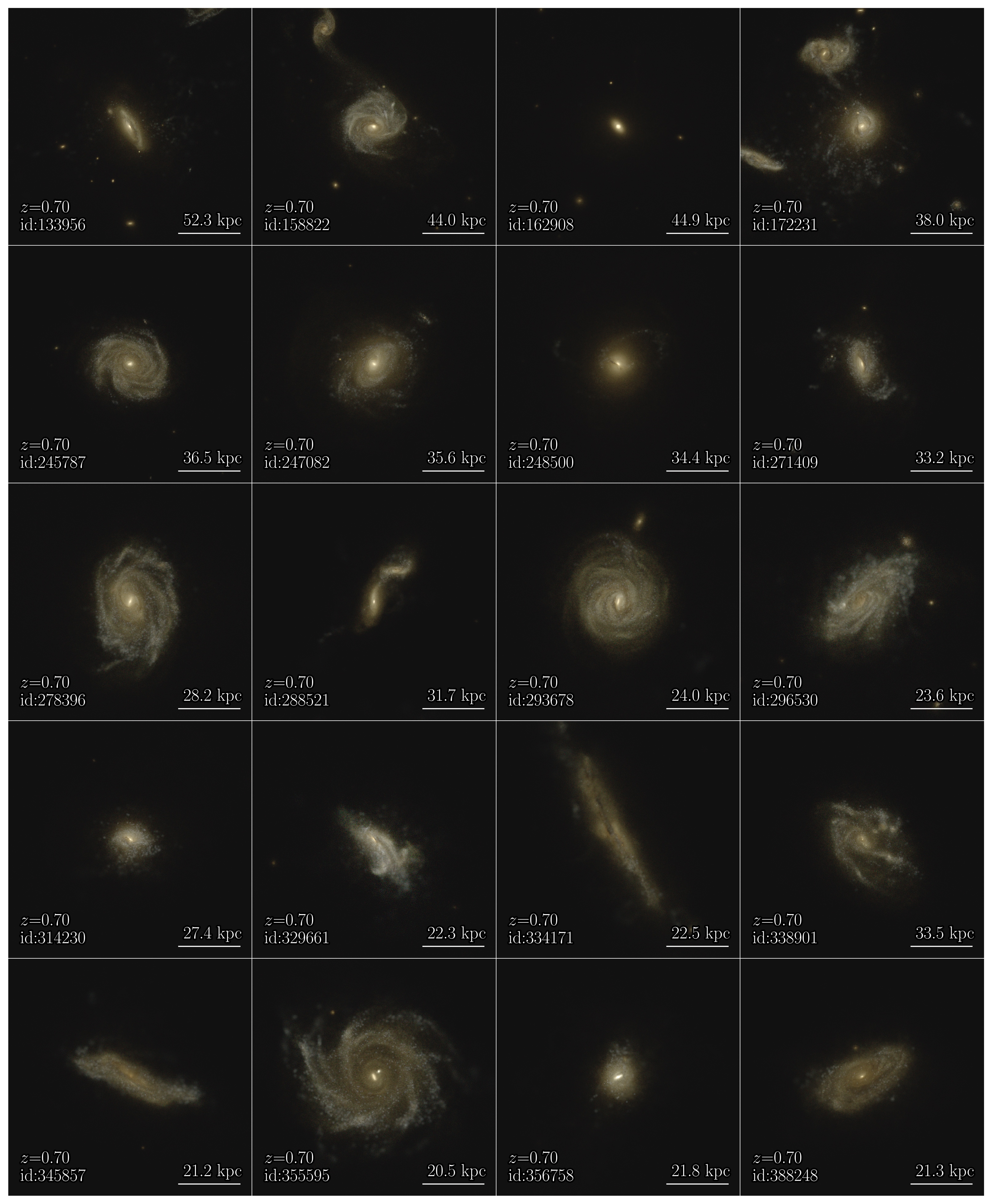}
   \caption[Idealized HSC $riz$ image mosaic $z=0.7$]{Mosaic of idealized HSC $riz$-colour images for 20 selected TNG50 galaxies with $\logMstar\geq10.5$ at $z=0.7$. See Figure \ref{fig:mosaic_idealized_91} caption for details on idealized images. The images are zoomed in by $150$ per cent from their original fields of view.}
    \label{fig:mosaic_idealized_59}
\end{figure*}

TNG galaxies are first forward-modelled into physically realistic but noise-free (idealized) synthetic images using \texttt{SKIRT}\footnote{\url{https://skirt.ugent.be}} Monte Carlo radiative transfer \citep[\texttt{SKIRT} version 9]{2020A&C....3100381C}. In our \texttt{SKIRT} simulations, we adopt the default wavelength bias factor of 0.5. In other words, photon packet wavelengths have a 50 per cent chance of being sampled from the source spectrum based on the source spectral energy distribution (SED) and a 50 per cent chance of being sampled uniformly from a wavelength range set by $\lambda_{\mathrm{min}}$ and $\lambda_{\mathrm{max}}$. These rest-frame emission wavelength limits are optimized for our set of observer-frame bandpasses and a given source redshift. For our broad-band images, we do not use the internal kinematics functionality of \skirt{} to account for Doppler shifts in emitted and scattered light. The spatial pixel scale for all idealized images is 100 pc.


For each galaxy in our sample, the transfer simulation is run using star and gas particle data taken from its friends-of-friends (FoF) group within a spherical volume. The radius of this sphere is specified by a scaling of either (1) the radius containing half of the galaxy's stellar mass, $R^{\star}_{\mathrm{half}}$, or (2) the radius containing half of the galaxy dark matter mass, $R^{\mathrm{dm}}_{\mathrm{half}}$. Specifically, the field-of-view (FoV) for the images is set to the largest of $(20 \times R^{\star}_{\mathrm{half}},\; 2 \times R^{\mathrm{dm}}_{\mathrm{half}},\; 50\; \mathrm{kpc})$ up to a maximum of 500 kpc. These large FoV options are chosen such that extended structure, satellites, and nearby group/cluster members are included in the transfer volume and resulting images. We therefore take $\sqrt{2}/2$ times the FoV as the radius of the sphere from which TNG star and gas particle data are extracted for the transfer simulations. We set the number of photon packets used in the transfer simulation to $100$ times the total number of stellar particles in the extracted volume -- requiring at least $10^8$ packets and up to at most $10^9$ packets based on our initial convergence tests. 

\subsubsection{HSC-SSP$^+$ broadbands} \label{sec:bands}

Idealized synthetic images of each galaxy are produced in the HSC $grizy$ bands \citep{2018PASJ...70...66K} and additionally the Canada-France-Hawaii Telescope (CFHT) large area deep survey (CLAUDS) $u$-band \citep{2019MNRAS.489.5202S} and CFHT MegaCam $r$-band \citep{2013A&A...552A.124B}. For TNG50 galaxies from snapshots corresponding to $z=\{0.1,0.2,0.3,0.4,0.5,0.6,0.7\}$, images are also generated using the United Kingdom Infrared Infrared Telescope (UKIRT) Infrared Deep Sky Survey (UKIDSS) $JHK$-bands  \citep{2006MNRAS.367..454H} and Spitzer Infrared Array Camera (IRAC) channels $1-3$ \citep{2004ApJS..154...10F}. Each galaxy is observed along four lines-of-sight aligned with the arms of a tetrahedron. The filter response functions of each instrument are taken from Spanish Virtual Observatory (SVO) Filter Profile Service\footnote{\url{http://svo2.cab.inta-csic.es/svo/theory/fps/}} \citep{2012ivoa.rept.1015R,2020sea..confE.182R}. The responses incorporate the instrument, filter, and atmospheric components where applicable\footnote{Note that the ground-based instruments provide system response curves at slightly different fiducial airmasses: 1.2  (HSC), 1.25 (CLAUDS, MegaCam) and 1.3 (UKIDDS).}. These \emph{response} curves, $R( \lambda )$, are necessarily converted to \emph{transmission} curves via $T( \lambda ) = \lambda R(\lambda)$ as a preprocessing step following the \skirt{} documentation. 

Photon packets from stellar particles (both young and old) are drawn stochastically from a volume defined by cubic spline kernel whose characteristic size is set by the distance to the $32^{\mathrm{nd}}$ nearest neighbouring stellar particle \citep{1992ARA&A..30..543M}. The hard limits for the rest-frame emission wavelengths of these packets are $(\lambda_{\mathrm{min}}, \lambda_{\mathrm{max}}) = (0.3, 5.0) \micron$. We therefore do not model dust re-emission because it is computationally expensive and contributes negligibly at $\lambda\lesssim5 \micron$ \citep{2020MNRAS.497.4773S}. Any observer-frame bandpass (laid out Section \ref{sec:bands}) that has partial non-zero response outside of $(\lambda_{\mathrm{min}}, \lambda_{\mathrm{max}})\times(1+z)$ is omitted from the bandpass list for that redshift, $z$. As a result, only Spitzer IRAC C3 is commonly omitted from the lower redshift transfer simulations. Lastly, the rest-frame range $(\lambda_{\mathrm{min}}, \lambda_{\mathrm{max}})$ is further clipped to ensure that no observer frame photon packets arrive outside the non-zero response range of the remaining bandpasses (i.e. the left edge of the CLAUDS $u$-band and the right edge of IRAC C2/C3). 

The idealized images are in Flexible Image Transport System (FITS) format, include comprehensive headers, and are archived at \url{https://www.tng-project.org/data}. The full sets of idealized synthetic images for TNG50 and TNG100 are 7.5 TB and 41 TB, respectively, and can be accessed individually. Figure \ref{fig:mosaic_idealized_91} (and \ref{fig:mosaic_idealized_59}) shows a mosaic of idealized synthetic HSC $gri$ ($riz$) colour images for 20 TNG50 galaxies from $z=0.1$ ($0.7$). Details of the model for emission, scattering, and attenuation in our \skirt{} simulations are laid out in the following two subsections. 

\subsubsection{Stellar light and birth clouds}

The source spectra of unresolved stellar populations (stellar particles) older than $10$ Myr are modelled with \cite{2003MNRAS.344.1000B} stellar population template spectra using a \cite{2003PASP..115..763C} initial mass function. Radiation from the birth clouds of younger stellar populations ($<10$ Myr) is modelled with the \texttt{MAPPINGS III} SED library \citep{2008ApJS..176..438G} and include dust-attenuated stellar light from HII and photo-dissociative regions (PDRs), as well as nebular and dust continuum and line emission. The \texttt{MAPPINGS III} SED of a given stellar particle is set by the (1) SFR of the HII region, (2) birth cloud metallicity, (3) birth cloud compactness, (4) ambient ISM pressure, and (5) PDR dust covering fraction. The SFR for each young stellar particle is computed as SFR$=m_i / (10$ Myr$)$, where $m_i$ is the initial (birth) mass of the stellar particle. The metallicity of the stellar particle at formation time is adopted as the birth cloud metallicity. The density of the birth cloud is characterized by a compactness parameter, $C$. The compactness of the birth cloud affects the dust temperature and subsequent emission at $\lambda\gtrsim5\micron$. Changes to compactness have a reportedly negligible effect on rest-frame UV/optical/NIR emission and we therefore adopt a constant fiducial value of $\log C = 5$ \citep{2008ApJS..176..438G}. Similarly, \cite{2008ApJS..176..438G} show that the ambient pressure of the ISM on the birth cloud has little affect on UV/optical/NIR spectra and we again adopt a fiducial constant $\log(P_{\mathrm{ISM}}/k_{\mathrm{B}}/ \mathrm{cm}^{-3}\mathrm{K}) = 5$. 

Our prescriptions for items (1)-(4) of the \texttt{MAPPINGS III} SED library parameters are identical to those adopted by previous works forward modelling TNG galaxies with \texttt{SKIRT} version 8 (e.g. \citealt{2019MNRAS.483.4140R,2020MNRAS.492.5167V,2020MNRAS.497.4773S})\footnote{see Appendix A of \citet{2020MNRAS.497.4773S} for a comparison of various \texttt{SKIRTv8} runs for TNG galaxies.}. The exception is our choice for (5) the PDR dust covering fraction, $f_{\mathrm{PDR}}$: the time-averaged fraction of the HII region that is covered by the PDR. When $f_{\mathrm{PDR}}=0$, the HII region is completely uncovered. When $f_{\mathrm{PDR}}=1$, the HII region is assumed to be completely surrounded by the PDR. By adopting $f_{\mathrm{PDR}}>0$, it is implicitly assumed that the PDR (and corresponding dust) is partially internal to the volume from which \skirt{} photon packets are emitted and that attenuation from this PDR dust is accounted for in the \texttt{MAPPINGS III} source spectrum. In this case, the metal mass of PDR component is commonly computed as ten times the birth mass of the young stellar particle \citep{2008ApJS..176..438G,2010MNRAS.403...17J,2017MNRAS.470..771T}. However, by adopting this prescription, metal mass is not conserved and dust is subsequently double-counted: first in the shape of the \texttt{MAPPINGS III} spectrum when $f_{\mathrm{PDR}}>0$ and second when the emitted light is attenuated by dust in the gas cells surrounding birth cloud emission volume.

\cite{2017MNRAS.470..771T} mitigate double-counting of dust by introducing `ghost' particles overtop each young stellar particle that contribute negatively to the local dust density. Similarly, \cite{2022MNRAS.510.3321P} downscale the metal content of gas cells surrounding young stellar particles such that metal mass is conserved. However, both (1) rely sensitively on the assumption of how the PDR mass scales with the mass of young stars in the birth cloud and (2) alter the dust medium for radiation emitted by other sources in the transfer simulation. To avoid both of these issues, we therefore set $f_{\mathrm{PDR}}=0$ and model all dust attenuation for young stellar particles (including the PDR) via the transfer medium. In other words, \texttt{MAPPINGS III} handles emission lines and nebulae and the \skirt{} medium system handles the dust. 

We compared our $f_{\mathrm{PDR}}=0$ approach to the approach taken by  \cite{2022MNRAS.510.3321P} which adopts $f_{\mathrm{PDR}}=0.2$ and uses local metal mass-downscaling for a few high-SFR galaxies in TNG50 and found that: (1) nebular emission lines are up $4$ per cent brighter using our approach; (2) continuum emission across $0.4\micron<\lambda<5\micron$ is approximately $1$ per cent lower using our approach; and (3) our approach produces UV continuum emission that is brighter for $\lambda<0.4\micron$, with the offset gradually rising to $4$ per cent at $0.1\micron$ -- which was the limit of our emission range for these tests. Given the small scale of these offsets, we proceed with $f_{\mathrm{PDR}}=0$ for particles representing the birth clouds of young stars in our transfer simulations.

\subsubsection{Dust attenuation and scattering}

The evolution of dust is not explicitly tracked in the TNG hydrodynamical simulations (however, see \citealt{2016MNRAS.457.3775M,2017MNRAS.468.1505M} for efforts in this area). A post-processing model for the relationship between the dust and gas properties is therefore needed. While the detailed behaviour of this relationship is still unclear, observational results converge on a positive correlation between the dust-to-gas mass density ratio (DTG) and gas-phase metallicity -- albeit with pronounced scatter \citep{1998ApJ...493..583V,2007ApJ...663..866D,2011A&A...532A..56G,2014A&A...563A..31R,2014ApJ...792...75Z}. As in \cite{2022MNRAS.510.3321P}, we use a redshift-independent dust post-processing model in which the dust-to-metal mass ratio (DTM) in each gas cell scales with its metallicity. The model is motivated by empirical scalings between these properties derived for local galaxies by \cite{2014A&A...563A..31R}. It is assumed that the solar oxygen abundance and total metal mass fraction are $12+\log($O/H$)_{\odot} = 8.69$ and $Z_{\odot}=0.014$, respectively \citep{2009ARA&A..47..481A}, and that $\log(\mathrm{O/H})_{\mathrm{cell}} -  \log(\mathrm{O/H})_{\mathrm{\odot}} = \log(Z_{\mathrm{cell}}/Z_{\odot})$. The gas cell metallicity can then be expressed:
\begin{align}
12 + \log(\mathrm{O/H})_{\mathrm{cell}} = \log(Z_{\mathrm{cell}}/0.014) + 8.69
\end{align}
and converted directly into a DTG via the empirical broken power law from \citet[Table 1, Column 3]{2014A&A...563A..31R}:
\begin{align}
\log(\mathrm{DTG}) = 
    \begin{cases}
       12 + \log(\mathrm{O/H})_{\mathrm{cell}} < 8.1:\\
        \quad\quad-0.96 - 3.10\; (8.69 - 12 + \log(\mathrm{O/H})_{\mathrm{cell}}) \\
       12 + \log(\mathrm{O/H})_{\mathrm{cell}} \geq 8.1: \\
        \quad\quad-2.21 - 1.00\; (8.69 - 12 + \log(\mathrm{O/H})_{\mathrm{cell}})
    \end{cases}
\end{align}
The dust mass density for each gas cell is then computed as $\rho_{\mathrm{dust}} = \mathrm{DTG}\; \rho_{\mathrm{gas}}$. As in \cite{2020MNRAS.497.4773S} and \cite{2022MNRAS.510.3321P}, the dust mass density is set to zero for cells in which $T_{\mathrm{cell}}>75000$ K and $\mathrm{SFR} =0\; \mathrm{M}_{\odot}\mathrm{yr}^{-1}$ as these cells contribute negligibly to dust attenuation and scattering and only serve to add computational load. We use a \cite{2001ApJ...548..296W} Milky Way dust grain composition and size distribution which comprises graphite, silicate, and polycyclic aromatic hydrocarbon (PAH) grains. Finally, considering that emission is restricted to $\lambda\leq5\micron$ where the stellar continuum dominates the dust continuum, dust self-absorption is not modelled in the transfer simulations.

\subsection{Injection into the HSC-SSP}\label{sec:realism}

\begin{figure*}
\centering
	\includegraphics[width=\linewidth]{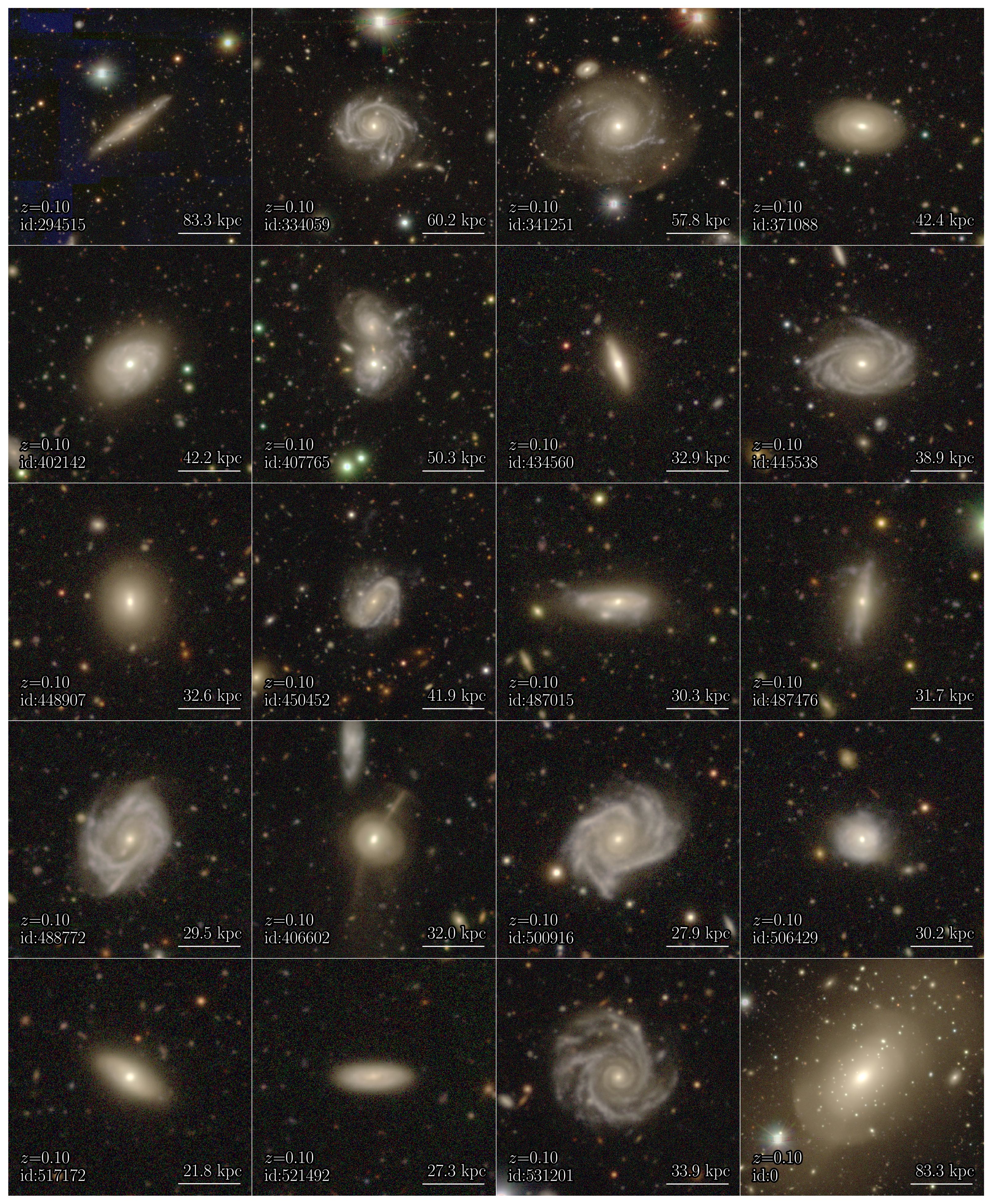}
   \caption[PDR3 HSC $gri$ image mosaic $z=0.1$]{Mosaic of survey-realistic HSC-SSP PDR3 Wide Layer $gri$-colour images for the same TNG50 galaxies at $z=0.1$ shown in Figure \ref{fig:mosaic_idealized_91} (also viewable in the Infinite Gallery on the TNG project website: \url{https://www.tng-project.org/explore/gallery}). The insertion sites are chosen to match statistics of sky brightness, seeing, and projection effects to a reference sample of HSC galaxies as a function of redshift. These images incorporate reconstructed seeing and real HSC-SSP skies -- including Milky Way stars, quasars, and physically unrelated galaxies in projection. The corresponding files also include the variance images, masks, and seeing kernels in each band. }
    \label{fig:mosaic_pdr3_91}
\end{figure*}

\begin{figure*}
\centering
	\includegraphics[width=\linewidth]{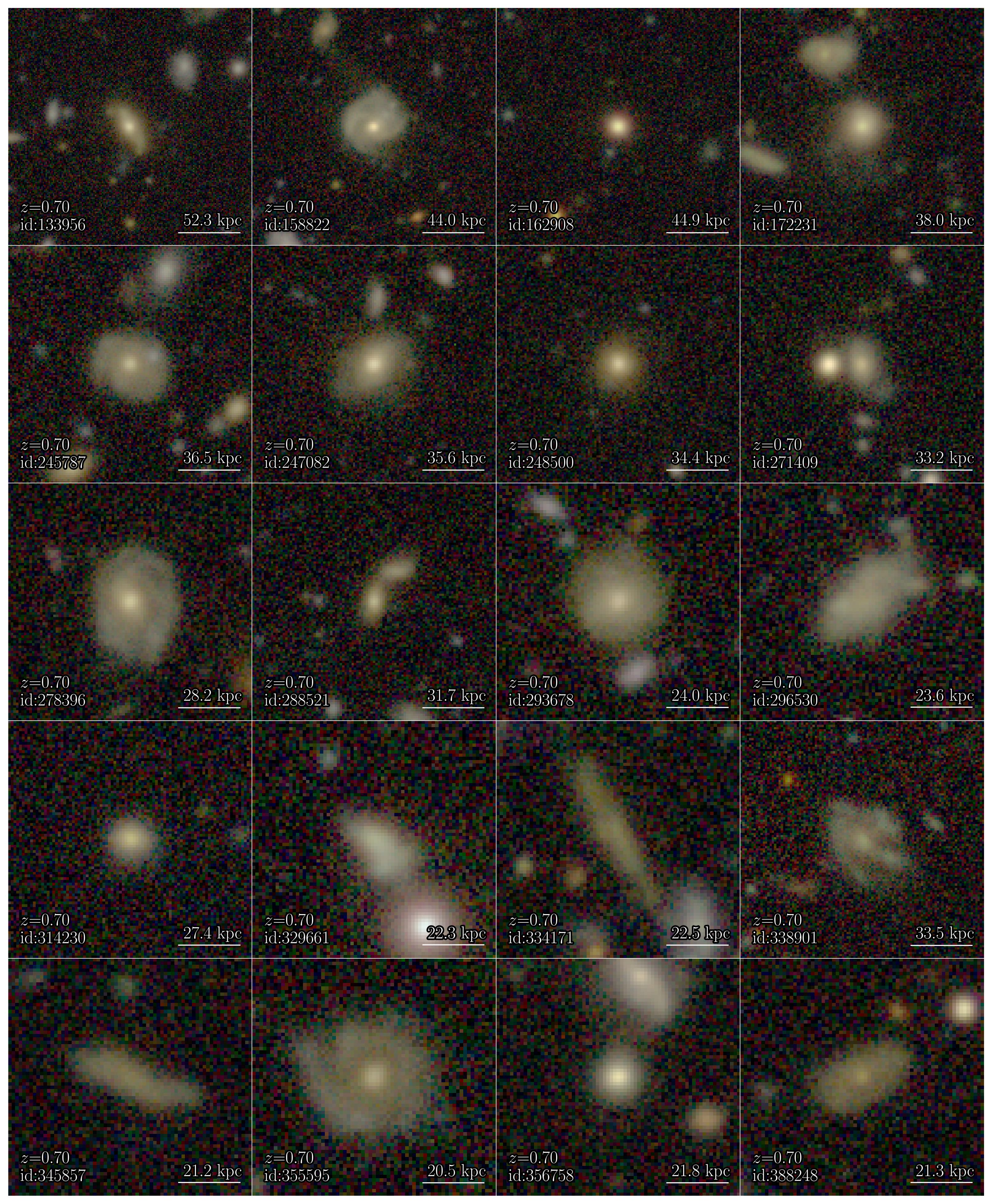}
   \caption[PDR3 HSC $riz$ image mosaic $z=0.7$]{Mosaic of survey-realistic HSC-SSP PDR3 Wide Layer $riz$-colour images for the same TNG50 galaxies at $z=0.7$ shown in Figure \ref{fig:mosaic_idealized_59}. See Figure \ref{fig:mosaic_pdr3_91} caption for details on survey-realistic images.}
    \label{fig:mosaic_pdr3_59}
\end{figure*}

The idealized images from the \skirt{} radiative transfer simulations only contain Monte Carlo (MC) noise from the finite number of photon packets. They do not incorporate survey-realistic limitations that come from observing with real ground or space-based instruments. Including these effects is crucial to any self-consistent, image-based prediction or comparison of galaxy structures in simulations with observations (e.g. \citealt{2015MNRAS.454.1886S,2017MNRAS.467.1033B,2017MNRAS.467.2879B,2018ApJ...853..194D,2019MNRAS.483.4525I,2021MNRAS.501.4359Z,2022MNRAS.511.2544D}; Eisert et al. in prep). Therefore, we make survey-realistic images using the HSC $grizy$ bands taken from the Wide Layer of the HSC-SSP Public Data Release 3 (PDR3; \citealt{2022PASJ...74..247A}) -- leaving extensions to the other instruments, depths, and bandpasses in future work. 

The HSC-SSP is a 330 night observing program using the Hyper Suprime-Cam instrument mounted at the prime focus the Subaru 8.2m telescope at the summit of Mauna Kea in Hawaii  \citep{2012SPIE.8446E..0ZM,2018PASJ...70S...1M,2018PASJ...70S...2K,2018PASJ...70S...3F}. The survey program comprises three layers: Wide; Deep; and UltraDeep -- with premier ground-based $i$-band $5\sigma$ point-source detection depths (26.2, 26.9, 27.7) mag, median seeing (0.61, 0.66, 0.73) arcseconds, and final survey area (1200, 27, 3.5) square degrees, respectively \citep{2022PASJ...74..247A}. All layers obtain dithered exposures in $5$ broad-band filters (\emph{grizy}; \citealt{2018PASJ...70...66K}) which are then co-added to construct the final-depth images with pixel scale 0.168 arcsec \citep{2018PASJ...70S...1M}. Co-addition is carried out by the hscPipe software which (1) reduces and calibrates individual exposures, (2) handles mapping and co-addition of calibrated imaging data into a regularized grid of \texttt{tracts} and \texttt{patches}, astrometric and photometric measurements, and the construction of variance images and processing flag maps (masks images) \citep{2018PASJ...70S...5B,2018PASJ...70S...6H}.

We use an adapted version of \realsim\footnote{\url{https://github.com/cbottrell/RealSim}} \citep{2019MNRAS.490.5390B} in conjunction with the HSC Data Access Tools\footnote{\url{https://hsc-gitlab.mtk.nao.ac.jp/ssp-software/data-access-tools/}} to create survey-realistic synthetic HSC-SSP images from the idealized \skirt{} output. The code handles: (1) assignment of insertion locations and flux calibration; (2) spatial rebinning to the HSC charge-coupled device (CCD) angular scale; (3) convolution of the idealized images with reconstructed HSC point-spread functions (PSFs); and (4) insertion into HSC images:

\begin{itemize}
\item \emph{Insertion statistics}: Insertion locations are selected based on a large reference catalogue of PDR3 galaxies from final-depth Wide Layer images with photometric redshifts computed with the \texttt{mizuki} SED template fitting code \citep{2015ApJ...801...20T,2018PASJ...70S...9T}. We restrict the reference sample to HSC objects for which: (1) the number of coadded exposures meet the final depth requirements in each bandpass: $N_{\mathrm{exp}}(g$, $r$, $i$, $z$, $y)$ = $(4,4,5,5,5)$; (2) \texttt{mizuki} photometric redshifts $0.01<z<1.0$ to cull foreground stars and objects that exceed the redshift range drawn from the simulation, (3) the SED template fit to the broadband fluxes by \texttt{mizuki} has $\chi_{\nu}^2<5$. For each TNG galaxy image, a single object from the HSC reference catalogue is randomly drawn from a narrow redshift slice about the galaxy's simulation snapshot redshift, $\delta z_{\mathrm{snap}}<0.01$. We then extract the full final-depth HSC-SSP patch ($11.7 \times 11.7$ arcmin$^2$) containing this reference galaxy (including the coadd, mask, and variance data in all bands). The $i$-band mask data are used to identify an insertion site for which the centroid contains only sky and is not flagged for any of the more troublesome processing flags (e.g. bright star contamination, deblending error, rejected/suspect pixel, sensor edge, etc.). As laid out in \cite{2017MNRAS.467.2879B}, this distribution-matching process guarantees that the redshift-dependent statistics for depth, seeing, and crowding by nearby sources in the synthetic HSC images are the same as for real HSC galaxies. 

\item \emph{Rebinning to the HSC angular scale}: The angular scale of the HSC camera is 0.168 arcseconds per pixel \citep{2018PASJ...70S...1M}. The physical FoV of the idealized image $(50\; \mathrm{kpc} \leq \mathrm{FoV} \leq 500\; \mathrm{kpc})$ is converted to an angular FoV at the galaxy's snapshot redshift, $z_{\mathrm{snap}}$. The convolved image is then rebinned to the HSC camera angular scale. For the redshifts we consider, this process always degrades the original 100 pc per pixel resolution of the idealized images.

\item \emph{Convolution with reconstructed seeing}: The HSC-SSP Software Pipeline \citep{2018PASJ...70S...5B} uses \texttt{PSFEx} \citep{2011ASPC..442..435B} to model the localized PSFs of isolated stars and the variation in the PSF as a second-order polynomial function of position in each exposure. The model PSFs for the individual exposures are then combined using the same coordinate transformations and weights used to produce the co-add images to produce reconstructed PSFs at any location in the HSC-SSP footprint. Each of our idealized $grizy$ images is convolved with the reconstructed seeing at the insertion site.
\end{itemize}

The notable omission among the realistic noise modelling functionalities of \realsim{} is that we do not model Poisson noise (shot noise) for simulated sources in the images. The HSC-SSP imaging data comprise stacks of co-added exposures to reach their targeted depth. Therefore, the source Poisson noise is $\sigma_\mathrm{src}^2 = I/G_{\mathrm{eff}}$, where $I$ is the intensity in analog to digital units (ADU) and $G_{\mathrm{eff}}$ is the effective gain (electrons/ADU) for converting arbitrary image intensities to integrated electron counts. In other words, $G_{\mathrm{eff}}$ includes the CCD conversion of electrons to digital intensity but also envelops the photometric calibrations used to homogenize individual exposures for stacking (e.g. see \citealt{2014ApJ...794..120A,2019MNRAS.486..390B}). The effective gain therefore varies (a) survey-wide and (b) pixel-by-pixel in HSC-SSP images. While the per-pixel effective gain is used to construct the variance maps corresponding to each HSC coadd image, the $G_{\mathrm{eff}}$ maps themselves were not stored as an HSC data product. We therefore neglect the Poisson noise from simulated sources due to the challenge of reconstructing $G_{\mathrm{eff}}$ in each pixel of the HSC-SSP footprint. 

After rebinning and convolution, an image is then added to a cutout of the same size at the insertion site in each band. These cutouts are already background-subtracted by the HSC data processing software. Figure \ref{fig:mosaic_pdr3_91} (and \ref{fig:mosaic_pdr3_59}) shows a mosaic of the resulting images for our sample at $z=0.1$ ($0.7$). The pixel scale of all synthetic HSC images is 0.167 arcseconds -- which corresponds to 0.32 (1.2) kpc at $z=0.1$ $(0.7)$. The images are archived at \url{www.tng-project.org/bottrell23} and available for visual classifications via Galaxy Cruise\footnote{\url{https://galaxycruise.mtk.nao.ac.jp/en/}} \citep{2023PASJ..tmp...75T}. Each image file includes the synthetic HSC co-adds and corresponding variance maps and bit-masks in each bandpass. Additionally, an \emph{inexact} PSF is provided for each bandpass that is drawn from a random position within 30 arcseconds of the insertion centroid. These are provided to support PSF-convolved modelling of galaxy light with a realistically mismatched PSF -- because having a \emph{exact} reconstruction of the PSF is unrealistic. These PSFs are provided for convenience but can easily be replaced with PSF reconstructions selected at larger or smaller offsets (e.g. \citealt[Section 6.4.2 of ]{2022MNRAS.516..942C}).

\section{Measurements and methods}\label{sec:measurements}

To investigate the origin of relation between galaxy SFRs, morphological asymmetries, and assembly via mergers, we first conduct a parametric and non-parametric quantitative morphology analysis of the synthetic HSC-SSP images with \texttt{GALIGHT}\footnote{\url{https://pypi.org/project/galight}} \citep{2020ApJ...888...37D}. Next, we measure the SFMS in each redshift snapshot following the approach laid out by \citet{2019MNRAS.485.4817D} to qualitatively examine patterns between the offset from the star forming main sequence and our image-derived structural parameters, as in \citet{2021ApJ...923..205Y}. 

\subsection{Quantitative morphologies} 

Parametric and non-parametric quantitative morphologies for all galaxies in the sample are measured using \texttt{GALIGHT} \citep{2020ApJ...888...37D,2021MNRAS.501..269D}: a user-friendly image analysis software whose parametric modelling component is built upon the \texttt{Lenstronomy}\footnote{\url{https://lenstronomy.readthedocs.io/en/latest/}} multi-purpose surface brightness modelling suite \citep{2018PDU....22..189B}. Galaxy structural parameters can exhibit strong wavelength dependences (e.g. \citealt{2012MNRAS.421.1007K,2013MNRAS.435..623V,2013MNRAS.430..330H,2014MNRAS.441.1340V,2022A&A...664A..92H}). Therefore, modelling in each $grizy$ band is done independently. The advantage of this approach is that structural and residual properties in each band are mutually exclusive of each other and there are no forced connections where none may physically exist. The disadvantage, compared to a more sophisticated simultaneous multi-wavelength fit (e.g. MegaMorph, \citealt{2013MNRAS.430..330H,2022A&A...664A..92H}), is that low signal-to-noise structures in a given bandpass get no support from their counterparts in other bands where the signal-to-noise may be greater. The morphological analysis and catalogues we present herein is not definitive by any means. We make our images public with the expressed intention of seeing many surface-brightness characterization tools exploit them (e.g. \texttt{ISOFIT} \citealt{2015ApJ...810..120C}; PROFIT \citealt{2017MNRAS.466.1513R}; \texttt{STATMORPH} \citealt{2019MNRAS.483.4140R}; \texttt{AUTOPROF} \citealt{2021MNRAS.508.1870S}; \texttt{GALAPAGOS-2}/\texttt{GALFITM} \citealt{2010AJ....139.2097P,2022A&A...664A..92H}, \texttt{PROFUSE} \citealt{2022MNRAS.513.2985R}).

\subsubsection{Source detection, deblending, and moment calculations}\label{sec:deblending}

We first run \texttt{sep}\footnote{\url{https://sep.readthedocs.io/en/v1.1.x}} \citep{2016JOSS....1...58B}, a Pythonized version of \texttt{SExtractor} \citep{1996A&AS..117..393B}, on the survey realistic science images. Although fitting is done independently in each band, source detection and deblending are done using the $i$-band images in all cases because these offer (1) the highest data quality and (2) reduced sensitivity to large, discrete variations in mass-to-light ratios on small spatial scales (i.e. tracing structure, not colour). We use the corresponding propagated HSC-SSP variance maps to set signal-to-noise based detection thresholds with \texttt{sep}. We devised an iterative strategy for detection and separation of contiguous sources (deblending) while preventing over-segmentation. This strategy was verified empirically based on visual inspections of several hundred segmentation masks and corresponding science images for each band and redshift slice. We initially take $1\sigma_{\mathrm{pix}}$ as the standard error above which pixels are flagged as sources (\texttt{detect\_thresh}) as long as at least $20$ contiguous pixels (\texttt{minarea}) also satisfy this threshold. We choose this number because it approximates the area, in pixels, contained within a typical PSF FWHM of the HSC-SSP, averaged over all bands. For deblending of sources meeting these criteria, we consistently use \texttt{deblend\_nthresh}$=32$ and \texttt{deblend\_cont}$=0.001$ in all iterations. 

After deblending, if the source footprint containing the central pixel of the image occupies over $80$ per cent of the science image or includes any pixel that borders the edge of the image, then the detection threshold is augmented by $+0.5\sigma_{\mathrm{pix}}$ to force segmentation based on pixels with higher signal-to-noise. The rationale is that if more than $80$ per cent of the (large) FoV of the cutouts belongs to the footprint of a single source, then the source is unlikely to have been adequately deblended from neighbouring sources in the image, projected or physical. By increasing the detection threshold, the extended, low surface brightness flux that can bridge physically distinct sources may not meet the signal-to-noise criterion for detection -- resulting in the proper separation of sources. 

Using the image moments for the central source in the FoV\footnote{The central source is the source whose footprint includes the central pixel of the image and therefore also the gravitational potential minimum of the inserted TNG galaxy.} are computed to be used as initial guesses for parametric fitting, as in \cite{2002ApJS..142....1S}. We compute:
\begin{align}
&M_{xx} = \frac{1}{M_{\mathrm{tot}}} \sum I_{ij} x_{ij}^2\\
&M_{yy} = \frac{1}{M_{\mathrm{tot}}} \sum I_{ij} y_{ij}^2 \\
&M_{xy} = \frac{1}{M_{\mathrm{tot}}} \sum I_{ij} x_{ij}y_{ij}\\
&M_{rr} = \frac{1}{M_{\mathrm{tot}}} \sum I_{ij} \sqrt{x_{ij}^2 + y_{ij}^2 }\\
&a_m^2 = 8(M_{xx}+M_{yy}+\sqrt{4M_{xy}^2 + (M_{xx} - M_{yy})^2}\\
&b_m^2 = 8(M_{xx}+M_{yy}-\sqrt{4M_{xy}^2 + (M_{xx} - M_{yy})^2}\\
&\phi_m = \frac{1}{2} \arctan{\left( \frac{2M_{xy}}{M_{xx} - M_{yy}}\right)} + \frac{\pi}{2} (M_{xx} < M_{yy})
\end{align}
where $M_{\mathrm{tot}} = \sum_{ij} I_{ij}$ over all pixels in the source footprint, $x_{ij}$ and $y_{ij}$ are the coordinates each pixel with respect to the intensity-weighted centre of the source footprint $(\bar{x},\bar{y})_m$, $M_{rr}$ is the intensity-weighted average radius, $q_m=b_m/a_m$ is the axis ratio of the semi-minor and semi-major axes, and $\phi_m$ is the position angle. Only the central source footprint and background pixels are used in subsequent fitting. All other sources in the \texttt{sep} segmentation map are masked. This approach (a) avoids unnecessary degeneracies between the model parameters of target galaxies and other sources in the images and (b) exhibits no significant systematics in inference of colours and magnitudes for galaxy pairs \citep{2011ApJS..196...11S} or parameter recovery analyses \citep{2014ApJS..210....3M}.

\subsubsection{Parametric S\'ersic fits}

In this study, the parametric fits to our synthetic galaxy sample comprise simple PSF-convolved S\'ersic function optimizations \citep{1963BAAA....6...41S,2005PASA...22..118G}. Our optimization strategy first uses Particle Swarm Optimization (PSO, \citealt{KennedyEberhartPSO}) to identify a candidate solution. The PSO results (which should already be near the global likelihood-maximizing solution, in principle) are then imported into a more rigorous Markov Chain Monte Carlo (MCMC) parameter and error inference using \texttt{emcee} \citep{2013PASP..125..306F}. Because the total intensity, $I_{\mathrm{tot}}$, is a linear scaling of the S\'ersic surface brightness distribution, it is inferred separately during likelihood estimation using semi-linear inversion \citep{2003ApJ...590..673W} and requires no initial estimate. The initial guess for S\'ersic index for the PSO fit is always $n_s=2$. Initial estimates for the remaining S\'ersic model parameters are taken from the image moments: the centroid position $(\hat{x}_c, \hat{y}_c) = (\bar{x},\bar{y})_{m}$, semi-major effective radius $\hat{r}_{\mathrm{eff}}$ = $M_{rr}$, axis ratio $\hat{q} = q_{m}$, and position angle $\hat{\phi} = \phi_{m}$. 

With the initial guess for a given galaxy in hand, we first run a PSO fit to identify a candidate solution. Initial guesses for each particle in the swarm are drawn from normally-distributed random variables centred on the each parameter's global guess. The result of the PSO optimization is imported as updated the initial guess for an MCMC fit. In principle, the result imported from the PSO fit should already be near the global $\chi^2_\nu$ solution. Consequently, the MCMC fit \emph{should} start in the convergence region and simply aggregate samples for parameter inference and uncertainty estimation. In our validation tests, this was true in all cases. The likelihood was not improved significantly by the MCMC fit and tracking of the MCMC chain revealed only small stochastic changes in model parameters. In practice, we still allow a short burn-in and convergence, and follow-up with $10^4$ samples of the convergence region. These samples are used to build the distribution $P(\boldsymbol{\theta} | D,M)$ and obtain $\hat{\boldsymbol{\theta}}$ and associated errors given the data and model. The median and $68$ per cent confidence interval of $P(\boldsymbol{\theta} | D,M)$ are computed independently for each parameter, $\theta_k$ (e.g. as in GIM2D, \citealt{2002ApJS..142....1S}) and these are the values reported in our catalogues.

For both PSO and MCMC optimizations, the HSC-SSP co-add variance maps are used as the noise model. These maps include all propagated noise from the individual exposures that go into a co-added region such as readout, dark current, Poisson noise (incl. sources and sky), and noise incurred from artefact removal. It is important to note that neither these variance maps nor the fitted images contain Poisson noise contributions from TNG galaxies. So although we could not incorporate realistic source Poisson noise for TNG galaxies into the images (Section \ref{sec:realism}), the variance maps also do not include this noise and therefore faithfully represent the per-pixel noise \emph{present} in the fitted images.

\subsubsection{Non-parametric morphological statistics}

\begin{figure*}
\centering
	\includegraphics[width=\linewidth]{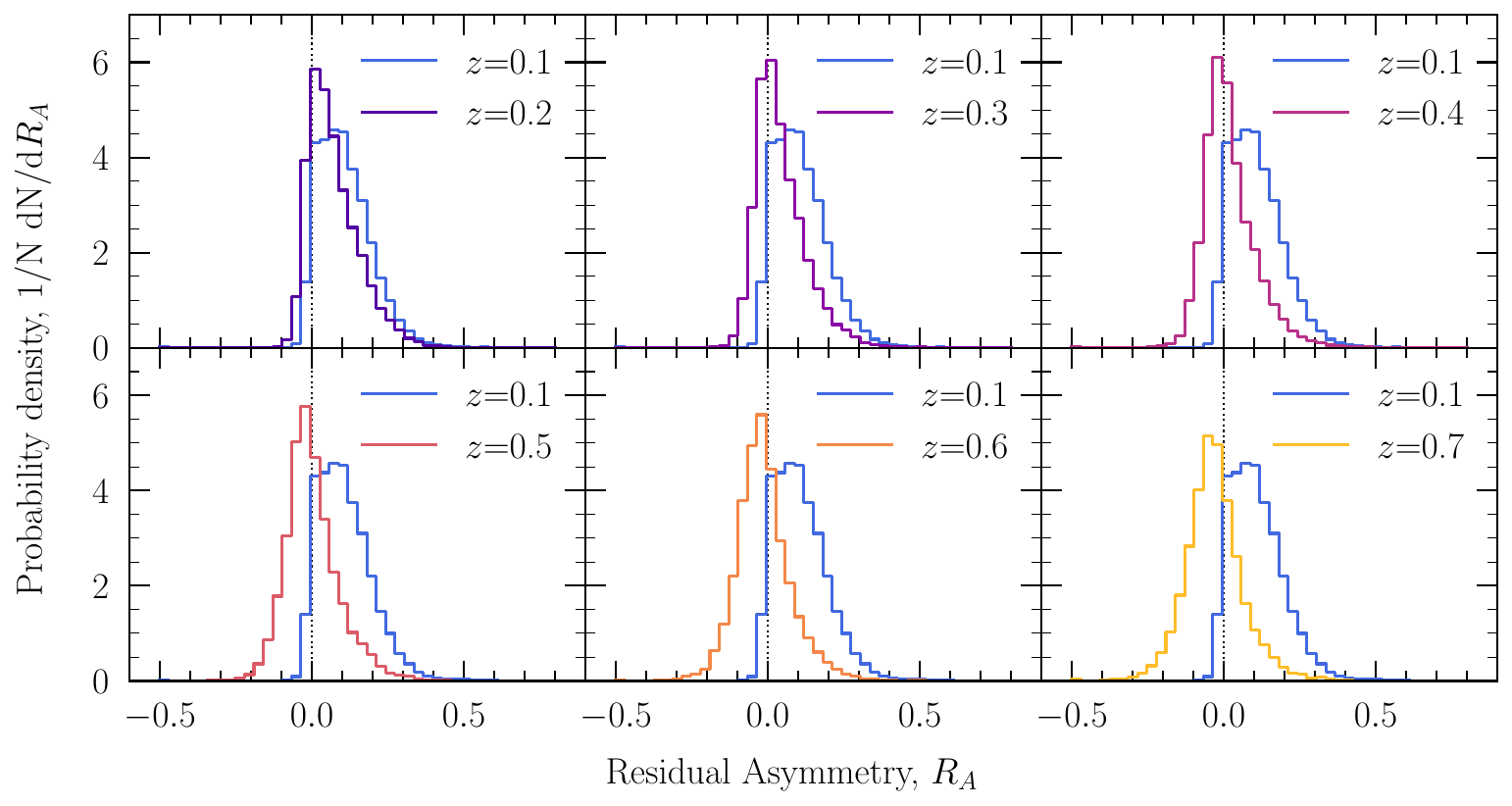}
   \caption[Asymmetry evolution]{Evolution in the distributions of $i$-band residual asymmetries, $R_A$, for synthetic HSC galaxies from TNG50 with $\logMstar \geq 9$. The $R_A$ distribution in each redshift slice from $z=0.2-0.7$ is compared with the $R_A$ distribution for the $z=0.1$ slice (blue), for reference. The distributions for conventional asymmetries, $A$, are very similar (not shown). Measured asymmetries are suppressed at higher redshifts due relatively large background asymmetries (e.g. \citealt{2000ApJ...529..886C,2021MNRAS.507..886T}) and limited capacity to capture asymmetric features at low-surface brightnesses (e.g. \citealt{2010MNRAS.406..382K,2019MNRAS.486..390B,2022MNRAS.516.4354W,2023MNRAS.521.5272T}). At $z\geq0.4$, the distributions peak at negative $R_A$; i.e. systematics in the background terms of Equations \ref{eq:asym} and \ref{eq:rasym} dominate. Even the differences between the distributions at $z=0.1$ and $z=0.2$ are stark. These issues highlight the need for a \emph{relative} approach to characterizing the asymmetries of galaxy populations at different redshifts (Section \ref{sec:controls}).}
    \label{fig:asy_evo}
\end{figure*}

\begin{figure*}
\centering
	\includegraphics[width=\linewidth]{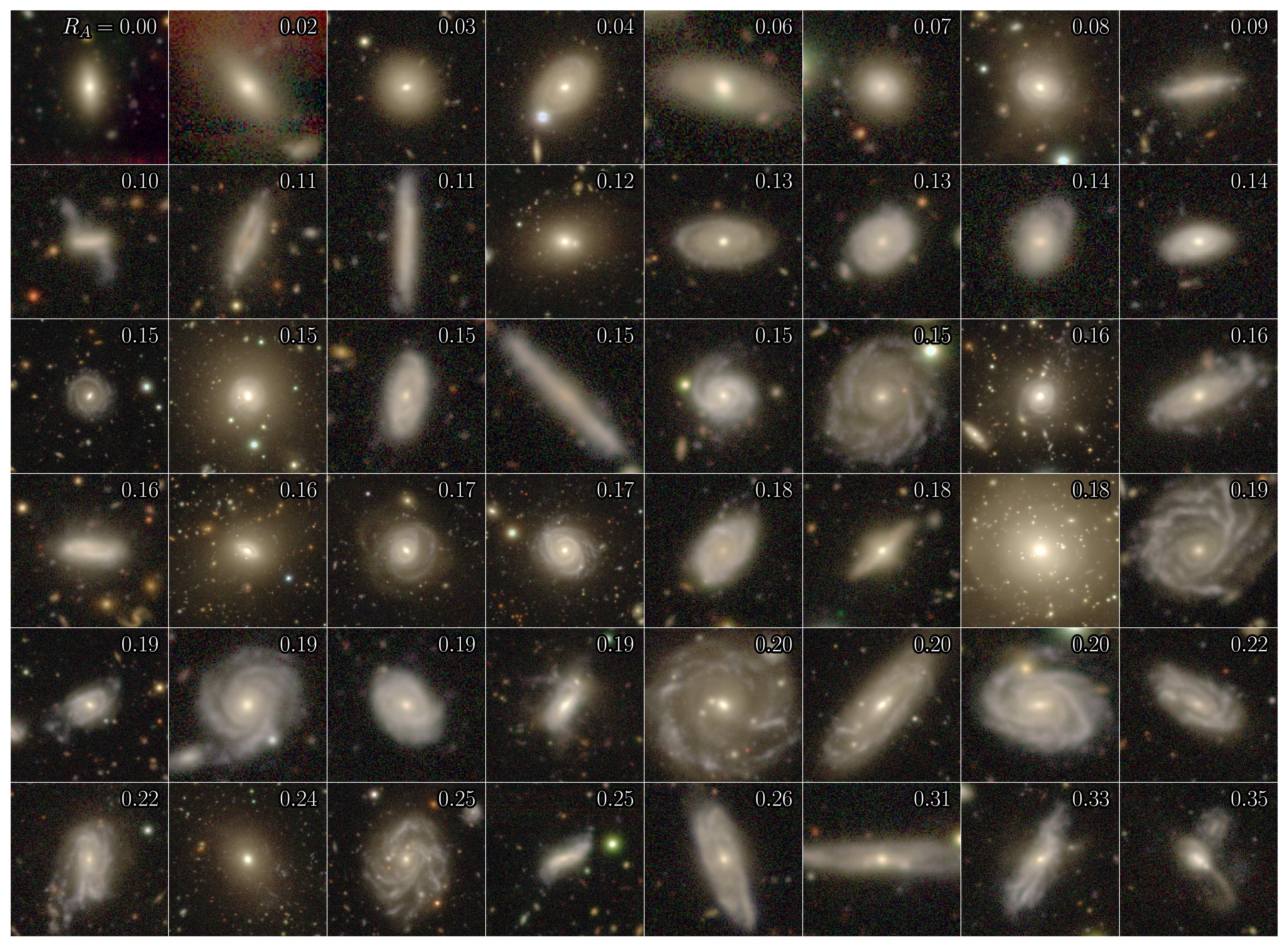}
   \caption[Asymmetry Zoo]{Menagerie of galaxy asymmetries for TNG50 at $z=0.1$ as seen by the HSC-SSP PDR3 Wide Layer in $gri$ colour. Residual asymmetries $R_A$ measured in the $i$-band are indicated at the upper right of each image. Galaxies are randomly sampled from the asymmetry distribution in Figure \ref{fig:asy_evo}.}
    \label{fig:asy_mosaic}
\end{figure*}

Our parametric fits are complemented by a number of summary statistics. These include: the Concentration, Asymmetry, and Smoothness/Clumpiness parameters which constitute the $CAS$ system \citep{1994ApJ...432...75A,1996ApJS..107....1A,2000AJ....119.2645B,2003ApJS..147....1C}; the Gini coefficient \citep{glasser1962variance,2003ApJ...588..218A,2004AJ....128..163L}; the $M_{20}$ statistic \citep{2004AJ....128..163L}; and the elliptical and circular Petrosian radii \citep{1976ApJ...209L...1P,2002AJ....123..485S}. Some of these statistics have been expressed in several forms since their inceptions. For the aforementioned parameters, we use the precise definitions compiled in the \texttt{Statmorph}\footnote{\url{https://statmorph.readthedocs.io}} code by \citep{2019MNRAS.483.4140R}. Most of these quantities are measured using only the pixels assigned to the central source by our detection and deblending strategy. 

For asymmetry measurements, which subtract a 180 degree rotated image from the original, we use all pixels in either (1) the source footprint in the original segmentation image or (2) the source footprint in the 180 degree rotated segmentation image (excluding the footprints of other sources in both the original and rotated segmentation map). Following \cite{2019MNRAS.483.4140R}, we find the pivot point in the image which minimizes asymmetry and use this pivot point (asymmetry centre) as the aperture centre for Concentration and Petrosian radius measurements. In addition to asymmetries measured directly on the science image, we also measure residual asymmetries, $R_A$ \citep{1995ApJ...451L...1S,2002ApJS..142....1S} and the residual flux fraction (RFF, \citealt{2012MNRAS.419.2703H}) from the S\'ersic model-subtracted residual image. The asymmetry and residual asymmetry are defined:
\begin{align}
&A = \frac{\sum | I_{ij} - I_{ij}^{180}|}{\sum I_{ij}} - \frac{\sum | B_{ij} - B_{ij}^{180}|}{\sum I_{ij}} \label{eq:asym} \\
&R_A = \frac{\frac{1}{2}\sum | R_{ij} - R_{ij}^{180}|}{\sum O_{ij}} - \frac{\frac{1}{2}\sum | B_{ij} - B_{ij}^{180}|}{\sum O_{ij}} \label{eq:rasym}
\end{align}
where $I_{ij}$ is a pixel intensity in the source footprint of the science cutout, $B_{ij}$ is a pixel intensity in the unmasked background, $R_{ij}$ is a pixel intensity in the source footprint of the S\'ersic model-subtracted residual image, and $O_{ij}$ is a pixel intensity in the source footprint of the S\'ersic model image. Background pixels are all those flagged as background in both the original and 180 degree rotated images. \cite{2021MNRAS.507..886T} show that measurement of accurate asymmetries relies sensitively on the measurement of this background term. We set a very low default detection threshold with \texttt{sep} to specifically mitigate systematics in background asymmetry measurements from extended galaxy light (Section \ref{sec:deblending}).

Figure \ref{fig:asy_evo} shows the distributions of measured $i$-band residual asymmetries for TNG50 galaxies out to $z=0.7$. The $R_A$ distribution for the $z=0.1$ snapshot is shown in all panels, for reference. The apparent evolution of asymmetries in Figure \ref{fig:asy_evo} is consistent with previous works demonstrating the sensitivity of asymmetry measurements to depth and physical resolution \citep{2000ApJ...529..886C,2019MNRAS.486..390B,2021MNRAS.507..886T,2022MNRAS.516.4354W,2023MNRAS.521.5272T}. Despite the fact that the $i$-band is probing increasingly young stellar populations at higher redshift (which tend to be more asymmetric in structure; \citealt{2000ApJ...529..886C,2019MNRAS.486..390B}), measured asymmetries still exhibit suppression in response to the $(1+z)^5$ cosmological attenuation of specific surface brightness. In other words, while the measurements imply a trend of decreasing asymmetry with redshift, this is also expected from the decreasing signal-to-noise ratio for intrinsically asymmetric structure. 

Whether intrinsic or apparent, Figure \ref{fig:asy_evo} highlights that the utility of asymmetry as a morphological indicator can be limited in the context of characterizing differences between galaxy populations across broad redshift ranges (or even for different intrinsic brightnesses at fixed redshift). However, asymmetry still has great utility in establishing physical differences between otherwise similar galaxies \emph{at fixed redshift}, as we will later show. As a visual guide of how measured galaxy asymmetries relate to their visual appearances, a mosaic of $z=0.1$ TNG50 galaxies and their $i$-band asymmetries are shown in Figure \ref{fig:asy_mosaic}.

\subsection{Star-forming main sequence fitting and galaxy sample}\label{sec:sfms}

\begin{table}\label{tab:sfms}
\caption{Linear regression results for the TNG50 star-forming main sequence described in Section \ref{sec:sfms} and by Equation \ref{eq:sfms}. Following \citet{2019MNRAS.485.4817D,2021MNRAS.506.4760D}, fits to the SFMS at each redshift are restricted to bins of $0.2$ dex in the range $9\leq \logMstar \leq 10.6$ due to the observed bending of the SFMS at $\logMstar\gtrsim10.5$ (e.g. \citealt{2011ApJ...730...61K,2014ApJ...795..104W,2015ApJ...801...80L,2016ApJ...817..118T,2018ApJ...853..131L}).}
\label{tab:sfms}
\begin{tabular*}{\linewidth}{c@{\extracolsep{\fill}} c c c c c}
\hline
Snapshot & $z$ & $\alpha$ & $\sigma_{\alpha}$ & $\beta$ & $\sigma_{\beta}$ \\
\hline
91 & 0.1 & 0.8774 & 0.0103 & -1.013 & 0.1475 \\
84 & 0.2 & 0.8863 & 0.0052 & -0.9369 & 0.0750 \\
78 & 0.3 & 0.8835 & 0.0153 & -0.8664 & 0.2198 \\
72 & 0.4 & 0.8704 & 0.0138 & -0.8363 & 0.1990 \\
67 & 0.5 & 0.9079 & 0.0138 & -0.7311 & 0.1986 \\
63 & 0.6 & 0.8943 & 0.0233 & -0.6866 & 0.3351 \\
59 & 0.7 & 0.8807 & 0.0244 & -0.6424 & 0.3512 \\
\hline
\end{tabular*}
\end{table}

Quantitative morphologies are measured for all TNG50 galaxies with $\logMstar \geq9$ from snapshots corresponding to $z=(0.1,0.2,0.3,0.4,0.5,0.6,0.7)$. Since this work concerns the connection between asymmetry and star formation, we restrict our analysis to SFGs. We follow an iterative approach laid out by \citet{2019MNRAS.485.4817D} to (1) measure the star forming main sequence (SFMS) and (2) separate star-forming and passive galaxies (e.g. \citealt{2021MNRAS.506.4760D}). We define the SFMS using masses and SFRs of SFGs in the simulations and not from observations because we are primarily interested in offsets \emph{relative} to the SFMS. We therefore do not want to enfold any systematic differences in the SFMS normalization or slope between observations and TNG.

Using the total galaxy stellar masses\footnote{The total galaxy stellar mass is the sum of the masses of all stellar particles belonging to a given galaxy's subhalo identified by the \texttt{SUBFIND} algorithm \citep{2001MNRAS.328..726S}.} and instantaneous SFRs for a given snapshot, we bin galaxies over the range $9\leq \logMstar \leq 10.6$ in bins of $0.2$ dex and measure the median specific star formation rate (SSFR) in each bin. Galaxies with SSFRs that are 1 dex lower than the median SSFR in each bin are labeled as ``passive'' and removed from subsequent iterations of median-estimation. This process is iterated in each bin until the change in the median SSFR between iterations is less than $10^{-5}$ dex. The resulting median SSFRs and corresponding bin centres are fitted with the relation:
\begin{align} \label{eq:sfms}
\log (\mathrm{SSFR/Gyr}^{-1}) = (\alpha-1) [ \logMstar - 10.5 ] + \beta
\end{align}

Table \ref{tab:sfms} shows the results of the SFMS fits at each redshift and corresponding standard errors in $\alpha$ and $\beta$. The SFMS slope, $\alpha-1$, is stable over the redshift range we consider but the normalization, $\beta$ at $\logMstar=10.5$ increases with redshift. Meanwhile, the standard errors in $\beta$ also increase significantly over this redshift interval. We therefore restrict comparisons between individual galaxies to those from the same redshift (including assignment of controls for calculating asymmetry and SFR offsets). For each redshift, all galaxies that are below $1$ dex from the corresponding star-forming main sequence, $\dsfms = \log(\mathrm{SSFR}) - \langle \log(\mathrm{SSFR})\rangle_{\mathrm{MS}} < -1$, are labelled as passive and removed from the star-forming galaxy (SFG) samples used in subsequent analyses.

\subsection{Characterizing the merger histories of galaxies}\label{sec:mergers}

\begin{figure}
\centering
	\includegraphics[width=\linewidth]{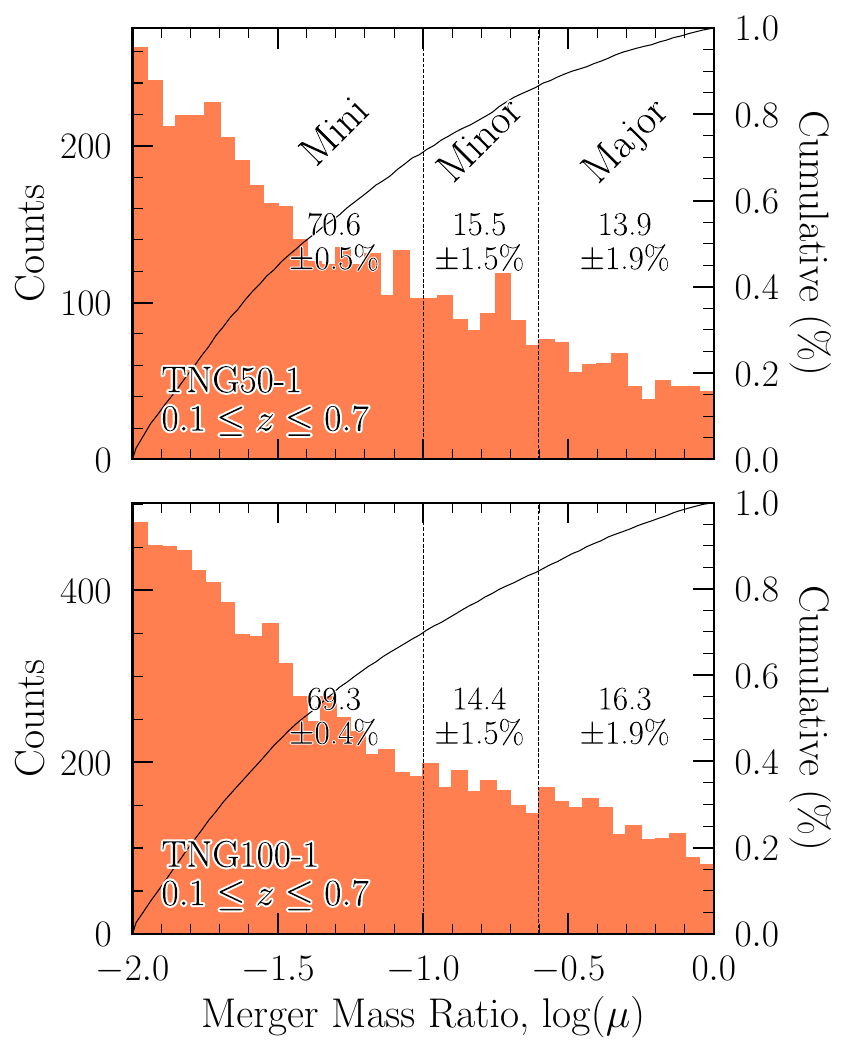}
   \caption[Merger $\mu$ distributions]{Merger mass ratio distributions drawn from the assembly histories of all $z=0.1$ galaxies with $\logMstar\geq9$ in TNG50 (upper panel) and $\logMstar\geq10$ in TNG100 (lower panel) out to $z=0.7$. Mini mergers, $0.01\leq \mu < 0.1$, dominate the merger mass ratio distributions of massive galaxy assembly histories in the local Universe ($77.7$ per cent for $\logMstar\geq9$ and $87.6$ per cent for $\logMstar\geq10$). The frequency of mini mergers in the assembly histories of $z=0.1$ galaxies with respect to minor, $0.1\leq \mu < 0.25$, and major mergers, $\mu \geq 0.25$, increases for galaxies with higher $z=0.1$ stellar masses. The solid lines show the cumulative distributions in each panel. Uncertainties are derived from binomial statistics in each mass ratio category.} 
    \label{fig:mus}
\end{figure}

Galaxy interactions and mergers are known mechanisms for simultaneously accelerating SFRs and producing asymmetric structure (e.g. \citealt{1977egsp.conf..401T,1996ApJ...464..641M,2013MNRAS.435.3627E,2016MNRAS.461.2589P}). To examine the role of mergers in driving this observed connection, we build a comprehensive summary of each TNG galaxy's merger history \emph{and} forecast using the TNG \texttt{SubLink} merger trees \citep{2015MNRAS.449...49R}.

For each galaxy from TNG50 with $\logMstar \geq9$ and TNG100 with $\logMstar \geq 10$, we move backward and forward along its main progenitor branch and descendent line and identify all mergers that satisfy $\mu \geq 0.01$, where $\mu=M^{(2)}_{\star} / M^{(1)}_{\star}$ is the stellar mass ratio of the merging galaxies and $M^{(1)}_{\star}$ is always the stellar mass of the more massive galaxy. These mergers are then divided into three categories: 
\begin{description}
\item \emph{Major}: $\mu \geq 0.25$
\item \emph{Minor}: $0.1 \leq \mu < 0.25$
\item \emph{Mini}: $0.01 \leq \mu < 0.1$
\end{description}
To identify mergers satisfying these conditions, we first obtain the maximum stellar mass along the main progenitor branch of every TNG galaxy, $M_{\star}^{\mathrm{max}}$ (specifically, the mass enclosed within twice their stellar half-mass radii). We also keep track of the snapshot at which each $M_{\star}^{\mathrm{max}}$ occurs. These masses are used to correct for numerical stripping effects in stellar mass ratio estimates for simulated galaxy mergers \citep{2020MNRAS.493.3716H,2020MNRAS.494.4969P}. Then, at each step backward along a galaxy's main progenitor branch (most massive history), we compare the main progenitor $M_{\star}^{\mathrm{max}}$ to the $M_{\star}^{\mathrm{max}}$ of all next progenitors (less massive histories; omitting progenitors with non-cosmological origin). Specifically, at a given snapshot, we compare the $M_{\star}^{\mathrm{max}}$ of each next progenitor to the $M_{\star}$ of the main progenitor evaluated at the snapshot where the next progenitor's $M_{\star} = M_{\star}^{\mathrm{max}}$. In other words, the merger stellar mass ratio between the more massive progenitor and less massive progenitor is always evaluated at the snapshot where the less massive progenitor's stellar mass is maximized \citep{2022MNRAS.516.5404S,2023MNRAS.519.2199E}. 

This procedure guarantees robust estimates of galaxy mass ratios which are not strongly affected by overlapping stellar components in late-stage mergers. We then follow the same procedure forward in time, following the descendent pointers for each galaxy to its root descendent. Future mergers are only counted if either (a) the main progenitor of the future remnant is the direct descendent of the target galaxy or (b) the merger is between the direct descendent of the target galaxy and the future remnant's main progenitor. In other words, if the descendent of a galaxy is the main progenitor of a future post-coalescence remnant, all mergers with next progenitors of that main progenitor that satisfy the mass ratio criteria are counted. If not, then the galaxy's descendent is a next progenitor to the merger remnant's main progenitor and only the merger between these two galaxies is counted (provided it also satisfies the mass ratio criteria). 


In the steps described above, all mergers satisfying $\mu \geq 0.01$ are recorded -- including multiple mergers at the same snapshot. We then compute a suite of summary statistics based on these merger histories and forecasts for each galaxy. Figure \ref{fig:mus} shows the distributions of merger mass ratios in the assembly histories of all galaxies from TNG50 with $\logMstar\geq9$ (upper panel) and TNG100 with $\logMstar\geq10$ (lower panel) at $z=0.1$ since $z=0.7$. Dashed vertical lines delineate mini, minor, and major mergers according to our definitions. In terms of frequency of merger events, mini mergers dominate the assembly histories of galaxies above both $z=0.1$ descendent stellar mass thresholds. $70.6$ ($69.3$) per cent of mergers in the assembly histories of $z=0.1$ TNG50 (TNG100) galaxies are in the mini regime, $0.01 \leq \mu < 0.1$. Minor and major mergers contribute relatively modest merger counts $15.5$ ($14.4$) and 13.9 ($16.3$) per cent, respectively. The fractions in each mass ratio regime are consistent between the two simulation volumes and corresponding stellar mass ranges apart from the mini merger regime -- which is slightly lower in TNG100. Merger frequency steepens with decreasing mass ratios, as expected \citep{1991ApJ...379...52W,1993MNRAS.261..921K,1993MNRAS.262..627L}. On average in TNG, there are $2.4$ ($2.3$) mini mergers in a galaxy's assembly history for every major or minor merger combined.

Throughout our analysis, we regularly use the time since/until a galaxy's nearest merger (past and future), $|\dtcoal|$ -- which derives from the time since/until coalescence parameters for mergers from each mass ratio range:
\begin{align}
&\mathrm{Time\;Since\;Merger} = T - T^{\text{Past}}_{\mathrm{Coal}} \label{eq:since} \\ 
&\mathrm{Time\;Until\;Merger} = T^{\text{Future}}_{\mathrm{Coal}} - T \label{eq:until} \\
&|\dtcoal| = \min(\mathrm{Time\;Since\;Merger, Time\;Until\;Merger})\label{eq:dtc} 
\end{align}
where $T$ is the age of the Universe at the galaxy's snapshot, $T^{\text{Past}}_{\mathrm{Coal}}$ is the age of the Universe at the snapshot of the most recent post-coalescence merger remnant along its main progenitor branch, and $T^{\text{Future}}_{\mathrm{Coal}}$ is the age of the Universe at the snapshot of the next future post-coalescence merger remnant along the line of descendants. We take the post-coalescence snapshot as an approximation for the time of coalescence. By these definitions, a galaxy that is the post-coalescence remnant of a merger between one or more progenitors in the previous snapshot has Time Since Merger$=0$ (i.e. at the post-coalescence snapshot). In contrast, we define the time until a future merger such that Time Until Merger$>0$ provided that a future merger exists. Therefore, $|\dtcoal|$ measures the time to the nearest post-coalescence merger remnant's snapshot, forward or backward in time. The absolute value denotes that this is an unsigned quantity, as both Time Since Merger and Time Until Merger are always positive by definition. We also use a signed $\dtcoal$ is negative for pre-mergers provided Time Until Merger is also less than Time Since Merger. 

\subsection{Control matching procedure} \label{sec:controls} 

Both asymmetries and SFRs are connected to a number of physical characteristics and mechanisms other than mergers. These include gas accretion rates and interstellar gas fraction (e.g. \citealt{2005A&A...438..507B,2016MNRAS.462.1749S,2016MNRAS.457.2790T,2019ApJ...884L..33L,2021MNRAS.504.1989W}), local and large scale environments (e.g. \citealt{1980ApJ...236..351D,2004ApJ...615L.101B,2016MNRAS.462.2559B,2020MNRAS.493.5987O,2022ApJ...936..124Y}), stellar mass (e.g. \citealt{2004MNRAS.351.1151B,2007ApJS..173..267S,2007ApJ...660L..43N}) and redshift (e.g. \citealt{2014ARA&A..52..415M,2014ApJ...795..104W,2018A&A...615A.146P,2019MNRAS.485.4817D}). To isolate the role of mergers, we carefully control each of these areas by assigning each sample galaxy a matching control galaxy in all properties simultaneously (e.g. \citealt{2013MNRAS.435.3627E,2016MNRAS.461.2589P,2021MNRAS.504.1888Q,2022MNRAS.511..100B,2023MNRAS.519.2119Q}). Meanwhile, characteristics such as SFRs, asymmetries, and merger history/forecast statistics are allowed to vary between sample galaxies and their matched controls. In other words, we examine how differences in merger histories/forecasts, asymmetries, and SFRs affect each other between \emph{otherwise similar galaxies}.

In this work, we make use of two separate controlled experiments to investigate the role of mergers as drivers of asymmetry and star formation. Selection of (a) sample galaxies and (b) the control \emph{pool} pertaining to each experiment is described in the corresponding sections (\ref{sec:sfg_merg} and \ref{sec:iso_merg}). However, for a given sample and its control pool, the matching procedure is identical. For each galaxy in the sample, we assign controls from same simulation snapshot (redshift) that match in stellar mass, environment, and gas fraction. We start by computing the gas fraction for every galaxy as:
\begin{align}\label{eq:fgas}
f_{\mathrm{gas}} = \frac{M_{\mathrm{gas}}(R<2R_{\star\mathrm{half}})}{M_{\mathrm{gas}}(R<2R_{\star\mathrm{half}})+M_{\star}(R<2R_{\star\mathrm{half}})}
\end{align}
where $R_{\star\mathrm{half}}$ is the half stellar mass radius of the galaxy and $M_{\mathrm{gas}}$ is the gas mass with no restriction on gas phase. We then compute a number of environmental characteristics which summarize environment from local to large scales. We use both the distance to the $2^{\mathrm{nd}}$ nearest neighbour, $R_2$, as a measure of local environment (isolation, \citealt{2020MNRAS.494.4969P}) and the number of galaxies within $2$ Mpc, $N_{2}$, as a measure of large-scale environment. These environmental parameters consider companions containing at least 0.1 times the target galaxy's stellar mass. 

Next, for each galaxy in the sample, we iteratively search the control pool for matches in stellar mass, gas fractions, and our two factors for environment at the same redshift. We start with initial tolerances of $\delta = 0.05$ dex for $\logMstar$ and $\delta = 5$ per cent for $R_2$, $N_2$, and $\fgas$ in the differences between samples and controls. To estimate per cent differences for $N_2$, and $\fgas$ which can be zero or very small, we use $\max(N_2,1)$ and $\max(\fgas,0.01)$ as denominators, respectively. Following the approach from \citep{2016MNRAS.461.2589P}, while the number of matches is less than $N_{\mathrm{controls}}$, we incrementally increase the tolerances, $\delta$, by $n_{\mathrm{grow}} \times 0.05$ dex and $n_{\mathrm{grow}} \times 5$ per cent, where $n_{\mathrm{grow}}$ is matching iteration. If the number of matches does not reach $N_{\mathrm{controls}}$ by $n_{\mathrm{grow}}=5$, for which the tolerances, $\delta$, are $0.25$ dex and $25$ per cent, the sample galaxy is rejected from subsequent analysis. 

Each matching control galaxy assigned a weight based on its offset from its sample galaxy in the matching parameters, $X$:
\begin{align}\label{eq:weights}
W_{i} = \prod_{X\in\{X_1,...,X_n\}} \frac{1-|X_{\mathrm{sample}}-X_{\mathrm{control},i}|}{ n_{\mathrm{grow}} \times \delta}
\end{align}
using $n_{\mathrm{grow}}$ from the matching iteration in which the target $N_{\mathrm{controls}}$ matches is satisfied and the matching parameters are $X \in \{ \logMstar, R_2, N_2, \fgas \}$. The weights of all controls for a given SFG are then normalized:
\begin{align}
w_{i} = W_{i} / \sum^{N_{\mathrm{controls}}}_{j=1} W_{j}
\end{align}
The results of the control-matched experiments shown in Section \ref{sec:results} use $N_{\mathrm{controls}}=1$ in all experiments (i.e. best-match) but agree qualitatively for all $1\leq N_{\mathrm{controls}} \leq 3$ -- though with increasing uncertainties due to the deceasing number of samples with $N_{\mathrm{controls}}$ controls satisfying the matching criteria. For the small number of galaxies that satisfy $N_{\mathrm{controls}}>3$, the uncertainties in offsets become too large to make statistically confident statements.

\section{Results with TNG50: Star formation, Asymmetry, and Mergers}\label{sec:results}

In this section, we present the results of our investigation into the role of mergers as physical drivers for the connection between SFRs and asymmetry galaxies using TNG50. First, in Section \ref{sec:sfms_ra}, we examine the relationship between asymmetry and offsets from the SFMS for star-forming galaxies from TNG50. Next, in Section \ref{sec:sfms_merg} we offer a qualitative comparison to a strikingly similar trend between SFMS offsets and incidence of mini, minor, and major mergers for TNG50 galaxies. Then, in Sections \ref{sec:sfg_merg} and \ref{sec:iso_merg}, we present the results of two controlled experiments designed to isolate the quantitative relationship between asymmetry, SFRs, and mergers from known correlated factors such as redshift, stellar mass, environment, and gas content. 

\subsection{Asymmetries and offsets from SFMS}\label{sec:sfms_ra}

\begin{figure*}
\centering
	\includegraphics[width=\linewidth]{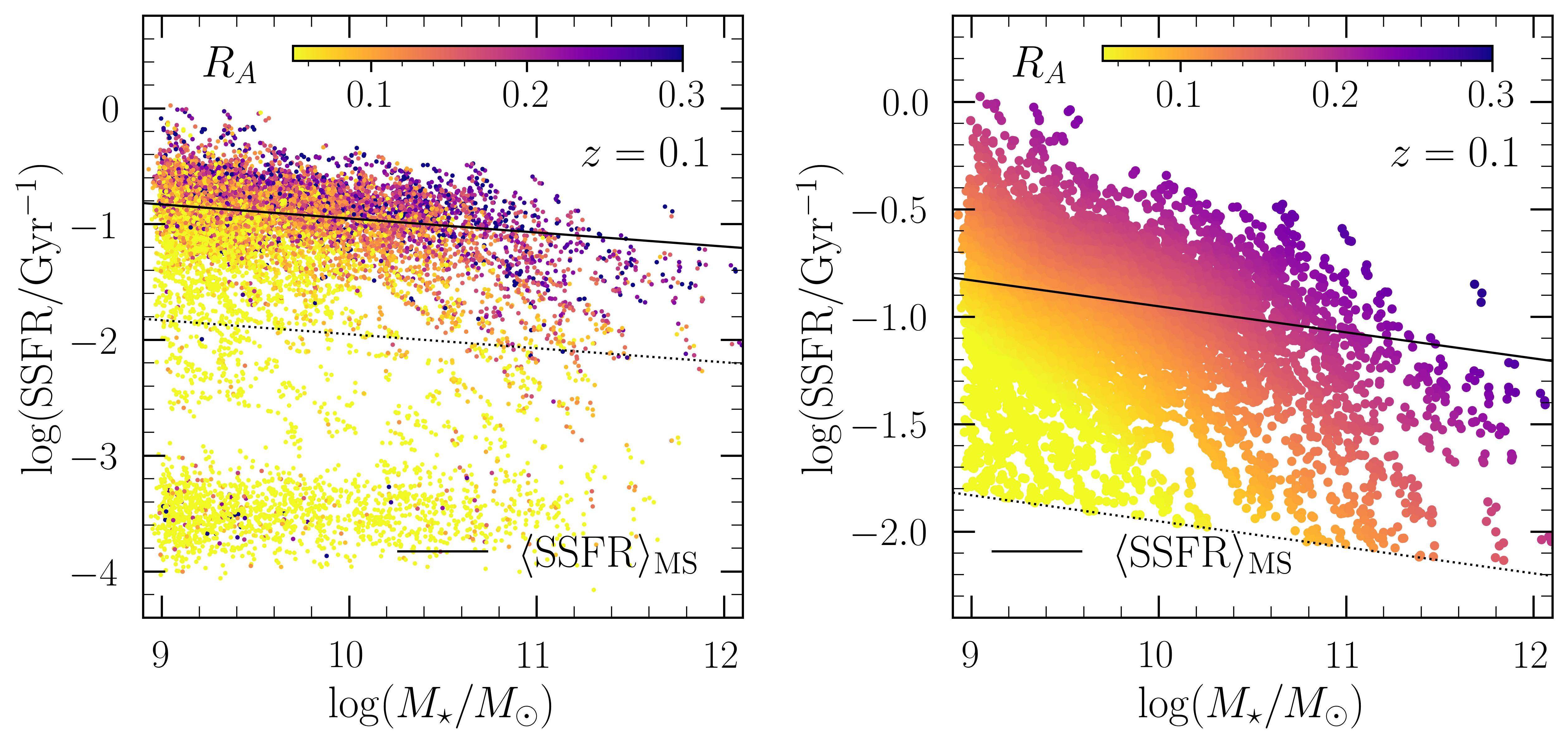}
   \caption[SFMS and Residual Asymmetry]{Star-forming main sequence (SFMS) fitting, star forming galaxy (SFG) selection, and the relationship between SFMS offset and galaxy residual asymmetries for TNG50 galaxies at $z=0.1$. The left panel shows the full sample of 12,088 galaxies ($3,047$ galaxies $\times 4$ sightlines). The solid black line shows the linear fit to the SFMS described in Section \ref{sec:sfms} and in Table \ref{tab:sfms}. The dotted line shows $\dsfms = \log(\mathrm{SSFR}) - \langle \log(\mathrm{SSFR})\rangle_{\mathrm{MS}} = -1$ dex. We define SFGs as those above this line. Each galaxy is coloured by its residual asymmetry, $R_A$ (measured after removal of the best-fitting S\'ersic model). For visual purposes, an artificial $0.05$ dex scatter in SFR and $M_{\star}$ is introduced such that different orientations of the same intrinsic galaxy are not directly atop each other. The right panel shows 2D locally weighted regression in $R_A$ for the star-forming sample (LOESS: \citealt{1988Cleveland2DLoess,2013MNRAS.432.1862C}). The trend between $\dsfms$ and $R_A$ is qualitatively similar to that shown by \cite{2021ApJ...923..205Y}.}
    \label{fig:sfms_ra}
\end{figure*}

The left panel of Figure \ref{fig:sfms_ra} shows the SSFRs and masses of $3,047$ TNG50 galaxies from the $z=0.1$ snapshot (each observed along 4 sightlines for a total of $12,088$). The solid and dashed lines show the best-fit SFMS and star-forming galaxy (SFG) cut, respectively, as described in Section \ref{sec:sfms}. Each galaxy is coloured according to its residual asymmetry -- measured in the S\'ersic model-subtracted residual image as defined in Equation \ref{eq:rasym}. The left panel shows the strong trend between galaxy asymmetry and the offset from the star-forming main sequence at fixed stellar mass, $\dsfms = \log(\mathrm{SFR}) - \langle \log(\mathrm{SFR})\rangle_{\mathrm{MS}}$. The right panel of Figure \ref{fig:sfms_ra} shows the 2D locally weighted regression fitting to the SFG sample (LOESS: \citealt{1988Cleveland2DLoess,2013MNRAS.432.1862C}). At fixed stellar mass, the morphological asymmetries of TNG50 galaxies serve as a strong predictors of their SFRs.

This monotonic trend between $\dsfms$ and $R_A$ for SFGs in TNG50 is very similar to that presented by \cite{2021ApJ...923..205Y} for observed galaxies. Using a spectroscopic sample of face-on SFGs from the Sloan Digital Sky Survey (SDSS) Stripe 82 co-added scans \citep{2014ApJ...794..120A,2016MNRAS.456.1359F,2014ApJS..213...12J} with morphological measurements from \cite{2019MNRAS.486..390B}, \cite{2021ApJ...923..205Y} identified residual asymmetry as the structural parameter with the strongest link to $\dsfms$ at fixed stellar mass compared to: concentration; stellar velocity dispersion; half-light radius; S\'ersic index; and mass surface density. Though we do not conduct a more rigorous mutual information analysis as done by \cite{2021ApJ...923..205Y}, we have examined the relationships between $\dsfms$ and our other structural parameters and find that residual asymmetry and asymmetry exhibit the strongest trend with $\dsfms$. 

The trend between $\dsfms$ and $R_A$ shown in Figure \ref{fig:sfms_ra} also exhibits stellar mass dependence. The average asymmetries of TNG50 galaxies along the SFMS increase with stellar mass. On some level, this result is dependent on the approach to the main sequence fit. We estimate the SFMS using galaxies with $\logMstar<10.6$ and extrapolate beyond. In contrast, \citet{2021ApJ...923..205Y} adopt their SFMS slope from \citet{2014ApJS..214...15S} -- which makes no such restriction in stellar mass. Consequently, the downward bend in the SFMS at $\logMstar>10.5$ can greatly steepen the slope measurement. Our $z\approx0.1$ slope estimate for the specific star forming main sequence, $(\alpha-1) =  -0.12$, is considerably shallower than the $(\alpha-1) =  -0.52$ slope used by \cite{2021ApJ...923..205Y}. Therefore, our trend of increasing average asymmetry along the SFMS is stronger than theirs. 


\subsection{Mergers and offsets from the SFMS}\label{sec:sfms_merg}

\begin{figure*}
\centering
	\includegraphics[width=\linewidth]{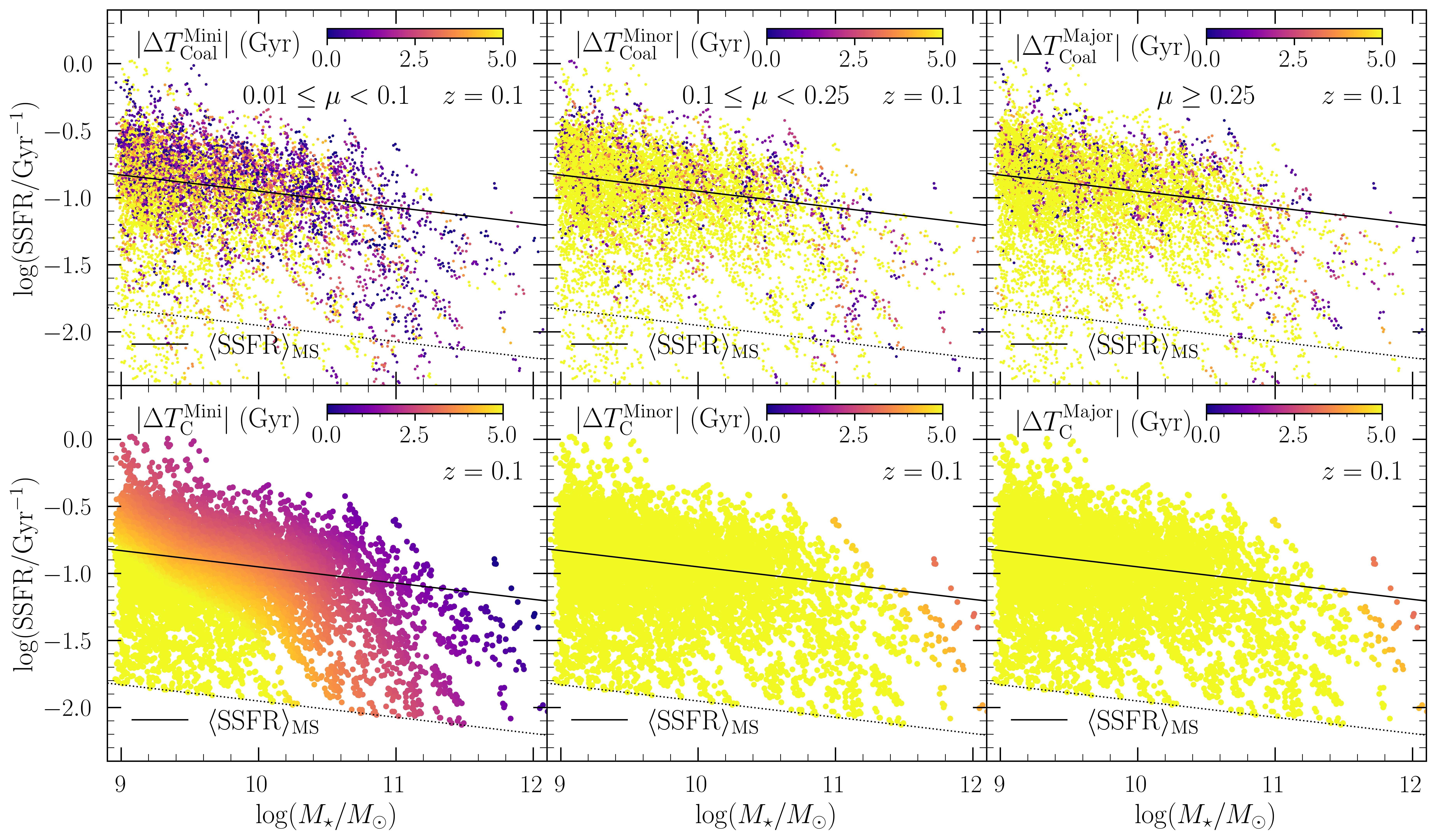}
   \caption[SFMS and $\Delta T_\mathrm{C}$]{Relationship between star-forming main sequence offsets of $\logMstar\geq9$ TNG50 galaxies at $z=0.1$ and time since/until coalescence, $|\dtcoal|$, for mini mergers ($0.01 \leq \mu < 0.1$; left panels), minor mergers ($0.1 \leq \mu < 0.25$; middle panels), and major mergers ($\mu \geq 0.25$; right panels). The $|\dtcoal|$ parameter is the time to the closest merger for a given mass ratio, tracing both forward and backward in time. Immediate remnants of a merger have $\dtcoal=0$. Galaxies which have no mergers in their past nor future are are given $\dtcoal$ equal to the age of the Universe ($[0.16, 3.1, 2.4]$ per cent for [mini, minor, major] mergers). Upper panels show the raw data. Lower panels show the LOESS smoothed data as in Figure \ref{fig:sfms_ra}. Recent and impending mini merger activity strongly correlates with offsets from the SFMS. Minor and major mergers are too rare to yield a visual trend under locally-weighted regression (see also Figure \ref{fig:mus} for merger mass ratio frequencies).}
    \label{fig:sfms_dtc}
\end{figure*}

The observational trend between asymmetry and SFR at fixed stellar mass is reproduced in our TNG50 mocks. This result gives us the impetus to investigate the origins of this relation in connection to known merger histories and forecasts from the simulations. In this section, we offer a comparison between the trend shown in Figure \ref{fig:sfms_ra} and the SFMS colourized by the coalescence time offsets, $|\dtcoal|$, of each galaxy's nearest merger (past or future) within a given stellar mass ratio range: mini, minor, and major (Equation \ref{eq:dtc}).

Figure \ref{fig:sfms_dtc} shows the relationship between $\dsfms$ and $|\dtcoal|$ for mini mergers (left panels), minor mergers (middle panels), and major mergers (right panels). At fixed stellar mass, TNG50 galaxies that are close to coalescence for mini mergers are elevated in on the SFMS, on average. Galaxies that are far from coalescence for mini mergers have suppressed SFRs. Meanwhile, galaxies with low major and/or minor merger $|\dtcoal|$ do exhibit elevated SFRs along the SFMS in the scatter plots, but are too infrequent to be significant contributors to the intrinsic SFMS scatter in TNG50. This mild role of major and minor mergers in TNG50 is consistent with \citet{2020MNRAS.494.4969P} using TNG100, who showed that the average contribution of major and minor mergers, $\mu \geq 0.1$, to SFR enhancements is $\sim0.06$ dex at fixed mass -- and smaller still for EAGLE \citep{2015MNRAS.446..521S,2015MNRAS.450.1937C} and Illustris \citep{2013MNRAS.436.3031V,2014MNRAS.444.1518V,2014MNRAS.438.1985T}. In contrast, the coalescence offsets we show for \emph{mini mergers alone} provide a clear trend with $\dsfms$ at any given stellar mass due to their relatively high frequency (Figure \ref{fig:mus}). Mini mergers dominate the trend.

The qualitative similarity of the trends in $\dsfms$ with $R_A$ and $\dtcoal$ shown in Figures \ref{fig:sfms_ra} and \ref{fig:sfms_dtc} is remarkable. This correlation suggests that mergers, particularly \emph{mini} ones, are significant drivers of the morphological asymmetries that are simultaneously tied to the intrinsic scatter in the SFMS (or, at least, \emph{probing} a significant driver). These results are qualitatively consistent with observational and computational works identifying \emph{sub-major} mergers as significant contributors to the cosmic star formation budget (e.g. \citealt{1996ApJ...473L..21S,2004A&A...423..481K,2014MNRAS.437L..41K,2014MNRAS.440.2944K,2019MNRAS.489.4679J}) and to the more modest morphological asymmetries and lopsidedness exhibited by the majority share of SFGs in the local Universe (e.g. \citealt{1997ApJ...477..118Z,2005A&A...438..507B,2008MNRAS.388..697M,2009PhR...471...75J,2022MNRAS.511.5878G,2023MNRAS.526..567D,2023MNRAS.523.5853V}). Therefore, while SFR or $R_A$ enhancement from individual minor or mini mergers is expected to be less than that of major mergers with otherwise similar initial conditions (e.g. \citealt{2003ApJ...597..893N,2005A&A...437...69B,2008MNRAS.384..386C,2010MNRAS.404..575L,2014MNRAS.445.1157C}) low-$\mu$ mergers greatly outnumber high-$\mu$ mergers (e.g. \citealt{2001MNRAS.326..255C,2006ApJ...647..763M,2008ApJ...683..597S,2010A&A...518A..20L,2011ApJ...742..103L,2014MNRAS.445.2198O,2019ApJ...876..110D,2021ApJ...919..139W,2022ApJ...940..168C}) -- resulting in a significant integrated effect.

\subsection{Connection between asymmetry and star formation}\label{sec:sfg_merg}

\begin{figure}
\centering
	\includegraphics[width=\linewidth]{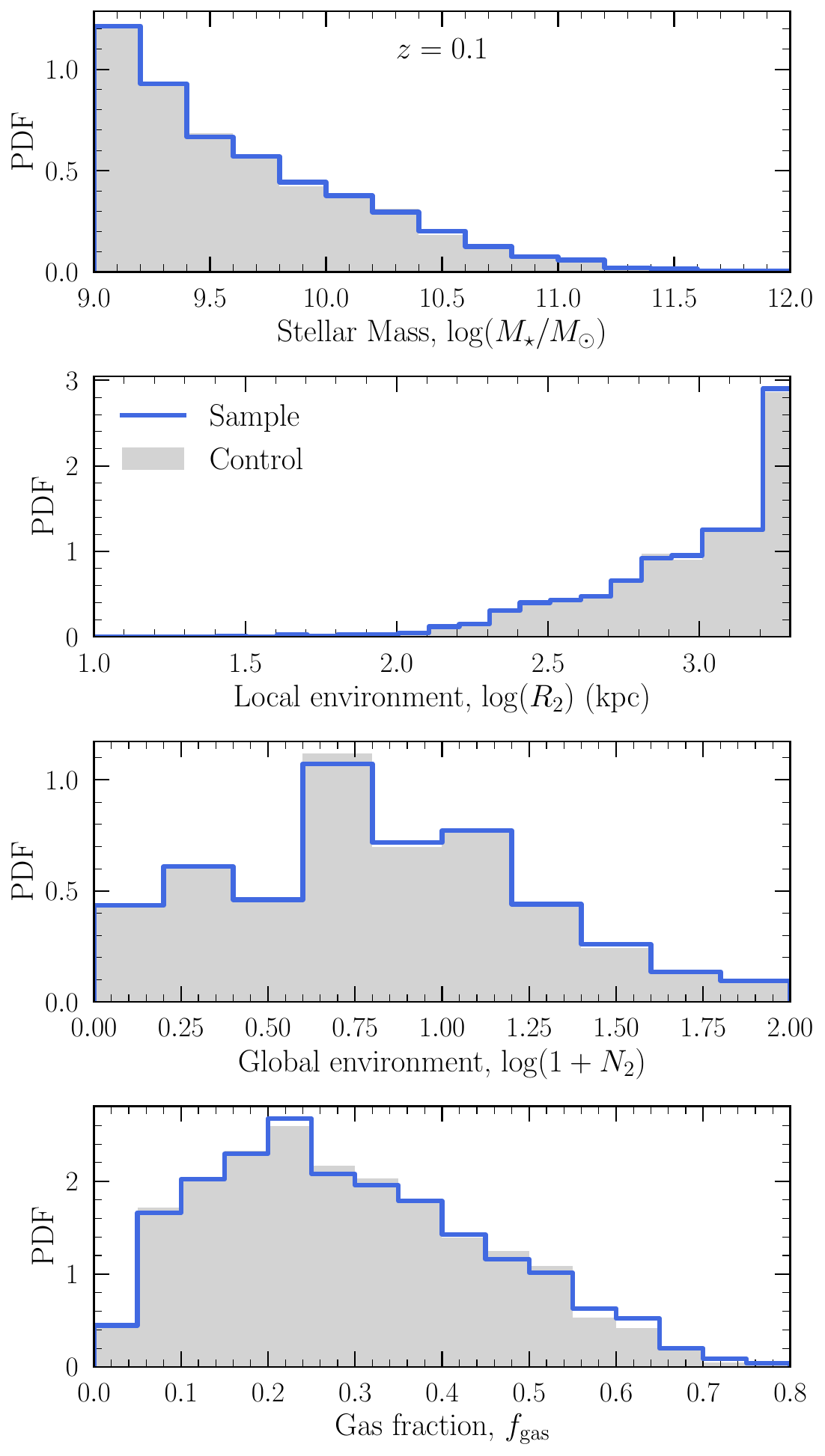}
   \caption[Control matching]{Probability distribution functions of SFGs and their statistical controls matched in stellar mass, local environment, large scale environment, and gas fraction. Only the $z=0.1$ snapshot is shown. Each sample SFG is matched to a set of $N_{\mathrm{controls}}$ controls from the same redshift. By examining offsets between SFR, asymmetry, and $\dtcoal$ between samples and their matched controls, we aim to remove the influence of the controlled parameters. We highlight that the specific control matching scheme used in Section \ref{sec:sfms_merg} is agnostic to whether samples or controls are mergers. We have asserted the quantitative robustness of our results for $N_{\mathrm{controls}}\leq3$. The distributions for $N_{\mathrm{controls}}=1$ are shown above.}
    \label{fig:control_matching}
\end{figure}

The qualitative connection between SFRs, asymmetries, and merger coalescence time offsets shown in Figures \ref{fig:sfms_ra} and \ref{fig:sfms_dtc} warrants a more rigorous examination of this relationship. In this subsection, we use the control-matching framework laid out in Section \ref{sec:controls} to examine how differences in asymmetry and time-to-coalescence relate to differences in SFR \emph{between otherwise similar SFGs}.

The selection of the sample and control pools used in this specific experiment is as follows. At each snapshot (redshift), we take the full SFG population as both the sample \emph{and} control pool. Each SFG is matched to a set of other SFGs in stellar mass, local and global factors of environment, and gas fraction. Meanwhile, the matching is agnostic to the asymmetry, time-to-coalescence, and SFR offsets between matched SFGs. Figure \ref{fig:control_matching} shows the resulting distributions of stellar mass, environments, and gas fractions for the $z=0.1$ snapshot SFG sample and their matched controls -- which are unsurprisingly similar on two counts: (1) the control matching itself and (2) the fact that the sample and control pool are the same.

Using the weighting scheme from Section \ref{sec:controls}, we then compute the offsets in asymmetry, time-to-coalescence, and SFR between matched SFGs:
\begin{align}
&\dra = (R_{A})_{\mathrm{sample}} - \sum_{i=1}^{N_{\mathrm{controls}}} w_i (R_{A})_{i,\mathrm{control}}\\
&\dlogsfr = \logsfr_{\mathrm{sample}} - \sum_{i=1}^{N_{\mathrm{controls}}} w_i \logsfr_{i,\mathrm{control}}\\
&\dtcoaloffset = |\dtcoal|_{\mathrm{sample}} - \sum_{i=1}^{N_{\mathrm{controls}}} w_i |\dtcoal|_{\mathrm{control}} \label{eq:dtcoffset}
\end{align}
By these definitions, a positive (negative) $\dra$ or $\dlogsfr$ mean that the sample SFG has an asymmetry or SFR that is enhanced (suppressed) with respect to the mean of its controls. Similarly, offsets in $\dtcoaloffset$ can be generalized as follows:
\begin{align}\label{def:dtcoffset}
\dtcoaloffset
    \begin{cases}
       < 0: \text{sample is closer to a merger than controls} \\
       > 0: \text{sample is farther from a merger than controls} \\
       = 0: \text{equally close to a merger}
    \end{cases}
\end{align}
One notable caveat is that when considering mergers of all types, $\dtcoaloffset$ is more likely to correspond to an offset to lower mass ratio merger than higher mass ratio mergers (Figure \ref{fig:mus}). The $\dtcoaloffset$ parameter can also be used to specify coalescence offsets for only mini, minor, or major mergers -- ignoring mergers of other types in the coalescence offset calculation. However, restricting $\dtcoaloffset$ to lower mass ratio mergers (mini and minor) ignores any higher mass ratio merger for the purposes of computing $\dtcoaloffset$ -- even if it is closer in time. In such cases, any impact of a higher mass ratio merger on SFR and asymmetry may be erroneously ascribed to the lower mass ratio merger for which $\dtcoaloffset$ is computed -- contributing both upward and downward scatter in $\dra$ and $\dtcoal$. It important to consider this point in the context of the results presented in the following sections.

\begin{figure}
\centering
	\includegraphics[width=\linewidth]{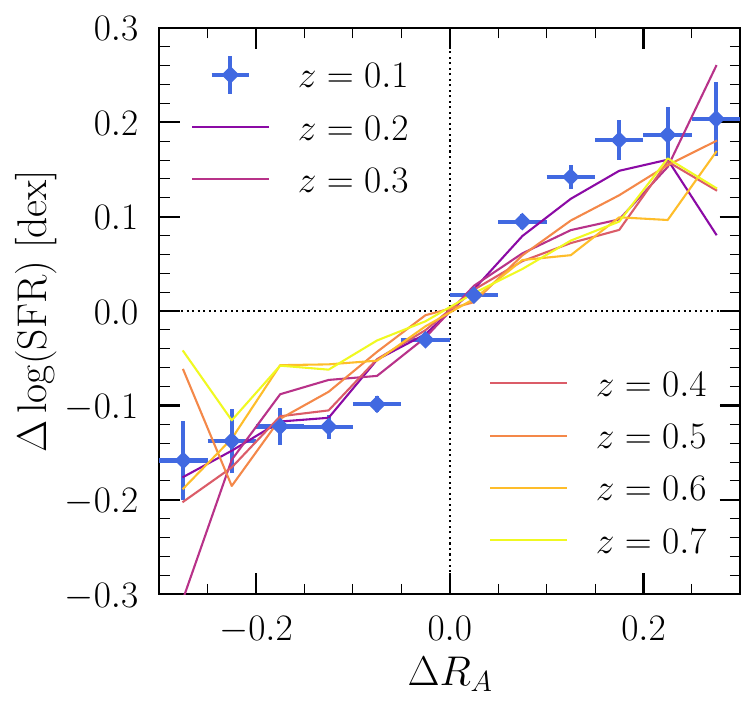}
   \caption[$\dlogsfr$ vs $\dra$]{Relationship between the enhancements/suppression in star formation rates, $\dlogsfr$, and residual asymmetries, $\dra$, between star-forming galaxies (SFGs) matched in redshift, stellar mass, environment, and gas fraction for $z=0.1-0.7$ (see Sections \ref{sec:controls} and \ref{sec:sfg_merg} for details). SFGs that are enhanced (suppressed) in asymmetry compared to their matched SFGs exhibit clear enhancement (suppression) in SFRs. The trend persists at all redshifts we consider, though with slightly mildly decreasing sensitivity of $\logsfr$ to $\dra$ with increasing redshift. Standard deviations in the mean of $\dlogsfr$ in each $\dra$ bin are shown for the $z=0.1$ sample only.  A minimum of 50 samples per bin are required. }
    \label{fig:delsfr_delra}
\end{figure}

Armed with $\dra$, $\dlogsfr$, and $\dtcoaloffset$ for TNG50 SFGs relative to \emph{otherwise similar} SFG controls, we examine the relationships between these offsets. Figure \ref{fig:delsfr_delra} shows the response of $\dlogsfr$ to $\dra$ for SFGs between $z=0.1-0.7$. SFGs that are enhanced in asymmetry compared to their controls exhibit clear enhancements in SFRs at all redshifts. For galaxies with the largest asymmetry offsets from their controls, $\dra \approx0.3$, the mean enhancement is $\dlogsfr \approx 0.2$ dex or a factor of 1.6. This quantitative trend between carefully controlled offsets in SFRs and asymmetry confirms the more qualitative trend shown in Figure \ref{fig:sfms_ra}, also identified by \cite{2021ApJ...923..205Y} for real galaxies. 

Enhancements in $\dlogsfr$ as a function of $\dra$ are slightly less pronounced with increasing redshift. This lack of sensitivity may owe to increasing systematic and random errors in asymmetry measurements at higher redshifts (e.g. Figure \ref{fig:asy_evo}). The trend of decreasing sensitivity of $\dlogsfr$ to $\dra$ with increasing $z$ in Figure \ref{fig:delsfr_delra} is expected to arise as a result of larger systematic and random errors on asymmetry measurements. Nonetheless, the trend is visible at all redshifts considered. So while systematic and random errors on $R_A$ measurements increase with redshift, they are not so large as to wash out the connection to SFRs for a galaxy population \emph{at fixed redshift}. In other words, even if measured asymmetries systematically decrease at higher redshifts, relative asymmetries at fixed redshift are still useful predictors of relative SFRs between SFGs. 


\subsection{Connection between asymmetry and mergers}

\begin{figure*}
\centering
	\includegraphics[width=\linewidth]{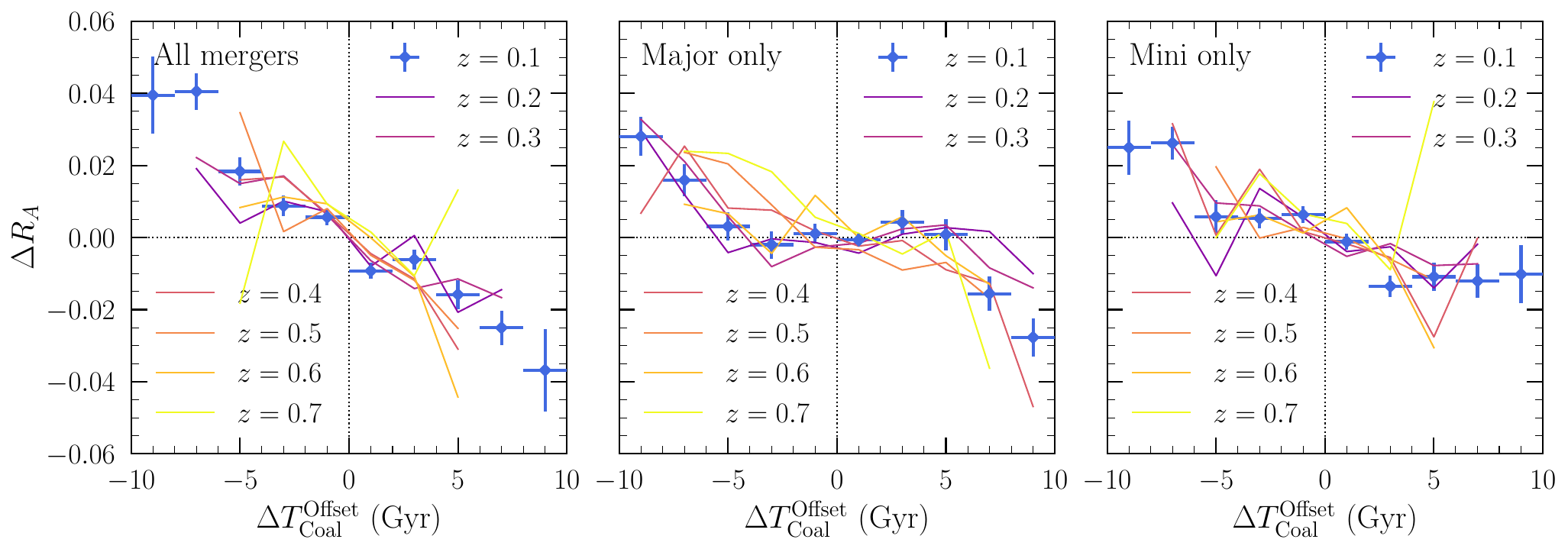}
   \caption[$\dra$ vs $\dtcoal$]{Relationship between the enhancements/suppression of morphological asymmetries, $\dra$, as a function of differences in merger coalescence time, $\dtcoaloffset$ (Eqs. \ref{eq:dtcoffset} and \ref{def:dtcoffset}), between SFGs matched in redshift, stellar mass, environment, and gas fraction for $z=0.1-0.7$. $\dtcoaloffset$ is calculated using the coalescence time offsets of samples and controls considering all $\mu \geq 0.01$ mergers (left panel), major mergers only (middle), and mini mergers only (right). SFGs that are closer to a merger of any mass ratio ($\dtcoaloffset<0$) exhibit clear enhancement in asymmetry compared to \emph{otherwise similar SFGs} that are farther from coalescence. When $\dtcoaloffset$ considers major mergers only, the asymmetries of SFGs are consistent with their matches within $-5<\dtcoaloffset<5$ Gyr -- whereas mini mergers exhibit statistically significant enhancements in this regime.}
    \label{fig:dra_dtc}
\end{figure*}

Figure \ref{fig:sfms_dtc} showed that, at fixed stellar mass, $\dsfms$ is correlated with merger activity -- specifically mini mergers with $0.01 \leq \mu < 0.1$ -- in a trend that mirrors its correlation with structural asymmetry. We now use our control-matched SFG samples spanning $z=0.1-0.7$ to isolate the role of mergers as drivers of asymmetry in TNG50 SFGs.

\subsubsection{Asymmetry offsets and merger timescale offsets in SFGs}

The left panel of Figure \ref{fig:dra_dtc} shows that $\dra$ between samples and controls exhibit a clear sensitivity to their $\dtcoaloffset$ when considering all mass ratios $\mu \geq 0.01$. The offsets in asymmetry peak at $\dra=0.039\pm0.010$ for SFGs that are $8-10$ Gyr closer to a merger than their matched controls (essentially guaranteeing that the controls have quiescent merger histories and forecasts). This average $R_A$ enhancement is consistent, for example, with the $\dra = 0.040\pm0.011$ shown by \cite{2016MNRAS.461.2589P} for minor and major spectroscopic merging pairs at separations less than $10$ kpc and rest-frame line-of-sight velocity offsets less than $300$ km$\;$s$^{-1}$.

The middle panel of Figure \ref{fig:dra_dtc} shows the relationship between $\dra$ and major merger $\dtcoaloffset$ only -- i.e. ignoring mini and minor mergers in the calculation of $\dtcoaloffset$ between samples and controls. For $|\dtcoaloffset|<5$ Gyr, the asymmetries of SFGs and their controls are consistent. At wider offsets in $\dtcoaloffset$, asymmetries are enhanced and reach $\dra=0.027$ in comparisons between SFGs which are very close to major merger coalescence and otherwise similar SFGs with no major merger in their forecast or history. The insensitivity of $\dra$ to $\dtcoaloffset$ for $-5<\dtcoaloffset<5$ Gyr is due to the high average major merger $|\dtcoal|$ for SFGs in this redshift interval. The average $|\dtcoal|$ to a minor or major merger for a given SFG from our sample is $5.8$ and $6.4$ Gyr, respectively ($4.2$ Gyr combined). Consequently, within $-5<\dtcoaloffset<5$ Gyr for minor and major mergers, neither the sample nor its controls are likely to be near coalescence. Conversely, for $|\dtcoaloffset|>5$ Gyr, the boundary set by the age of the Universe forces either the sample or controls to be near coalescence and its counterpart to be far from coalescence, yielding statistically significant offsets in asymmetry.

The right panel of Figure \ref{fig:dra_dtc} shows the same relation considering only mini mergers in the calculation of $\dtcoaloffset$. The average mini merger $|\dtcoal|$ for our SFG sample is 1.9 Gyr. Consequently, $\dra$ is more sensitive to small changes in $\dtcoaloffset$ -- including those in $-5<\dtcoaloffset<5$ Gyr where no significant enhancement or suppression is seen with respect to minor or major merger $\dtcoaloffset$. Given the relatively high frequency of mini mergers compared to minor/major, small offsets in asymmetry between otherwise similar galaxies (redshift, mass, environment, gas fraction) are much more likely to arise from differences in mini merger history and forecast. Additionally, for $|\dtcoaloffset|>5$ Gyr, mini mergers appear to yield $\dra$ that rival those of major mergers. This apparent similarity is a consequence of the small number minor and major mergers amongst SFG samples that are (1) near coalescence and (2) matched to controls which are far from coalescence. In other words, when considering only major mergers in the calculation of $|\dtcoal|$, nothing prohibits a sample or control from being very close to coalescence in a minor or mini merger -- which are relatively frequent and drive statistically significant asymmetries (right panel). The consequence is an artificial suppression of the amplitudes of asymmetry offsets between samples and controls as a function of major merger $\dtcoaloffset$ due to the ``noise'' in $\dra$ from interloping sub-major mergers. Later in Section \ref{sec:iso_merg}, we conduct a separate analysis examining changes in morphological asymmetry in response to individual mergers in each mass ratio range compared to controls with no $\mu \geq 0.01$ mergers within a wide time interval. 

\subsubsection{Merger statistics as a function of asymmetry in SFGs}\label{sec:ra_statistics}

\begin{figure}
\centering
	\includegraphics[width=\linewidth]{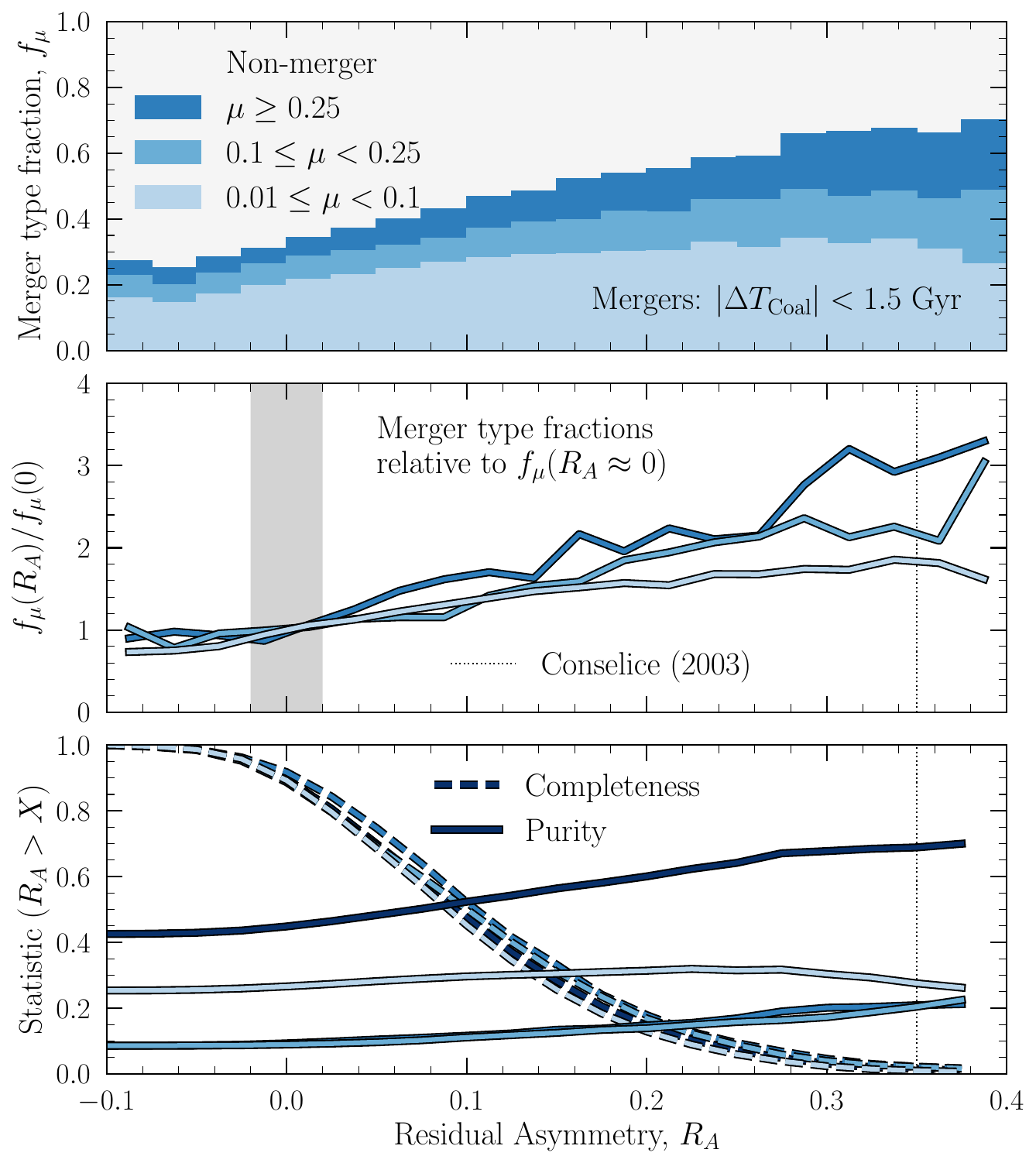}
   \caption[$R_A$ by merger type]{Statistics of merging SFGs near coalescence, $| \dtcoal |<1.5$ Gyr, as a function of HSC residual asymmetry for $z\leq 0.3$. \emph{Upper panel}: The merger fraction in each bin, $f_\mu$, split into major, minor, and mini regimes. The unshaded area shows the fraction of SFGs that have no $\mu\geq0.01$ merger within $1.5$ Gyr. Compared to minor and major mergers combined, mini mergers dominate for $R_A<0.3$. $68$ per cent of galaxies at $R_A>0.3$ are close to a merger -- of which $47$ per cent are mini mergers with no major or minor merger within $1.5$ Gyr. $32$ per cent of galaxies with asymmetries $R_A>0.3$ are non-mergers. \emph{Middle panel}: Merger type fractions relative to $f_\mu(R_A\approx0)$ (grey band). The fraction of high-$\mu$ mergers increases more rapidly with $R_A$ than low-$\mu$ mergers. \emph{Lower panel}: Completeness and purity curves as a function of asymmetry threshold, $R_A>X$. Dark lines show statistics for all $\mu>0.01$ mergers.}
    \label{fig:ra_mergers}
\end{figure}

Figure \ref{fig:ra_mergers} examines merger statistics amongst SFGs as a function of asymmetry. The upper panel shows the fractions of SFGs, $f_\mu$, near coalescence for mini, minor, and major mergers, respectively, using a $| \dtcoal |<1.5$ Gyr cut. A given galaxy can satisfy this condition for multiple mergers. Therefore, for a given galaxy, only the most massive merger satisfying this cut is counted such that total galaxy counts in each bin are conserved. The unshaded area shows the fractions of non-mergers -- which do not satisfy $| \dtcoal |<1.5$ Gyr for any mass ratio range\footnote{Figure \ref{fig:ra_mergers} is not examining statistical offsets in $R_A$ between matched controls. Therefore, we restrict our focus to $z\leq0.3$, where asymmetry measurements are most robust to changes in redshift (Figure \ref{fig:asy_evo}).}. The relative frequencies of mini mergers dominates minor and major mergers for all $R_A<0.3$. For $R_A>0.3$, approximately $68$ per cent of galaxies satisfy $| \dtcoal |<1.5$ Gyr for a $\mu\geq0.01$ merger -- of which $47$ per cent are mini mergers with no major or minor merger satisfying $|\dtcoal|<1.5$ Gyr (i.e. $32$ per cent of all SFGs with $R_A>0.3$; see also precision curves in lower panel).  

The middle panel of Figure \ref{fig:ra_mergers} compares the frequencies of mergers as a function of $R_A$ to their \emph{respective} frequencies at $R_A\approx0$ (grey band). The incidence of $\mu\geq0.01$ mergers amongst SFGs approximately doubles from $R_A=0$ to $R_A=0.25$. At $R_A>0.3$, a SFG is three times more likely to be involved in a major merger than at $R_A=0$. The frequency ratio of minor and major to mini mergers, $(f_{\mathrm{minor}}+f_{\mathrm{major}})/f_{\mathrm{mini}}$, is constant at around $60$ per cent for all $R_A<0.1$ (upper panel). This ratio then increases rapidly at higher asymmetries: $(104, 123, 150, 167)$ per cent for $R_A>(0.25, 0.30, 0.35, 0.375)$, respectively. Taken together, these results show that with increasing asymmetry, major and minor mergers become increasingly likely culprits as the physical driver of the asymmetry compared to mini mergers. 

The lower panel of Figure \ref{fig:ra_mergers} shows the completeness (recall, dashed) and purity (precision, solid) of galaxies satisfying $| \dtcoal |<1.5$ Gyr for each mass ratio range above a given cut in $R_A$ (dark curve for all $\mu \geq 0.01$). The selection completenesses for mini, minor, and major mergers differ only modestly -- all of which decline rapidly with increasing asymmetry. However, higher mass ratios have generally higher completeness for a given asymmetry cut, as expected. The purity curves, however vary greatly, as their respective normalizations are dominated by the intrinsic fractions of mergers in each mass ratio range (e.g. \citealt{2021MNRAS.504..372B,2022MNRAS.511..100B,2022A&A...661A..52P,2023arXiv230211288P,2023MNRAS.521.3861D}). Still, the increase in mini merger purity with increasing $R_A$ closely matches the increases in major and minor merger purity. Above $R_A=0.25$, the mini merger purity is suppressed as major and minor mergers frequencies become comparable. But even for $R_A>0.35$ (analogous to the \citealt{2003ApJS..147....1C} asymmetry cut), approximately $30$ per cent of SFGs have no major or minor mergers within $|\dtcoal|<1.5$ Gyr but have \emph{at least one} mini merger in this window (lower panel, precision). Consecutive mini mergers may contribute to the high asymmetries in these galaxies. It is important to note, however, that only $2.0$ per cent of SFGs have $R_A>0.3$ -- whereas $62$ per cent of SFGs with $|\dtcoal|<1.5$ Gyr for any merger (predominantly mini) have asymmetries $R_A<0.1.$ 


\begin{figure*}
\centering
	\includegraphics[width=\linewidth]{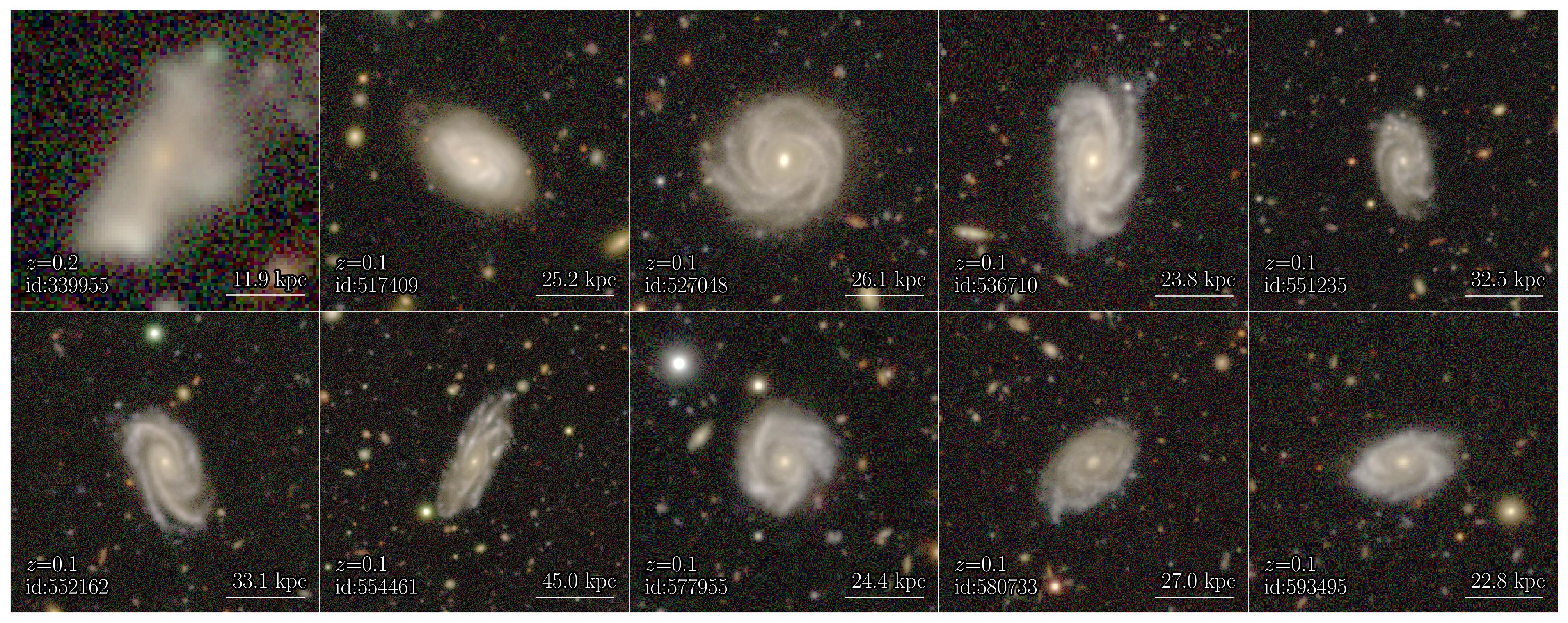}
   \caption[Non-mergers with high $R_A$]{Randomly selected SFGs from $z\leq 0.3$ with $R_A>0.25$ and no merger in their history or forecast satisfying $| \dtcoal |<1.5$ Gyr for any mass ratio $\mu \geq 0.01$. $35$ per cent of SFGs with $R_A>0.25$ have no major/minor/mini merger satisfying $| \dtcoal |<1.5$. The majority of high-$R_A$ galaxies with no recent/forthcoming merger appear to be star-forming late-type galaxies with asymmetric spiral structure -- though they may also be early pairs, fly-bys, or late merger remnants with long-lasting lopsidedness / stellar halo substructure.}
    \label{fig:ra_nonmergers}
\end{figure*}

Meanwhile, approximately $35$ per cent of TNG50 SFGs with $R_A>0.25$ do not satisfy $|\dtcoal|<1.5$ Gyr for any $\mu\geq0.01$ merger. Figure \ref{fig:ra_nonmergers} shows a random selection of $10$ such galaxies. All exhibit \emph{bona fide} structural asymmetries. After visually inspecting many other images of galaxies with $| \dtcoal |>1.5$ Gyr and $R_A>0.25$, we also found (1) some badly deblended systems in the presence of nearby field stars, (2) a number of apparent fly-by systems which may yet merge, and (3) merger remnants with long-lasting stellar halo structures such as shells (e.g. \citealt{1983ApJ...274..534M,1989ApJ...342....1H,1999MNRAS.307..495H,2008ApJ...689..936J,10.1093/mnras/sty1932,2020MNRAS.492.2075B}) However, the large majority were simply asymmetric late-types with no \emph{clear} evidence of a past/ongoing interaction, such as those in Figure \ref{fig:ra_nonmergers}.


\subsection{Connection between star formation and mergers}

\begin{figure*}
\centering
	\includegraphics[width=\linewidth]{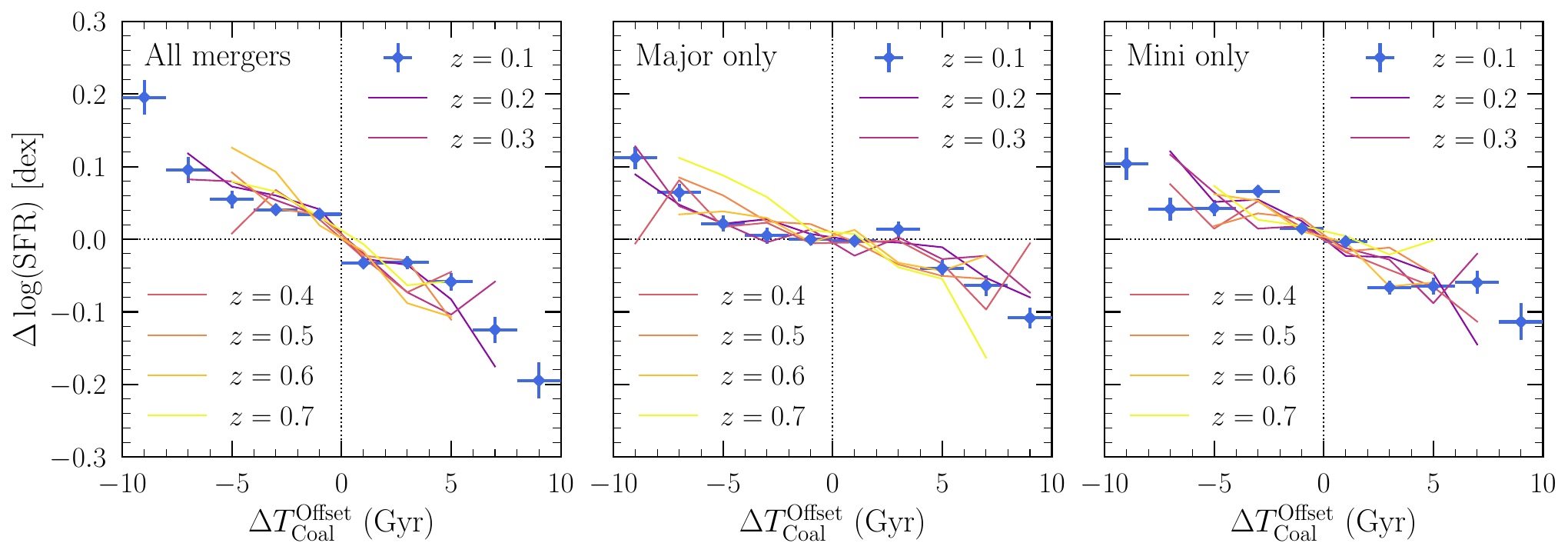}
   \caption[$\dlogsfr$ vs $\dtcoal$]{Relationship between the enhancements/suppression of star formation rates, $\dlogsfr$, as a function of merger coalescence time offsets, $\dtcoaloffset$ (Eqs. \ref{eq:dtcoffset} and \ref{def:dtcoffset}), between SFGs matched in redshift, stellar mass, environment, and gas fraction for $z=0.1-0.7$. As in Figure \ref{fig:dra_dtc}, the three panels show the relationship using $\dtcoal$ calculated using the closest merger of any mass ratio (left panel), the closest major merger (middle panel), and the closest mini merger (right panel). Galaxies that are closer to merger of any mass ratio ($\dtcoaloffset<0$) exhibit enhanced SFRs compared to \emph{otherwise similar SFGs} that are farther from coalescence. When $\dtcoaloffset$ considers major mergers only, sample SFRs are consistent with their matches within $-5<\dtcoaloffset<5$ Gyr -- whereas mini mergers exhibit statistically significant enhancements in this regime.}
    \label{fig:dsfr_dtc}
\end{figure*}

\begin{figure}
\centering
	\includegraphics[width=\linewidth]{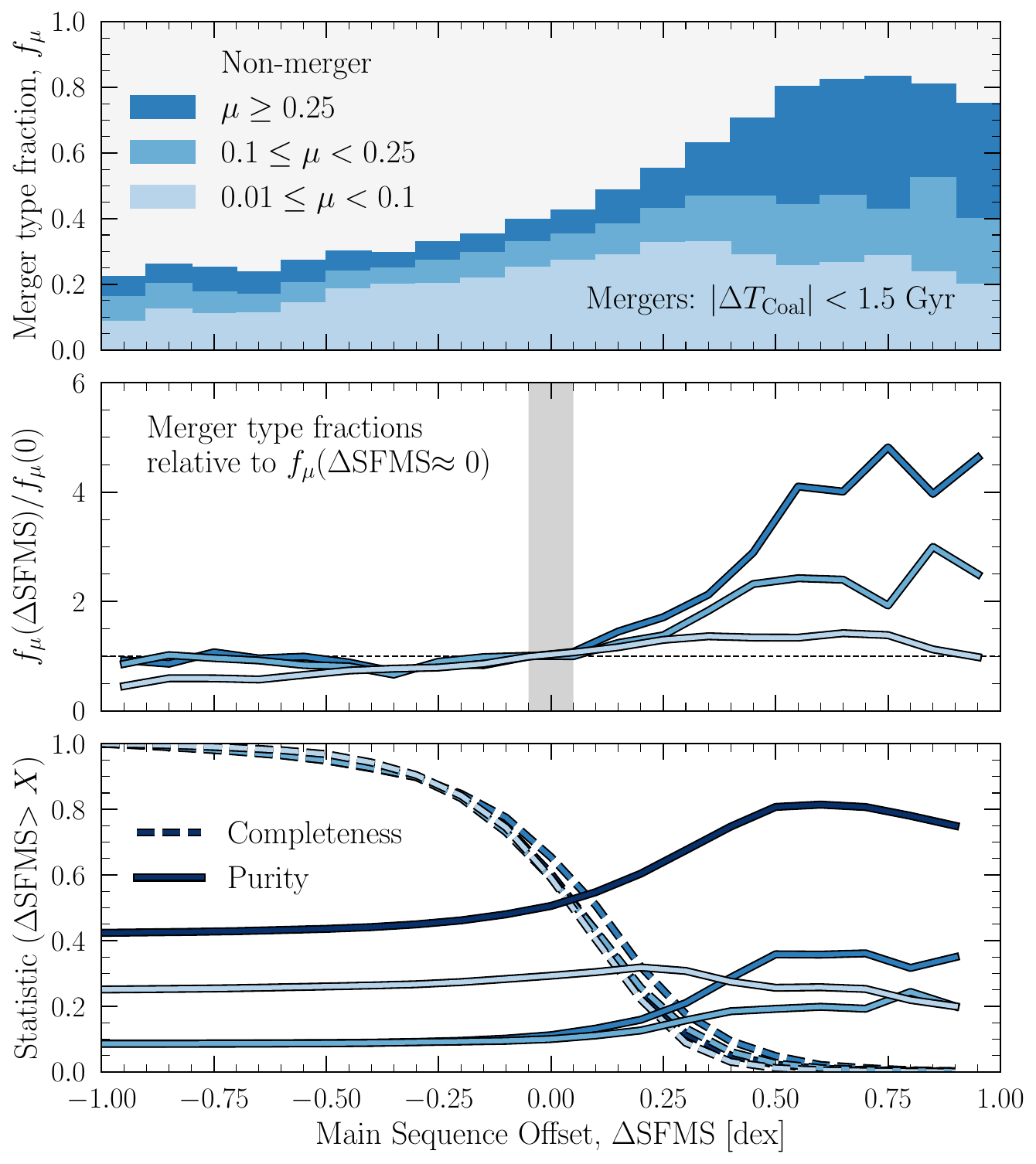}
   \caption[$\dsfms$ by merger type]{Statistics of merging SFGs near coalescence, $| \dtcoal |<1.5$ Gyr, as a function of SFMS offset, $\dsfms$, for $z\leq 0.3$. \emph{Upper panel}: The merger fraction in each bin, $f_\mu$, split by mass ratio: major ($\mu \geq 0.25$); minor ($0.1\leq \mu < 0.25$); and mini ($0.01 \leq \mu < 0.1$). The unshaded area shows the fraction of SFGs that do not satisfy the $| \dtcoal |<1.5$ Gyr criterion for any merger type. According to TNG50, $81$ per cent of starburst galaxies ($\dsfms>0.5$) are mergers. Of these, $32$ per cent are mini mergers with no minor or major merger within $1.5$ Gyr ($26$ per cent of all SFGs with $\dsfms>0.5$). The incidence of mini mergers dominates for $\dsfms<0.4$ dex compared to minor and major mergers combined. \emph{Middle panel}: Merger type fractions relative to $f_\mu(\dsfms \approx0)$. Minor and major mergers double in frequency amongst SFGs from $\dsfms=0$ dex to $\dsfms=0.35$ dex. Mini mergers are $50$ per cent less frequent at $\dsfms = - 0.7$ dex and $50$ per cent more frequent at $\dsfms = 0.7$ dex compared to $\dsfms=0$ dex. \emph{Lower panel}: Completeness (recall) and purity (precision) curves as a function of main sequence offset threshold, $\dsfms>X$ dex. Dark lines show these statistics considering all mass ratio ranges (completeness is hidden behind other mass ratio curves).}
    \label{fig:sfms_mergers}
\end{figure}


In the last section, we showed that TNG50 SFGs which are nearer to coalescence compared to their matched controls exhibit enhanced morphological asymmetries in their HSC-SSP synthetic imaging data. In particular, we showed that mini mergers contribute strongly to the asymmetries of galaxies due to their higher frequency relative to minor and major mergers. We now examine the relationship between coalescence proximity and specific SFR offsets between SFGs and matched controls. 

\subsubsection{SFR offsets and merger timescale offsets in SFGs}

Figure \ref{fig:dsfr_dtc} shows the relationship between enhanced/suppressed SFRs, $\dlogsfr$, and coalescence time offsets, $\dtcoaloffset$, between SFGs and their matched controls. The trend seen in Figure \ref{fig:dra_dtc} is mirrored for SFRs. Considering all $\mu\geq0.01$ mergers in the calculation of $\dtcoaloffset$ (left panel), galaxies that are nearer to coalescence $|\dtcoaloffset|<0$ Gyr have enhanced SFRs -- peaking at $\dlogsfr=0.21$ dex for SFGs near coalescence whose controls have relatively quiescent merger histories and forecasts. The enhancements are more modest for smaller $\dtcoaloffset$, but nonetheless statistically significant. For example, SFGs with $\dtcoaloffset=-1$ Gyr have a $0.03$ dex enhancement in SFRs, on average. The middle panel shows the same trend considering only major mergers. As in the previous section, the relatively low frequency of major mergers is responsible for the lack of statistical offsets in $\dlogsfr$ for $|\dtcoaloffset|<5$ Gyr. Beyond, $\dlogsfr$ peaks at $0.12$ dex. However, as also seen for the asymmetries, mini mergers (right panel) produce comparable $\dlogsfr$ offsets to major mergers as a function of their respective $\dtcoaloffset$. Mini mergers drive statistically significant $\dlogsfr$s even for small $\dtcoaloffset$.

We note once again that when computing $\dtcoaloffset$ for major mergers only, nothing prohibits a sample or control from being near coalescence for mini or minor mergers -- regardless of the major merger $\dtcoaloffset$. As with the asymmetries in Figure \ref{fig:dra_dtc}, mini and minor mergers amongst the controls could elevate their SFRs and artificially suppress the sensitivity of $\dlogsfr$ to major merger $\dtcoaloffset$ as a result. Meanwhile, this suppression is less expected of the mini mergers. For a given galaxy, the mini merger $|\dtcoal|$ is likely to be smallest considering the averages are $|\dtcoal| = (1.9, 5.8, 6.4)$ Gyr to mini, minor, and major mergers respectively. Consequently, for mini merger $\dtcoaloffset$, the nearest merger for both samples and controls \emph{are indeed} most likely to be mini mergers (regardless of the restriction on what mass ratio range $\dtcoaloffset$ is computed). As a result, the sensitivity of $\dlogsfr$ to mini merger $\dtcoaloffset$ are relatively \emph{unlikely} to be artificially suppressed by the incidence of mergers from other mass ratio ranges. Later in Section \ref{sec:iso_merg}, we identify individual mergers of each type and compare with non-merging controls to examine the role of individual mergers in driving enhanced star formation and asymmetries.

\subsubsection{Merger statistics as a function of SFMS offsets in SFGs}

Similar to Figure \ref{fig:ra_mergers} for asymmetries, Figure \ref{fig:sfms_mergers} shows the statistics of $|\dtcoal|<1.5$ Gyr merger incidence as a function of offset from the star-forming main sequence at fixed stellar mass. The upper panel shows the fractional incidence of mini, minor, major, and non- mergers in bins of $\dsfms$. There is a clear trend of increasing merger incidence with increasing $\dsfms$. At $\dsfms<-0.5$ dex, only $27$ per cent of galaxies are involved in a mini, minor, or major merger. Meanwhile, $81$ per cent of starburst galaxies ($\dsfms>0.5$ dex) are within $1.5$ Gyr of a merger -- $32$ per cent of which are mini mergers with no overlapping minor/major merger in the same 1.5 Gyr window ($26$ per cent of all starburst galaxies). Between $-0.2<\dsfms<0.2$ dex, the fraction of galaxies near coalescence increases by $17$ per cent. Over this range, the major/minor merger fractions only increase by $5$ per cent -- while the mini merger fraction increases by $12$ per cent. The middle panel shows that minor and major merger fractions are stable as a function of $\dsfms$ up to $\dsfms=0$, at which point both begin increasing rapidly. Minor mergers are at least a factor of $2$ more frequent $\dsfms>0.5$ dex than at $\dsfms=0$ dex. Major mergers are $4$ times more frequent at $\dsfms>0.5$ dex than at $\dsfms=0$ and dominate SFG fractions above this threshold. $36$ per cent of galaxies with $\dsfms>0.5$ dex are in major mergers satisfying $|\dtcoal|<1.5$ Gyr (e.g. \citealt{2012MNRAS.426..549S}).

The incidence of mini mergers increases monotonically with increasing $\dsfms$ for $\dsfms<0$. This connection between mini merger frequency and $\dsfms$ contrasts with the curves for major and minor mergers at $\dsfms<0$ -- which are largely stable. Around $\dsfms>0.25$ dex, mini mergers appear to drop in frequency. This apparent suppression arises simply from our choice to conserve galaxy counts by counting only one $|\dtcoal|<1.5$ Gyr merger per galaxy (for which we take the merger highest mass ratio in the case of multiple mergers). By removing this restriction, the incidence of mini mergers peaks at $150$ per cent at $\dsfms=1$ dex -- yielding rotational symmetry with $\dsfms<0$. With this correction, mini mergers are the only class of $\mu\geq0.01$ mergers for which incidence increases monotonically with $\dsfms$. A TNG50 SFG that is $\dsfms=0.3$ dex above (below) the SFMS is $25$ per cent more (less) likely to be in a mini merger than an SFG at $\dsfms=0$.


\subsection{Isolating the effects of individual mergers}\label{sec:iso_merg}

\begin{figure*}
\centering
	\includegraphics[width=0.49\linewidth]{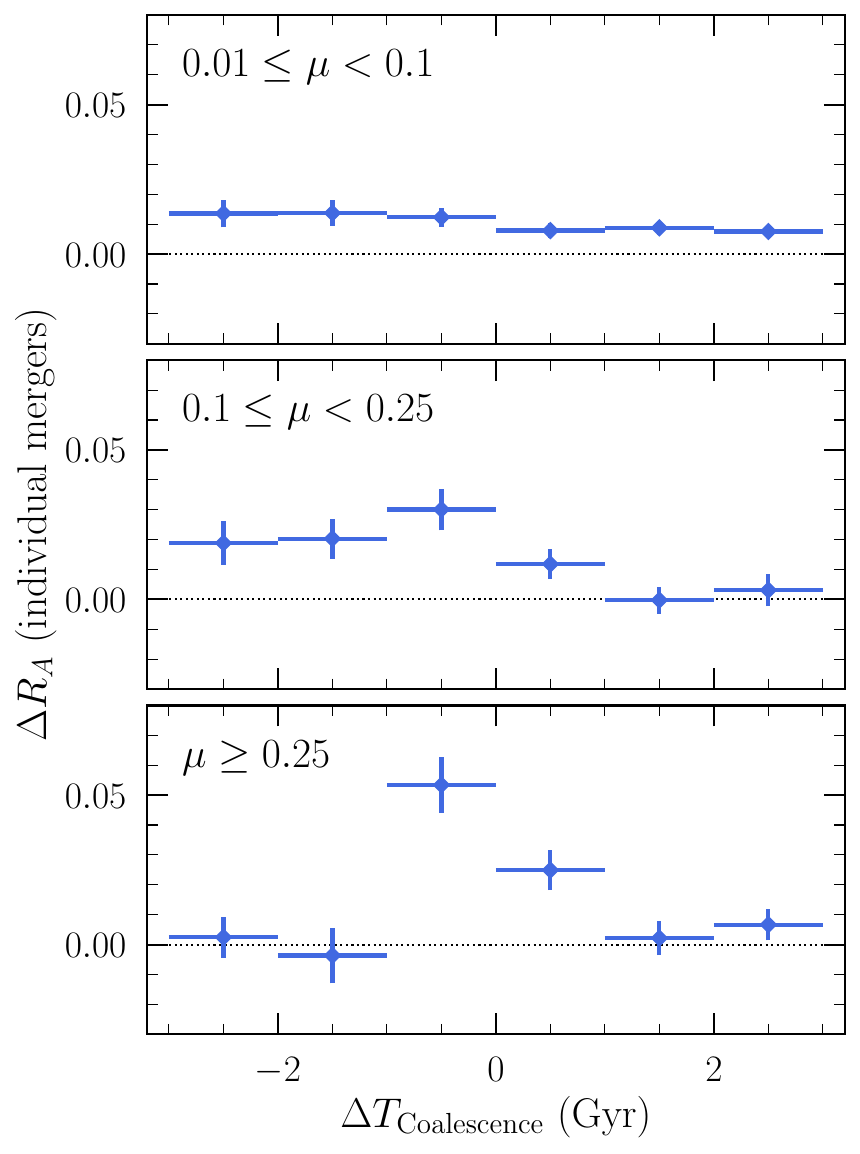}
	\hspace{10pt}
	\includegraphics[width=0.481\linewidth]{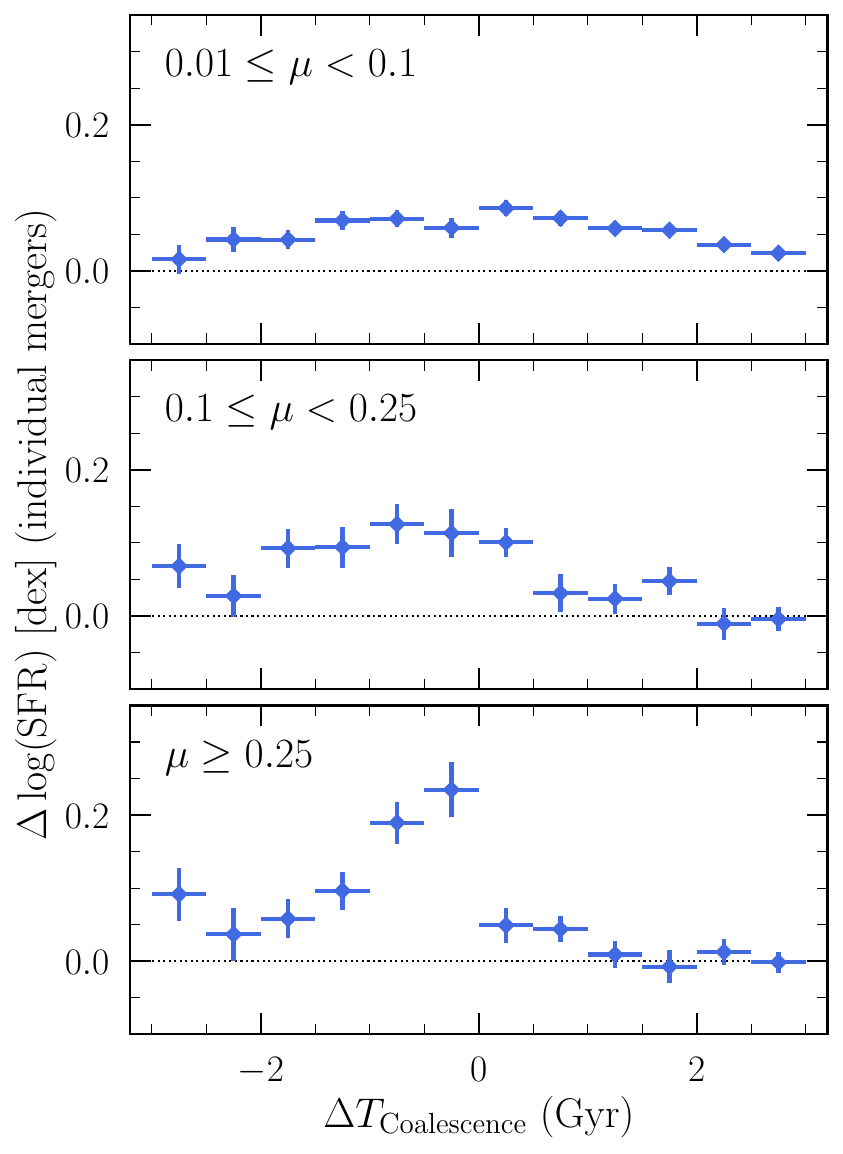}
   \caption[$\dra$ vs $\dtcoal$ in individual mergers]{Residual asymmetry enhancements, $\dra$, and SFR enhancements, $\dlogsfr$, in individual mergers compared to non-merging controls as a function of merger $\dtcoal$ in TNG50. Galaxies are considered individual mergers if they have \emph{exactly} one coalescence event within $|\dtcoal|<3$ Gyr and no other mergers from other mass ratio range. Each individual merger is matched to a control which has no coalescence event satisfying $|\dtcoal|<3$ Gyr (see Section \ref{sec:iso_selection}). Mini mergers exhibit statistically enhanced asymmetries and SFRs across the full range in $\dtcoal$. The peak enhancement amplitude in asymmetry and SFR increases with mass ratio. However, the enhancements in minor and major mergers are short-lived after coalescence. The asymmetries of individual minor and major merger remnants are consistent with their controls for $\dtcoal>1$ Gyr. In contrast, mini mergers yield long-lasting, modest asymmetry enhancements. Similarly, the longevity of SFR enhancement declines with increasing mass ratio. The SFRs of individual minor and major merger remnants are consistent with their controls for  $\dtcoal>2$ Gyr and $\dtcoal>1$ Gyr, respectively. \emph{Individual} mini mergers in TNG50 drive statistically significant changes to galaxy morphologies and SFRs. Mini mergers are not piggybacking on the enhancements driven by minor and major mergers.}
    \label{fig:iso_merg}
\end{figure*}

Our investigation into the connection between asymmetry, SFRs, and mergers identifies mini mergers as important players in driving both star formation and asymmetry in SFGs. However, the analysis thus far does not isolate the effects of \emph{individual} mergers of each type. In this section, we address this shortcoming by identifying \emph{individual} mergers of each type and comparing them with \emph{non-merging} controls. 

\subsubsection{Selection of individual mergers and non-merging controls}
\label{sec:iso_selection}
We start by selecting all SFGs with \emph{exactly one} merger satisfying $|\dtcoal|<3$ Gyr. This broad window is chosen to minimize contributions from other mergers to asymmetries and SFRs -- although they cannot be ruled out entirely. We create a pool of non-merging control galaxies with no $\mu \geq 0.01$ mergers satisfying $|\dtcoal|<3$ Gyr. Following the matching procedure described in Section \ref{sec:controls}, each SFG with a single merger satisfying $|\dtcoal|<3$ Gyr is assigned a single non-merging control. However, in this case, the control matching for pre-mergers and post-mergers is distinguished by the parametrization of local environment. For pre-mergers, we match on the distance to the second nearest physical companion, $R_2$, as before \citep{2020MNRAS.494.4969P}. For post-mergers, we match on the distance to the first nearest physical companion \citep{2020MNRAS.493.3716H,2021MNRAS.504.1888Q}. 

\subsubsection{Impact of individual mergers on asymmetries}
\label{sec:iso_ra}
The left panels of Figure \ref{fig:iso_merg} show the affect of individual mini, minor, and major mergers on morphological asymmetries as a function of $\dtcoal$ within 3 Gyr of coalescence. The upper left panel shows that individual mini mergers drive small but statistically significant enhancements in asymmetry compared to their non-merging controls. This enhancement is greatest in the pre-merger phase and has no statistically significant sensitivity to the pre-coalescence timescale -- with $\dra = 0.015$ on average across $-3<\dtcoal<0$ Gyr. Lower asymmetries are exhibited for SFGs in the post-merger phase for mini mergers, with $\dra = 0.007$ on average across $0\leq \dtcoal<3$ Gyr. 

The middle and lower left panels of Figure \ref{fig:iso_merg} show the enhancements in asymmetry arising from individual minor and major mergers, respectively. Both minor and major mergers exhibit much stronger sensitivity to $\dtcoal$ in the pre- and post-merger sequence relative to mini mergers. Compared to mini mergers, minor and major mergers trigger larger asymmetry enhancements on average -- peaking at $\dra=0.03$ and $\dra=0.05$ between $-1<\dtcoal<0$ Gyr directly before coalescence, respectively. Like the mini mergers, minor mergers have enhanced asymmetries extending to 3 Gyr pre-coalescence (average $\dra=0.024$). In contrast, the asymmetries of early major mergers, $\dtcoal<-1$ Gyr, are consistent with their controls. As for the post-coalescence phase, the asymmetries of minor and major merger remnants are enhanced by $\dra=0.013$ and $\dra=0.03$ for $0\leq \dtcoal<1$ Gyr, respectively. However, this enhancement is short-lived and minor and major remnant asymmetries are statistically consistent with their controls for $\dtcoal>1$ Gyr (e.g. as also shown by \citealt{2008MNRAS.391.1137L,2019ApJ...872...76N,2021ApJ...919..139W,2022MNRAS.515.3406M,2022MNRAS.511..100B}).

\subsubsection{Impact of individual mergers on star formation rates}
\label{sec:iso_sfr}

The right panels of Figure \ref{fig:iso_merg} compare the SFRs of individual mergers and their matched non-merging controls. Most interestingly, mini mergers drive small-amplitude but longer-lived enhancements in SFRs compared to minor and major mergers. Mini mergers have enhanced SFRs across the entire $-3 <\dtcoal<3$ Gyr sequence we consider -- peaking at $\dlogsfr=0.08$ dex directly following coalescence and declining symmetrically with increasing $|\dtcoal|$. Minor and major mergers also exhibit enhanced star formation efficiencies (SFEs) in the pre-coalescence phase. However, in contrast to mini mergers, these enhancements are shorter-lived after coalescence\footnote{We note that because our control-matching scheme includes matching in both gas fraction and stellar mass, individual mergers and their non-merging controls have similar total gas masses and offsets in SFRs are directly related to offsets in star formation efficiencies.} -- mirroring the relative enhancements in their asymmetries.

The results presented in Figure \ref{fig:iso_merg} support a trend between the longevity of enhanced SFRs after coalescence and merger mass ratio. Mini mergers have statistically enhanced SFRs for all $0<\dtcoal<3$ Gyr. Minor mergers are enhanced for $0<\dtcoal<2$ Gyr. And major mergers are enhanced for only $0<\dtcoal<1$ Gyr. Taken together, these results suggest that the remnants of smaller-$\mu$ mergers have more prolonged enhancements in their SFEs compared to the remnants of larger-$\mu$ mergers.  

The trend between mass ratio and the longevity of post-coalescence SFR enhancement may arise from the interplay between enhanced SFRs and subsequent stellar feedback. The gravitational and tidal forces from mini mergers drive SFR enhancements that are modest compared to minor and major collisions. Consequently, the corresponding stellar feedback may not be efficient at returning their SFRs to the level of their controls. Instead, the enhanced SFRs in mini mergers are sustained compared to minor and major mergers. The stronger stellar feedback following the main burst of star formation near coalescence for major mergers ($\dlogsfr=0.25$ dex, on average) is more likely to be efficient at curtailing sustained SFR enhancements (e.g. \citealt{2022MNRAS.516.4354W}). 

We also highlight that our results show that whatever process stifles prolonged SFR enhancements in major, minor, and mini mergers, it is only capable of returning post-merger SFRs to the same level as their controls. We find no evidence for a statistical \emph{suppression} in SFRs compared to matched controls for any mass ratio range. So, while TNG50 merger remnants are statistically more likely to be quenched than a sample of matched non-merging controls, quenching of SFGs via mergers is rare (e.g. \citealt{2015MNRAS.448..221E,2019MNRAS.490.2139R,2020MNRAS.493.3716H,2021MNRAS.504.1888Q,2023MNRAS.519.2119Q,2022MNRAS.517L..92E}). Alternatively, the quenching must be sufficiently fast that remnants do not have time to exhibit suppressed SFRs \emph{while still SFGs} before joining the quenched population (e.g. \citealt{2008ApJS..175..356H,2009ApJ...693..112P,2009MNRAS.395..144W,2010ApJ...721..193P,2021ApJ...911..116S,2023MNRAS.519.2119Q}).

\section{Discussion}\label{sec:discussion}

Our results with TNG50 and corresponding HSC-SSP synthetic imaging show that mini mergers are tied closely to galaxy asymmetries and SFRs. Given their high frequency relative to major and minor mergers, mini mergers may have an understated role as drivers of morphological disturbance and \emph{in-situ} stellar assembly in galaxies. In this section, we use the results of the previous section to discuss the role of mini mergers in these broader contexts and some additional considerations.


\subsection{The importance of mini mergers}

\begin{table}
\caption{\emph{In-situ} stellar mass growth triggered by mini, minor, and major mergers for SFGs from $0.1\leq z \leq0.7$. Estimates for the average \emph{in-situ} stellar mass growth, $\langle \Delta M^{\mathrm{merger}}_{\star,\mathrm{\emph{in-situ}}} \rangle$, are computed by integrating $\dsfr$s of individual mergers compared to non-merging controls within 3 Gyr of coalescence (Equation \ref{eq:intsfr}). $N^{0.1\leq z\leq 0.7}_{\mathrm{mergers}}$ is the number of merger coalescences for SFGs -- which are then used to estimate the fractional contribution of mergers from each mass ratio range to the total merger-driven \emph{in-situ} stellar assembly over this redshift interval. Calculations are shown for $\logMstar\leq9$ (TNG50) and $\logMstar\leq10$ (TNG100).}
\label{tab:insitu}
\begin{tabular*}{\linewidth}{l@{\extracolsep{\fill}} c c c}
\hline
\vspace{-5pt}\\
Type & $\langle \Delta M^{\mathrm{merger}}_{\star,\mathrm{\emph{in-situ}}} \rangle$ ($\times 10^{9} M_{\odot}$) & $N^{0.1\leq z\leq 0.7}_{\mathrm{mergers}}$ & $f^{\mathrm{merger}}_{\star,\mathrm{\emph{in-situ}}}$ \\
\vspace{-5pt}\\
\hline
\vspace{-5pt}\\
Mini & 0.437 / 2.04 & 3059 / 4325 & 0.551 / 0.601 \\
Minor & 0.570 / 2.62 & 697 / 864 & 0.164 / 0.155 \\
Major & 1.10 / 3.68 & 625 / 971 & 0.285 / 0.244 \\
\vspace{-5pt}\\
\hline
\end{tabular*}
\end{table}

\begin{figure}
\centering
	\includegraphics[width=\linewidth]{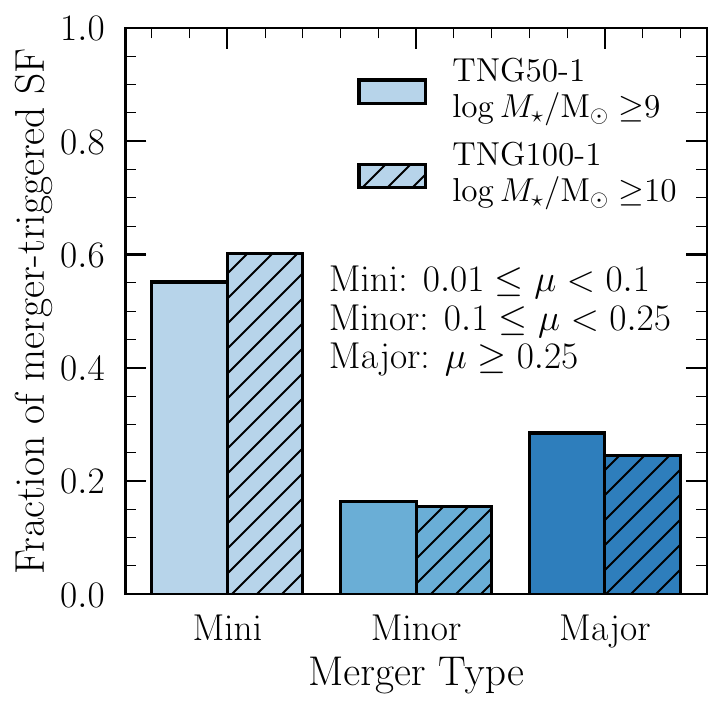}
   \caption[Fractional contributions to SFRs]{Fraction of merger-triggered \emph{in-situ} star formation contributed by mini, minor, and major mergers for TNG50 (open bars) and TNG100 (hatched bars) SFGs over $0.1\leq z \leq 0.7$. The merger-triggered \emph{in-situ} star formation fractions are estimated using the $\dsfr$s of individual mergers compared to non-merging SFG controls -- which are then scaled by the number of mergers that occur amongst SFGs in each mass range over this interval. $55$ per cent of \emph{in-situ} stellar mass assembly that occurs as a result of $\mu>0.01$ mergers is from mini mergers.}
    \label{fig:frac_sfrs}
\end{figure}

\begin{figure}
\centering
	\includegraphics[width=\linewidth]{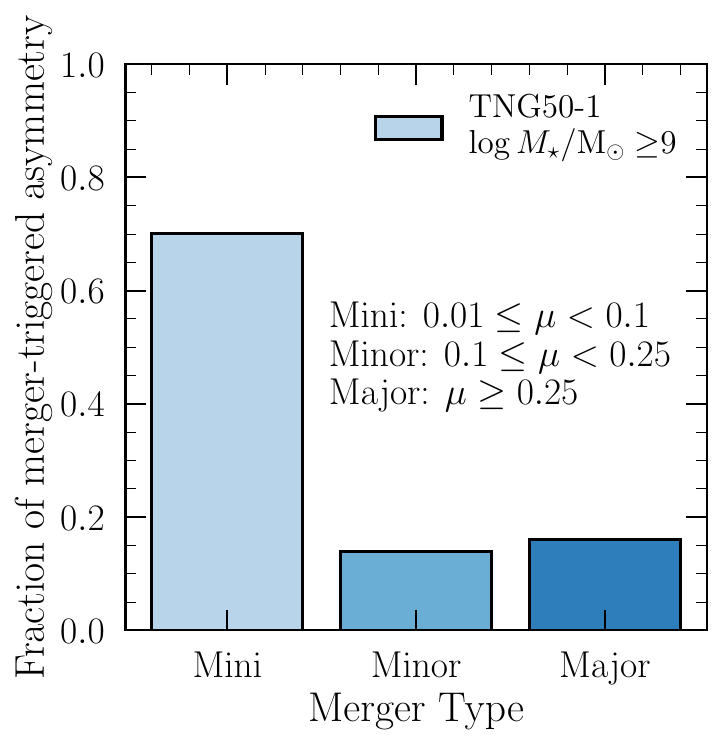}
   \caption[Fractional contributions to Asymmetries]{Same as Figure \ref{fig:frac_sfrs} but now comparing relative contributions of mini/minor/major mergers to light-weighted asymmetric structure in TNG50 SFGs with $\logMstar\geq9$, integrated over $0.01 \leq < z \leq 0.7$. $70$ per cent of the asymmetric structure driven by $\mu>0.01$ mergers is specifically driven by mini mergers. Mini mergers dominate over minor and major mergers in driving asymmetric structure in TNG50 SFGs.}
    \label{fig:frac_asy}
\end{figure}

\subsubsection{Relative contribution to in-situ stellar assembly}\label{sec:rel_sfrs}

In Sections \ref{sec:iso_ra} and \ref{sec:iso_sfr}, we showed that \emph{individual} mini mergers drive small but statistically significant enhancements in SFRs and asymmetries compared to matched non-merging controls. In particular, while the mean enhancement in SFRs near $\dtcoal=0$ for mini mergers is only 0.08 dex (20 per cent) compared to 0.25 dex (79 per cent) for major mergers, the duration of enhanced SFRs after coalescence is longer. Therefore, the integrated \emph{in-situ} formation of stars resulting from the average individual mini merger is enhanced by this greater longevity. So, considering also their increased frequencies, what is the aggregate effect of mini mergers on \emph{in-situ} stellar mass assembly compared to minor and major mergers? 

To answer this question, we first compute the non-logarithmic $\dsfr$s of individual mergers compared to non-merging controls, $\dsfr=$SFR$_{\mathrm{merger}} - $SFR$_{\mathrm{control}}$. Taking these controls as an ensemble, they represent the sub-population of SFGs which \emph{have not} merged and \emph{will not} merge within 3 Gyr (i.e. do not satisfy $|\dtcoal|<3$ Gyr). Consequently, the integrated $\dsfr$s of merging galaxies compared to these controls over some time interval about coalescence, e.g. $[T_{\mathrm{C}}-3,T_{\mathrm{C}}+3]$ Gyr, is the merger-triggered \emph{in-situ} stellar mass assembled over that interval:
\begin{align}\label{eq:intsfr}
\langle \Delta M^{\mathrm{merger}}_{\star,\mathrm{\emph{in-situ}}}\rangle = \int^{T_{\mathrm{C}}+T}_{T_{\mathrm{C}}-T} \dsfr(t) \,dt
\end{align}
Note that this deliberately excludes the \emph{ex-situ} material assembled in the merger via accretion of existing stars. Similar to Figure \ref{fig:iso_merg}, we take $\dtcoal \in [-3, 3]$ Gyr as the interval over which the merger-triggered stellar mass growth is integrated for mini, minor, and major mergers. The results of these calculations are shown in Table \ref{tab:insitu} for both the TNG50 and TNG100 simulations. As for Figure \ref{fig:mus}, the TNG50 and TNG100  stellar mass cuts are $\logMstar\leq9$ and $\logMstar\leq10$, and the differences in our estimates between TNG50 and TNG100 reflect relationship between merger frequency and stellar mass (e.g. \citealt{2014MNRAS.445.1157C,2014MNRAS.444.3986R,2015MNRAS.449...49R,2019MNRAS.490.2139R,2023MNRAS.519.4920G}). The fractions of merger-triggered \emph{in-situ} stellar assembly driven mini, minor, and major mergers are calculated assuming that every merger of a given type drives $\langle \Delta M^{\mathrm{merger}}_{\star,\mathrm{\emph{in-situ}}}\rangle$ new \emph{in-situ} stars -- ignoring potential overlap of multiple mergers over the integration timescale.

Figure \ref{fig:frac_sfrs} illustrates the calculations presented in Table \ref{tab:insitu}. According to our analysis of $0.1\leq z \leq0.7$ SFGs in TNG50 with $\logMstar\leq9$ (and TNG100 with $\logMstar\leq9$), mini mergers contribute $55$ $(60)$ per cent of the merger-driven \emph{in-situ} star formation driven by all mergers with mass ratios $\mu > 0.01$. In other words, mini mergers produce $1.23$ $(1.51)$ \emph{in-situ} units of stellar mass stars for every unit produced as a result of a minor or major merger, on average. Our approach to this calculation crudely neglects overlapping mergers and corresponding second order effects on merger-driven star formation. Nonetheless, this result presents a tantalizing case for mini mergers as vital players in the \emph{in-situ} assembly of stellar material in galaxies at $z\lesssim1$, leastwise compared to minor and major mergers.

Figure \ref{fig:sfms_mergers} showed that the frequencies of minor and major mergers for SFGs below the SFMS is constant as a function of $\dsfms$ (specifically, galaxies satisfying $\dtcoal<1.5$ Gyr). In contrast, SFMS offsets are sensitive to mini merger frequency for all $\dsfms$. This result is consistent with a scenario in which individual major (and to some extent minor) mergers amongst gaseous galaxies trigger high-amplitude but short-lived bursts of star formation -- with further star formation enhancement stifled by accordingly strong feedback from the burst \citep{1989Natur.340..687H,1994ApJ...431L...9M,2008MNRAS.384..386C,2019MNRAS.485.1320M,2020MNRAS.493.3716H}. Meanwhile, Figure \ref{fig:iso_merg} showed that the SFR enhancements triggered by mini mergers are mild in amplitude and long-lived. Consequently, the timescale of observability for enhanced SFRs from mini mergers is longer (or the suppressed SFRs in the absence of mini mergers). The combination of mild SFR enhancements, long observability timescales, and higher frequencies of mini mergers compared to minor and major mergers yield higher sensitivity of $\dsfms$ to mini mergers than their higher-$\mu$ competitors. 

\subsubsection{Relative contribution to galaxy asymmetries}\label{sec:rel_ras}

In Figure \ref{fig:iso_merg}, we showed that individual mini mergers produce statistically enhanced morphological asymmetries recoverable from HSC-SSP synthetic images. Taking the same approach as in the previous section, we now estimate the relative contribution of mini, minor, and major mergers to merger-driven asymmetric structure. 

Integrating the $i$-band luminosity-weighted HSC asymmetry offsets for individual mergers from Figure \ref{fig:iso_merg}, the fractions of time-averaged, merger-driven asymmetric structure attributed to mini, minor, and major mass ratio ranges in TNG50 are shown in Figure \ref{fig:frac_asy}. Mini mergers produce $70$ of the asymmetric structure generated in all mergers with mass ratios $\mu \geq 0.01$. Integrating over all mini mergers in the SFG population from TNG50, they produce more than double the amount of asymmetric structure compared to minor and major mergers combined (each approximately 15 per cent). Meanwhile, as for the SFR offsets, individual mini mergers do not generate strong peaks in asymmetry the way individual minor and major mergers do (Figure \ref{fig:iso_merg}). The role of mini mergers is critically tied to their relatively high frequencies. 

In Section \ref{sec:ra_statistics} and Figure \ref{fig:ra_mergers} we used the merger criterion $|\dtcoal|<1.5$ Gyr to obtain the statistics of mergers as a function of asymmetries recovered from HSC imaging. We found that while merger frequency statistics rise with observed asymmetries (as explicitly shown in our controlled experiments in Figures \ref{fig:dra_dtc} and \ref{fig:iso_merg}), there remains a large community of galaxies with modest-to-high asymmetries but without a coalescence event in this window. For galaxies with asymmetries $R_A>0.1$, only $52$ per cent satisfy the $|\dtcoal|<1.5$ Gyr merger criterion for any mass ratio (purities -- mini: $30$ per cent; minor: $11$ per cent; major: $11$ per cent). For the remainder, there are two options: (1) asymmetric structure which emerges in the absence of mergers (e.g. \citealt{1997ApJ...477..118Z,2004AJ....127.1900W,2005A&A...438..507B,2008MNRAS.388..697M,2013ApJ...772..135Z,2022ApJ...935...48Z}) and (2) merger-driven asymmetric structure that persists on timescales that extend beyond this narrow window around coalescence (e.g. \citealt{1994A&A...290L...9R,1994ApJ...421..481W,1998ApJ...496L..13L,2009PhR...471...75J}). 

Asymmetric surface brightness distributions are expected to arise in SFGs via mechanisms other than mergers. First, the spatial distribution of star formation sites in star-forming disk galaxies viewed face-on is generally asymmetric (e.g as traced by HI gas density and HII regions; \citealt{1994A&A...290L...9R,2003ApJS..147....1C,2012ApJ...747...34B}). The consequence is an intrinsic relationship between asymmetry and SFR in isolated, turbulent gas disks that will exist in the absence of external factors. Meanwhile, asymmetric gas accretion, secular gravitational instabilities, and asymmetry in the dark matter potential may also drive asymmetry and lopsidedness without requiring mergers (e.g. \citealt{1997ApJ...477..118Z,2004AJ....127.1900W,2005A&A...438..507B,2008MNRAS.388..697M,2013ApJ...772..135Z,2022ApJ...935...48Z,2023MNRAS.523.5853V}). Dust attenuation is another important source of asymmetry in the optical surface brightness distributions of inclined disk galaxies (e.g. \citealt{2008MNRAS.388.1708G}). Our dust radiative transfer methodology accounts for this source of asymmetry.

Asymmetry and lopsidedness in the stellar mass distributions of star-forming galaxies is ubiquitous in the low-$z$ Universe (e.g. \citealt{1980MNRAS.193..313B,1994A&A...288..365B,2013ApJ...772..135Z,2021ApJ...923..205Y}). This ubiquity requires a mechanism for driving asymmetries which cannot not depend on rare events producing short-lived enhancements -- such as those driven by individual major or minor mergers. In contrast, our results with TNG50 show that mini mergers are frequent and individually drive small, long-lived changes in asymmetry. \emph{In particular, the average mini merger $|\dtcoal|$ for SFGs ($1.9$ Gyr) is a factor of $\sim3$ shorter than the timescale over which asymmetries are enhanced by individual mini mergers in TNG50}. Consequently, these asymmetry enhancements would be expected to aggregate over consecutive merger events for a given galaxy. Mini mergers therefore present a more competitive pathway for the build-up of asymmetry and lopsidedness in galaxies than higher mass ratio mergers.

\subsection{Comparison with other simulation results} \label{sec:comp_sim}

Our results with TNG50 show that individual mini mergers trigger modest but statistically significant enhancements in SFR compared to non-merging statistical controls (Figure \ref{fig:iso_merg}). These SFR enhancements are ensemble averages over mergers in the range $0.01 \leq \mu < 0.1$. As expected, the merger mass ratio distributions in TNG are bottom-heavy \citep{1993MNRAS.262..627L,1993MNRAS.264..201K,1999ApJ...524L..19M,1999ApJ...522...82K}. Consequently, the SFR enhancements in Figure \ref{fig:iso_merg} correspond to the typical enhancement expected from the average mass ratio in each category, $\bar{\mu}(\mathrm{mini, minor, major})=(0.033, 0.16, 0.50)$. The enhancements in SFRs and asymmetries in Figure \ref{fig:iso_merg} can therefore be interpreted as the expected enhancements from individual mergers with these category-averaged mass ratios. With this in mind, we compare with other simulation results. 

Some of the most important investigations into the effect of merger mass ratios on the corresponding triggered star formation, gas consumption, and asymmetry have been carried out using hydrodynamical simulations of binary galaxy mergers in the absence of cosmological environment (e.g. \citealt{2008MNRAS.384..386C,2010MNRAS.404..575L}). Using a suite of binary, gas-rich disk merger simulations, \cite{2006MNRAS.373.1013C,2008MNRAS.384..386C} showed the integrated SFR enhancement and peak enhancement amplitude (at coalescence) are both sensitive to merger mass ratio. These results are consistent with ours including TNG50's cosmological context. However, unlike our result, \cite{2008MNRAS.384..386C} found no SFR enhancement in their $\mu = 0.02$ mini merger compared to an isolated, non-merging control. The cosmological context of TNG may play a significant role in this discrepancy. However, there are also some key differences between the baryonic physics models used by \citep{2008MNRAS.384..386C} and TNG which may explain this discrepancy on the impact of mini mergers. 

\citealt{2008MNRAS.384..386C} adopt a star formation and feedback prescription based on the heuristic cooling/heating model from \cite{2000MNRAS.312..859S}. In this model: (1) gas pressure follows a single-phase, ``stiff'' equation of state (EOS); (2) star formation density scales with gas density and inversely with the local dynamical timescale above a threshold gas density; and (3) kinetic energy from supernova feedback is retained locally in the surrounding gas as sub-grid turbulence and slowly dissipated as thermal energy -- thereby delaying rapid thermalization and preventing runaway star formation \citep{1993MNRAS.265..271N,1994ApJ...437..611M}. In contrast, in the TNG50 model: (1) gas follows a two-phase equilibrium EOS describing the evolution of cold clouds embedded in a hot ambient medium; (2) mass in a cold cloud phase is stochastically converted into stars; and (3) the kinetic component is capable of driving large-scale, supernova-driven winds resulting in galactic outflows and fountains \citep{2003MNRAS.339..289S,2019MNRAS.490.3234N}.\footnote{Specifically, to prevent over-pressurization of the ISM, the temperature of star-forming gas in TNG is determined by interpolating the the \cite{2003MNRAS.339..289S} two-phase EOS with an isothermal EOS at $10^4$ K \citep{2005MNRAS.361..776S,2010MNRAS.407.1529H,2013MNRAS.436.3031V}.} In particular, the retention of all kinetic energy from supernovae as sub-grid disk turbulence (high pressurization) may stifle the impact of low- mass ratio mergers on gas redistribution in the \citealt{2008MNRAS.384..386C} merger suite. In TNG, a fixed fraction of supernova kinetic energy is transported away from star forming gas via winds while the remainder contributes to the evaporation of cold clouds into the ambient hot phase. Consequently, star forming gas may be expected to be less pressurized in the TNG model and more susceptible to influence from smaller gravitational impulses and torques arising in mini mergers.

Using the \cite{2008MNRAS.384..386C} binary merger simulations, \cite{2010MNRAS.404..575L} further examined the sensitivity of morphological asymmetry and other structural parameters to merger mass ratio and stage. As we have seen using our HSC-SSP synthetic images of TNG50 galaxies (Figure \ref{fig:ra_mergers}), \cite{2010MNRAS.404..575L} used forward-modelling to show that \emph{individual} mini mergers are exceedingly unlikely to drive asymmetries that satisfy the empirical \cite{2003ApJS..147....1C} major merger cut -- but can still drive small asymmetry enhancements compared to non-merging controls before and after coalescence. Therefore, while the gas, star formation, and stellar feedback model from the \cite{2008MNRAS.384..386C} merger suite may not permit increased SFRs from a $\mu=0.02$ merger, they can still affect asymmetry/lopsidedness in stellar light. These hypotheses could be thoroughly explored in binary merger suites using comprehensive physics models extending to $\mu<0.1$ mergers (e.g. \citealt{2019MNRAS.485.1320M}).

One of our main results with TNG50 is that the timescale over which SFRs and asymmetries are enhanced after coalescence increases with decreasing mass ratio (Figure \ref{fig:iso_merg}). Using a suite of binary $\mu=0.1$ merger simulations spanning a large number of orbital configurations, \cite{2022MNRAS.511.5878G} showed that minor mergers can trigger prominent lopsidedness in the \emph{idealized} stellar components of galaxies in the pair phase but that persists only $\sim500-850$ Myr after coalescence (see also \citealt{2023MNRAS.523.5853V}). Our results using TNG50 galaxy asymmetries with full HSC-SSP realism considerations show a similar $\sim1$ Gyr post-merger enhancement timescale in minor mergers -- further establishing that the premier HSC-SSP image is sufficient to probe low surface brightness asymmetric structure and lopsidedness (e.g. \citealt{2018ApJ...866..103K,2022ApJS..262...39H,2023MNRAS.521.3861D,2023arXiv230807962D}). Meanwhile, the post-coalescence enhancement of asymmetries in TNG50 mini mergers are smaller and stable on even longer timescales ($\gtrsim3$ Gyr). Further investigation into the long timescale of enhanced asymmetries in individual mini mergers is needed and will be supported by the public image database we present here.

The literature on the impact of $\mu<0.1$ mini mergers on SFRs or asymmetries in cosmological hydrodynamical simulations is sparse. However, using the Horizon-AGN cosmological hydrodynamical simulation \citep{2014MNRAS.444.1453D,2017MNRAS.467.4739K} and definitions of minor and major mergers used here, \cite{2018MNRAS.480.2266M} examined the impact of major, minor, and mini mergers on the morphological transformation from disk to spheroids at $z\lesssim1$ -- adopting stellar particle kinematics as a proxy for morphology. \cite{2018MNRAS.480.2266M} showed that essentially all morphological transformation of present-day spheroids is driven by $\mu \geq 0.1$ mergers since $z=1$. They found that mini mergers did not yield significant redistribution of stellar mass between rotational and dispersion-supported components in the histories of present day disks and spheroids. These results are not in tension with ours. Indeed, combined with our results for morphological asymmetries in TNG50, the results from \cite{2018MNRAS.480.2266M} present an interesting framework in which the small, frequent gravitational impulses from mini mergers drive enhanced SFRs and asymmetries/lopsidedness but not broader transformation from disk to spheroid.

Using the SIMBA cosmological hydrodynamical simulations \citep{2019MNRAS.486.2827D}, \cite{2019MNRAS.490.2139R} showed that major mergers produce large SFR boosts at $z<0.5$ (factor of $\sim3$, on average). In addition, they showed that (1) major mergers in SIMBA contribute less than $1$ per cent of the cosmic SF budget and (2) major merger frequencies are too low to explain quenched galaxy fractions at $z<1.5$. While we do not characterize the contribution of mergers to the total cosmic SF budget in TNG50 and TNG100, we do estimate that major mergers produce only a 30 per cent share of merger-triggered star formation amongst mergers with $\mu\geq0.01$ for $z<0.7$. Adopting the $1$ per cent major merger contribution from \cite{2019MNRAS.490.2139R} and incorporating our merger-triggered SF fractions from Figure \ref{fig:frac_sfrs}, the contribution of $\mu\geq0.01$ mergers to the cosmic SF budget would still be only $\sim3$ per cent. Such a small fraction is in tension with the strong trends we see between $\dlogsfr$ and mini merger incidence, frequency, and coalescence offset (Figures \ref{fig:sfms_dtc} and \ref{fig:dsfr_dtc}). We defer the comparison of our $\mu\geq0.01$ merger-driven star formation estimates to the star formation histories of individual galaxies and populations in TNG to future work.

Lastly, we find no evidence for suppressed SFRs at any stage in the merger sequence compared to non-merging controls matched in redshift, stellar mass, environment, and gas fraction. This result is in agreement with \cite{2019MNRAS.490.2139R} with SIMBA and previous studies using different control-matching prescriptions in TNG \citep{2020MNRAS.493.3716H,2021MNRAS.504.1888Q,2023MNRAS.519.2119Q}. However, our sample selection targets SFGs. If a merger did quench a galaxy from our sample, the quenched descendent would not be retained for analysis against matched SFG controls. This selection bias was thoroughly examined by \cite{2020MNRAS.493.3716H} and \cite{2021MNRAS.504.1888Q} using control-matching techniques that do not exclude galaxies which are quenched during a merger. No statistically significant suppression of star formation was found in the ensemble of post-merger galaxies. In particular, \cite{2021MNRAS.504.1888Q} showed that while post-coalescence galaxies are up to $2$ times more likely to be quenched than their matched controls, the fraction of galaxies quenched by $\mu\geq0.1$ mergers is $\sim 5$ per cent. Accordingly, our ensemble-averaged SFRs would be only mildly suppressed by including quenched merger descendants in our SFG samples.

\subsection{Comparison with observational results}\label{sec:comp_obs}

Spectroscopic galaxy surveys can enable robust identification of merging pairs and measurement of their stellar mass ratios (e.g. \citealt{1997ApJ...475...29P,Barton_2000,2003MNRAS.346.1189L,2004ApJ...617L...9L,2007MNRAS.375.1017A}). However, the magnitude selection cuts used in such surveys constrain the stellar mass ratio range that is probed. For example, the SDSS main spectroscopic galaxy sample \citep{2002AJ....124.1810S} has $r$-band magnitude limit, $m_r\approx17.77$ mag. The brightest galaxy in the sample has $m_r=14.0$ mag. Assuming a constant mass to light ratio, this magnitude range allows mass ratios to cover $\mu\geq0.031$ -- which would be favourable for mini mergers. Unfortunately, the average brightness of galaxies in the sample is $m_r = 17.0$. So, for the average galaxy in the SDSS main galaxy sample, the expected mass ratio range to be probed under the selection cut is $\mu\geq0.5$. The spectroscopic selection cut thereby greatly favours identification of major merging pairs and misses most minor and mini mergers. An additional complication in extending to $\mu<0.1$ companions is the difficulty in distinguishing \emph{in-situ} star-forming clumps from late-stage mini mergers. 

Despite these limitations and using a range of merger identification techniques, observational works generally agree that merger frequency increases with decreasing merger mass ratio (e.g. \citealt{2011A&A...530A..20L,2012ApJ...747...34B,2014MNRAS.437L..41K,2014MNRAS.445.2198O,2022ApJ...940..168C,2023MNRAS.522....1N}). As for the role of sub-major mergers on star formation and asymmetry, there are few works providing observational constraints. Using morphological classifications of $\sim6500$ galaxies SDSS Stripe 82 deep images, \cite{2014MNRAS.440.2944K} presents a case that \emph{at least} $35$ per cent of the star formation budget in local galaxies is driven by minor mergers -- compared to $2.3$ in major mergers. Their approach to estimating this lower limit used the reasoned assumption that star formation in early-type galaxies (ETGs) is entirely driven by sub-major mergers -- thereby circumventing the actual selection of sub-major mergers spectroscopically or otherwise. By then selecting ETGs to match the mass and environment distributions of their late-type galaxy (LTG) sample and accounting for the stellar mass fractions in each class, \cite{2014MNRAS.440.2944K} further estimate that sub-major mergers account for \emph{at least} $24$ per cent of the LTG star formation budget.

Using the results from \cite{2014MNRAS.440.2944K}, the fraction of current merger-triggered star formation from sub-major mergers can be estimated as $35/(2.3+35) = 93$ per cent. This estimate assumes that the star formation driven by major mergers is entirely accounted by the SFRs of galaxies which are visually mergers/irregulars according to their Stripe 82 classifications. This result is qualitatively consistent with those we derived by integrating merger-driven star formation in the histories of TNG50 (and TNG100) galaxies. Our estimate for the relative contribution of sub-major mergers to the total number of stars formed as a result of $\mu\geq0.01$ mergers in Figure \ref{fig:frac_sfrs} is $72$ $(75)$ per cent. Quantitatively, however, there is a $\sim20$ per cent discrepancy in favour of stars formed as a result of major mergers in TNG. 

The key difference between our approaches and that \cite{2014MNRAS.440.2944K} is that we estimate the integrated merger-driven star formation since $z=0.7$, whereas \cite{2014MNRAS.440.2944K} measures the current star formation contributed by galaxies in each morphological type. Therefore, the main assumption in our comparison is that our merger-triggered star formation fractions do not evolve over $0.1\leq z \leq0.7$. In other words, we assume that the relative fractions of stars formed in mini, minor, and major mergers since $z=0.7$ are the same as their relative contributions to the \emph{current} star formation budget. Testing this assumption is beyond the scope of this study, but the relative rates of major, minor, and mini mergers as a function of redshift, mass, and environment may be an interesting subject of future work. In particular, recent observational work by \cite{2022ApJ...940..168C} suggests that the ratio of minor to major mergers increases with redshift -- which could be tested using our catalogues. 

\subsection{Ambiguity with accretion: what makes a merger?}\label{sec:accretion}

Our results with TNG show that mini mergers play an important role as drivers of star formation and asymmetry in galaxies. However, the mechanism by which mini mergers drive these changes is not directly elucidated in our study. Can a $\mu\sim0.01$ companion generate a sufficient tidal response to impart dynamical instability and subsequent structural change, internal gas inflows, central star formation, and nuclear activity as in the prevailing theoretical framework (e.g. \citealt{1989Natur.340..687H,1992ARA&A..30..705B,1996ApJ...464..641M})? Does its gas reservoir provide the fuel for the enhanced SFRs we see in individual mini mergers? Or are mini mergers simply entrained by large-scale, external gas inflows and serve only as tracers of SFR and asymmetry enhancements driven by that accretion?


\subsubsection{Requirement that mini companions are galaxies}

The variation in external gas accretion along cosmic filaments is a theoretically competitive process for explaining asymmetries in SFGs and the stratification the SFMS at fixed stellar mass \citep{2005MNRAS.363....2K,2005A&A...438..507B,2009ApJ...703..785D,2008A&ARv..15..189S,2009PhR...471...75J,2012A&A...544A..68L,2014ApJ...789L..21C}. Observational work shows that the average rate of cool gas accretion onto local SFGs is a factor of 5-10 times too low to sustain their current SFRs (e.g. \citealt{2008A&ARv..15..189S,2021ApJ...923..220D}). Nonetheless, simulation work predicts that local SFGs with high gas accretion rates have preferentially positive $\dsfms$ (e.g. \citealt{2018ApJ...859..109S,2021MNRAS.504.5702W}). Such inflows can entrain pockets of stellar material formed from cold gas clumps embedded in the flows \citep{2009Natur.457..451D,2018ApJ...861..148M}. If we were identifying such objects as the companion systems in mini mergers, then the role of mini mergers would be conflated with the flows in which they are entrained. However, if such stellar substructures have no dark matter excess, as predicted, then they would be excluded from our merger history/forecast calculations (non-cosmological halos). 

Our requirement that merger progenitors must be halos with cosmological origin (via the TNG SubhaloFlag parameter; \citealt{2019ComAC...6....2N}) is an attempt to strictly delineate \emph{mergers between galaxies} from accretion. We guarantee under this condition that mergers are only ever between galaxies which are embedded in unique cosmological dark matter halos. Nevertheless, this definition leaves much to be desired in controlling for external gas accretion. Even if our definition excludes accretion of cold gas clumps or the diffuse gas along which they are entrained, it does not exclude cases in which actual galaxies in cosmological halos are entrained along with these flows. In such cases, accretion-driven phenomena during an ongoing merger would be ascribed to the merger alone in our analysis. Furthermore, if the less-massive galaxies in low-$mu$ mergers are more likely to be entrained along accretion flows, then our estimates for the role of mini mergers as drivers asymmetries and SFRs would preferentially be compromised by the role of accretion.

\subsubsection{Assuming that mini mergers are only tracing accretion}

Intriguingly, if we assume that mini mergers themselves have no effect on galaxies and are instead just partial tracers of external gas accretion, then they allow us to place a lower limit on the relative role of accretion as a driver of asymmetry and SFRs in TNG50 SFGs compared to minor/major mergers. Comparing relative contributions of accretion and minor/major mergers, \emph{at least} $70$ per cent of the light-weighted asymmetric structure driven since $z=0.7$ would be due to accretion, with at most $30$ per cent ascribed to minor and major mergers (Figure \ref{fig:frac_asy}). For galaxies with asymmetries $R_A>0.1$, at least $60$ per cent of externally-driven asymmetries could be ascribed to accretion traced by mini mergers (Figure \ref{fig:ra_mergers}; $0.33$ mini / ($0.11$ minor + $0.11$ major + $0.33$ mini)). We also showed that $45$ per cent of $R_A>0.1$ SFGs that are not associated with any $\mu\geq0.01$ merger. Further extending to $\mu<0.01$ mergers as accretion tracers may therefore reveal an even larger part played by accretion. Meanwhile, the fraction of minor and major mergers with $R_A>0.1$ is capped at $22$ per cent. 

As for SFRs, the fraction of integrated star formation triggered by accretion relative to minor/major mergers since $z=0.7$ would be $55$ per cent in TNG50. Again, this number would be a lower limit for the role of accretion assuming: (1) mini mergers have no transformative effects other than gas transport; (2) mini mergers companions are always entrained in flows; and (3) not all flows necessarily contain these mini merger companions. Under these assumptions, our results with TNG suggest that minor/major mergers a play a sub-dominant role over accretion in driving the stratification in the SFMS -- contributing at most $45$ per cent to the SFMS scatter at fixed stellar mass. 

Following this interpretation, our results are similar to those presented by \cite{2021ApJ...923..205Y} using SDSS Stripe 82 coadded images. \cite{2021ApJ...923..205Y} found that only $\lesssim20$ per cent of SFGs with $R_A>0.1$ have a visually conspicuous signature of a recent merger or companion (minor/major) and that the remaining SFGs with $R_A>0.1$ ($\gtrsim 80$ per cent) were simply lopsided and/or had asymmetric spiral arms. Considering also other observational constraints (e.g. \citealt{1980MNRAS.193..313B,2011MNRAS.412.1081W,2011A&A...532A.117E,2018ApJ...852...94V}), \cite{2021ApJ...923..205Y} argued that accretion should have an important role in accounting for the asymmetries in this majority share of local SFGs. 

However, there are at least two key problems with this interpretation. The first is that the interpretation assumes that mini mergers trace gas accretion with reasonably high purity, i.e. mini mergers do not occur in the absence of large-scale gas accretion flows. It is beyond the scope of this work to test this assumption but should be the subject of future investigation. However, there is no reason that $\mu\sim0.01$ mergers should only occur in the presence of large scale gas accretion flows. The second problem is that we show that asymmetries and SFRs are sensitive to the $\dtcoal$ of individual mini mergers (Figure \ref{fig:iso_merg}). This result shows that the gravitational impulse and tidal forces from the mini companion do impart instabilities in the host. Otherwise, the enhancement in SFR near coalescence in mini mergers would need to be attributed only to an increase in cool gas supply. This interpretation would be in tension with the HI content in Local Group dwarf galaxies found inside the Milky Way and M31 virial radii -- which are found to be largely devoid of HI gas (e.g. \citealt{2009ApJ...696..385G,2021ApJ...913...53P}).

Considering these shortcomings, the interpretation that mini mergers \emph{only} track large-scale gas accretion is incomplete. It is more realistic that mini merger companions \emph{can} be entrained along gas flows like their minor/major counterparts -- but it is not a necessary condition. We defer computation of TNG galaxy accretion rates and an detailed investigation into the interplay between gas accretion and mergers in driving structural and physical change in galaxies to future work.

\subsection{Considering gas phase and distribution}

We control for gas fractions when comparing the properties of (1) SFGs to other SFGs or (2) mergers to non-mergers. But a caveat is that we do not control for the phase, distribution, or angular momentum of the gas. The gas definition we use in our TNG galaxy gas fraction calculations includes all gas contained within twice the stellar half-mass radius. Consequently, the phase and density distribution of the gas can vary greatly for TNG galaxies at fixed gas fraction. For our purposes, it is scientifically undesirable to control for such properties because mergers are known to redistribute gas and boost central cool/cold phase fractions compared to non-merging controls (e.g. \citealt{2018MNRAS.478.3447E,2019A&A...627A.107L,2021MNRAS.503.3113M,2023ApJ...950...56H}). On the other hand, while central boosts are observed and found in simulations, the global cool gas content in interacting galaxies is not expected to change based on atomic hydrogen observations of interacting pairs \citep{2009ApJ...698.1437K,2020A&A...635A.197D}. In future work, it may therefore be desirable to refine our approach from matching total gas fractions to matching mass in the cool (e.g. atomic) and cold (e.g. molecular) phases.

\section{Summary}

We present a public dataset of over $750$k synthetic images of galaxies from the TNG50 and TNG100 cosmological hydrodynamical simulations over $0.1 \leq z \leq 0.7$ computed using dust radiative transfer with SKIRT (Figures \ref{fig:mosaic_idealized_91} and \ref{fig:mosaic_idealized_59}). The filter set includes the HSC $grizy$ broad bands and several complementary bandpasses spanning $0.3-5 \mu$m. From the idealized HSC $grizy$ images (dusty, but noise/seeing free), we make survey-realistic HSC-SSP images by statistically inserting into real survey fields (Figures \ref{fig:mosaic_pdr3_91} and \ref{fig:mosaic_pdr3_59}). In this paper, we use these synthetic TNG50/HSC-SSP images and simple morphological measurements to examine the relative roles of major ($\mu \geq 0.25$), minor ($0.1\leq \mu < 0.25$), and mini mergers ($0.01 \leq \mu <0.1$) as drivers of the observed link between asymmetry and the stratification of SFRs at fixed stellar mass for SFGs. Our key results are as follows:

\begin{itemize}

\item \textbf{Asymmetry is ubiquitous in TNG50 galaxy light distributions, in agreement with observations of nearby galaxies}. Over 50 per cent of $z=0.1$ TNG50 galaxies with $\logMstar \geq 9$ have $i$-band asymmetries $R_A>0.1$ in their surface brightness profiles (Figure \ref{fig:asy_evo}). \\

\item \textbf{At fixed stellar mass, asymmetry is a strong predictor for the star forming main sequence offset, $\dsfms$}. The observational trend between stratification in SFRs and $R_A$ at fixed galaxy stellar mass identified at $z\approx0.1$ by \cite{2021ApJ...923..205Y} is reproduced using asymmetries measured in our TNG50 images (Figure \ref{fig:sfms_ra}).\\

\item \textbf{Major and minor mergers are too rare to be significant contributors to the vertical stratification of the SFMS}. At fixed stellar mass, no trend emerges between $\dsfms$ and the time since/until the nearest merger, $\dtcoal$, for both major or minor mergers in TNG50 (Figure \ref{fig:sfms_dtc}, middle and right panels). \\

\item \textbf{At fixed stellar mass, $\dsfms$ is strongly correlated with the $\dtcoal$ of mini mergers in TNG50} (Figure \ref{fig:sfms_dtc}, left panels) and mini merger frequency (count) over the past 3 Gyr -- mirroring the correlation with asymmetry. The evolution of this trend with galaxy stellar mass is also remarkably similar to the trend with asymmetry.\\

\item Controlling for redshift, stellar mass, environment, and gas fraction, asymmetry offsets $\dra$ between TNG50 SFGs are linked to average $\dsfms$ up to $\pm0.2$ dex (Figure \ref{fig:delsfr_delra}). There is no obvious redshift-evolution in this relationship over the $0.1 \leq z\leq0.7$ redshift interval we examine.\\


\item Offsets in asymmetries between control-matched SFGs are statistically linked to the difference in the time since/until coalescence of their nearest major/minor/mini merger events, $\dtcoaloffset$ (Figure \ref{fig:dra_dtc}). However, owing to higher merger frequencies, mini mergers offer sensitivity to smaller $\dtcoaloffset$ and smaller $\dra$. \\


\item Offsets in SFRs, $\dlogsfr$, between control-matched SFGs are also statistically linked their major/minor/mini merger $\dtcoaloffset$ (Figure \ref{fig:dsfr_dtc}). Again, mini merger $\dtcoaloffset$ is uniquely sensitive to small $\dlogsfr$, which make up the vast majority of SFR offsets between otherwise similar galaxies on the SFMS. \\

\item \textbf{$80$ per cent of TNG50 SFGs with $\dsfms>0.5$ dex (starbursts) satisfy $|\dtcoal|<1.5$ Gyr for \emph{at least} one $\mu\geq0.01$ merger (Figure \ref{fig:sfms_mergers})}. For $\dsfms>0$ the incidence of all $\mu\geq0.01$ mergers increases with $\dsfms$ -- but particularly for minor/major mergers. \\

\item \textbf{At fixed stellar mass, only mini merger incidence is correlated with $\dsfms$ for galaxies currently below the SFMS} (Figure \ref{fig:sfms_mergers}; middle panel)  -- showing a relatively stronger coupling between $\dsfms$ and mini merger activity within the SFMS compared to minor/major mergers.\\

\item \textbf{Individual mini merger events yield statistically enhanced $\dra$ and $\dlogsfr$ compared to matched non-merging controls} (Figure \ref{fig:iso_merg}). Mini mergers are not piggybacking off enhancements from higher-$\mu$ mergers. Nonetheless, the peak enhancement amplitude (near coalescence) increases greatly with mass ratio, as expected.\\

\item \textbf{The longevity of asymmetry/SFR enhancement after coalescence increases with decreasing mass ratio} (Figure \ref{fig:iso_merg}). While major mergers are consistent with their controls after just $1$ Gyr, on average, mini mergers are enhanced in SFR for as long as $3$ Gyr. Sustained SFE enhancements in low-$\mu$ mergers may be due to an interplay between merger mass ratio, peak SFR burst amplitude, and subsequent stellar feedback (Section \ref{sec:iso_sfr}). \\

\item \textbf{A series of consecutive mini merger events must be a viable pathway to high SFG asymmetries in the absence of major/minor mergers}. The average mini merger $\dtcoal$ for SFGs is 1.9 Gyr -- at least a factor of $\sim3$ shorter than the timescale over which asymmetries are enhanced by individual mini mergers in TNG50 (Figure \ref{fig:iso_merg}). Meanwhile, of the $68$ per cent of SFGs with $R_A>0.3$ and a $\mu>0.01$ merger satisfying $|\dtcoal|<1.5$ Gyr, only half are minor/major mergers (Figure \ref{fig:ra_mergers}).\\

\item \textbf{Mini mergers produce more asymmetric structure in SFGs than minor/major mergers (combined) by over a factor of 2} (Figures \ref{fig:ra_mergers} and \ref{fig:frac_asy}) -- integrating asymmetry enhancements from mini/minor/major mergers over $0.1\leq z \leq 0.7$.\\ 

\item \textbf{Over half ($55$ per cent) of the \emph{in-situ} stellar mass formed in mergers with $\mu>0.01$ was formed in mini mergers} (Figure \ref{fig:frac_sfrs}). According to TNG50, mini mergers dominate over major/minor mergers with respect to \emph{in-situ} stellar mass assembly at $z\leq0.7$. 

\end{itemize}

Ambiguity between mini mergers and large-scale gas accretion flows is possible. Assuming that (1) the gravitational impulse from mini mergers do nothing to their host galaxies and (2) they are pure but incomplete tracers of accretion, our mini merger estimates serve as lower limits to the relative role of accretion compared to minor/major mergers. In that case, our results with TNG50 show that major/minor mergers are sub-dominant to accretion in driving star formation and asymmetric structure at $z\leq0.7$. 

The expanded magnitude limits and corresponding merging pair mass ratio sensitivity in next-generation galaxy spectroscopic surveys such as the Wide Area Vista Extragalactic Survey (WAVES; \citealt{2019Msngr.175...46D}) and the 4MOST Hemisphere Survey (4HS; \citealt{2023Msngr.190...46T}) using the 4-meter Multi-Object Spectroscopic Telescope (4MOST; \citealt{2019Msngr.175....3D}) will provide a unique opportunity to compare the characteristics of observed mini pairs in the local Universe with ours in TNG50. As for observationally characterising mini merger remnants, the significant challenge is their identification. The synthetic imaging data set we provide may serve as a useful tool in this regard. If the morphological signatures of mini mergers are statistically distinguishable from those of higher-$\mu$ mergers and accretion, then deep/representation learning models trained on our images may make their classification possible and enable direct comparison with our results. Our synthetic images will be useful to tackle a broad range of problems in galaxy astrophysics.

\section*{Acknowledgements}
We acknowledge the cultural, historical, and natural significance and reverence that the summit of Maunakea has for the indigenous Hawaiian community. We are deeply fortunate and grateful to share in the opportunity to explore the Universe from this mountain. The authors thank Peter Camps for helpful consultations on using \textsc{SKIRT} and Brent Groves for advice on the \texttt{MAPPINGS III} template spectra. We also thank Lawrence Faria, Shoshannah Byrne-Mamahit, Douglas Rennehan, Sara Ellison, Danail Obreshkow, Connor Stone, Ignacio Ferreras, Marc Huertas-Company, Rachel Somerville, Robin Cook, and Chris Power for useful discussions and suggestions. CB gratefully acknowledges support from the Forrest Research Foundation and the Natural Sciences and Engineering Research Council of Canada (NSERC) and their post-doctoral fellowship program (PDF-546234-2020). HMY was supported by JSPS KAKENHI Grant Number JP22K14072 and the Research Fund for International Young Scientists of NSFC (11950410492). KO is supported by JSPS KAKENHI Grant Number JP23KJ1089. DN acknowledges funding from the Deutsche Forschungsgemeinschaft (DFG) through an Emmy Noether Research Group (grant number NE 2441/1-1). JEG acknowledges funding from NSF/AAG (grant numbers 1007094 and 1007052). JDS is supported by JSPS KAKENHI (JP22H01262) and the World Premier International Research Center Initiative (WPI), MEXT, Japan. Kavli IPMU is supported by World Premier International Research Center Initiative (WPI), MEXT, Japan. Computations were performed on the HPC system Vera at the Max Planck Computing and Data Facility as well as the idark HPC system at Kavli IPMU. We thank the Max Planck and Kavli IPMU computing support teams for their valued assistance. This research has made use of the Spanish Virtual Observatory project (\url{https://svo.cab.inta-csic.es}) funded by MCIN/AEI/10.13039/501100011033 through grant PID2020-112949GB-I00. Python packages used in this research include \texttt{Astropy} \url{http://www.astropy.org} \citep{astropy:2013, 2018AJ....156..123A} and Numpy \url{https://numpy.org} \citep{harris2020array}. 

The TNG50 simulation was run with compute time granted by the Gauss Centre for Supercomputing (GCS) under Large-Scale Projects GCS-DWAR on the GCS share of the supercomputer Hazel Hen at the High Performance Computing Center Stuttgart (HLRS)

The Hyper Suprime-Cam (HSC) collaboration includes the astronomical communities of Japan and Taiwan, and Princeton University. The HSC instrumentation and software were developed by the National Astronomical Observatory of Japan (NAOJ), the Kavli Institute for the Physics and Mathematics of the Universe (Kavli IPMU), the University of Tokyo, the High Energy Accelerator Research Organization (KEK), the Academia Sinica Institute for Astronomy and Astrophysics in Taiwan (ASIAA), and Princeton University. Funding was contributed by the FIRST program from the Japanese Cabinet Office, the Ministry of Education, Culture, Sports, Science and Technology (MEXT), the Japan Society for the Promotion of Science (JSPS), Japan Science and Technology Agency (JST), the Toray Science Foundation, NAOJ, Kavli IPMU, KEK, ASIAA, and Princeton University. This paper makes use of software developed for Vera C. Rubin Observatory. We thank the Rubin Observatory for making their code available as free software at \url{http://pipelines.lsst.io/}. This paper is based on data collected at the Subaru Telescope and retrieved from the HSC data archive system, which is operated by the Subaru Telescope and Astronomy Data Center (ADC) at NAOJ. Data analysis was in part carried out with the cooperation of Center for Computational Astrophysics (CfCA), NAOJ. 

\section*{Data and Code Availability}

The IllustrisTNG simulations, including TNG50, are publicly available at \url{www.tng-project.org/data} \citep{2019ComAC...6....2N}. The idealized and survey-realistic image datasets provided in this work are also available at \url{www.tng-project.org/bottrell23}. All other data and code related to this publication, including bespoke TNG subhalo, environment, merger history, and morphological measurement catalogues are available upon request to the corresponding author.



\bibliographystyle{mnras}
\bibliography{References} 

\bsp	
\label{lastpage}
\end{document}